\documentclass[l]{article}

\usepackage{amssymb,amsmath,array,graphicx,amsthm}
\usepackage{float}
\usepackage{soul}
\usepackage{pbox, color}
\usepackage{rotating}
\usepackage{tabularx,booktabs}
\usepackage [margin=0.75 in] {geometry}
\usepackage[utf8]{inputenc}
\usepackage{subcaption}
\usepackage{url}
\usepackage{fix-cm}

\begin{document}

\title{Living in a Low Density Black Hole, Non-Expanding Universe                         -- Perhaps a Reflecting Universe}
\author{Bernard McBryan}

\maketitle
\abstract{What is the average density of a black hole, assuming its spin can prevent it from collapsing into a singularity? For stellar black holes, the average density is incredibly dense and has over a trillion G force and tidal force that will rip almost anything apart at the black hole boundary.  

Surprisingly, the average density decreases dramatically for massive black holes.  A black hole of 387 million solar masses would have the average density of water and would be comparable to a giant water balloon extending from the sun almost to Jupiter. A black hole of 11 billion solar masses would have the average density of air and would be analogous to a giant air filled party balloon extending 2.5 times farther out than Pluto. The average mass density in space itself, however small, eventually can become a low density black hole.   If the average density of the universe matches the critical density of just 5.67 hydrogen atoms per cubic meter, it would form a Schwarzschild low density black hole of approximately 13.8 billion light years, matching the big bang model of the universe.  

A black hole can use rotation and/or charge to keep from collapsing.  The G and tidal forces become negligible for large low density black holes.  Thus one can be living in a large low density black hole and not know it.  Further analysis about the critical density rules out an expanding universe and disproves the big bang theory.  Higher densities could produce a much smaller reflecting universe.}
\section*{Overview}
A overview of this paper is as follows:
\begin{enumerate}
\item Black Hole Definition and Equations
\item Low Density Black Hole Calculations
\item Candidate Localized Black Holes
\item Expanding Universe and Disproof of the Big Bang Theory
\item Non-Expanding Universe Options
\item Non-Collapsing Black Hole via Rotation
\item Sources of Redshifts
\item Non-Reflective Universe
\item Reflective Universes
\item Quantized redshift
\item Conclusion
\item Future investigations
\item Appendix A. Background on Black Holes
\item Appendix B. Acronym Table 
\item References
\end{enumerate}

\section{Black Hole Definition and Equations}
There are many, slightly different, definitions of black holes (Classical Schwarzschild, Classical finite height, relativistic Schwarzschild, Reissner-Nordström charged, Kerr spinning/rotating, and Kerr-Newman charged spinning/rotating, and quantum black holes).  Some of these definitions are radically different but use identical names and abbreviations.  Thus, it is a good idea to define terms to achieve a common language and understanding. A literature survey of black holes is summarized in Figure \ref{BHRadii} and Table \ref{TableBlackHoleRadiiEquations1}, and discussed further in Appendix A.
\begin{figure}[H]
\begin{center}
\includegraphics[width=.9\textwidth]{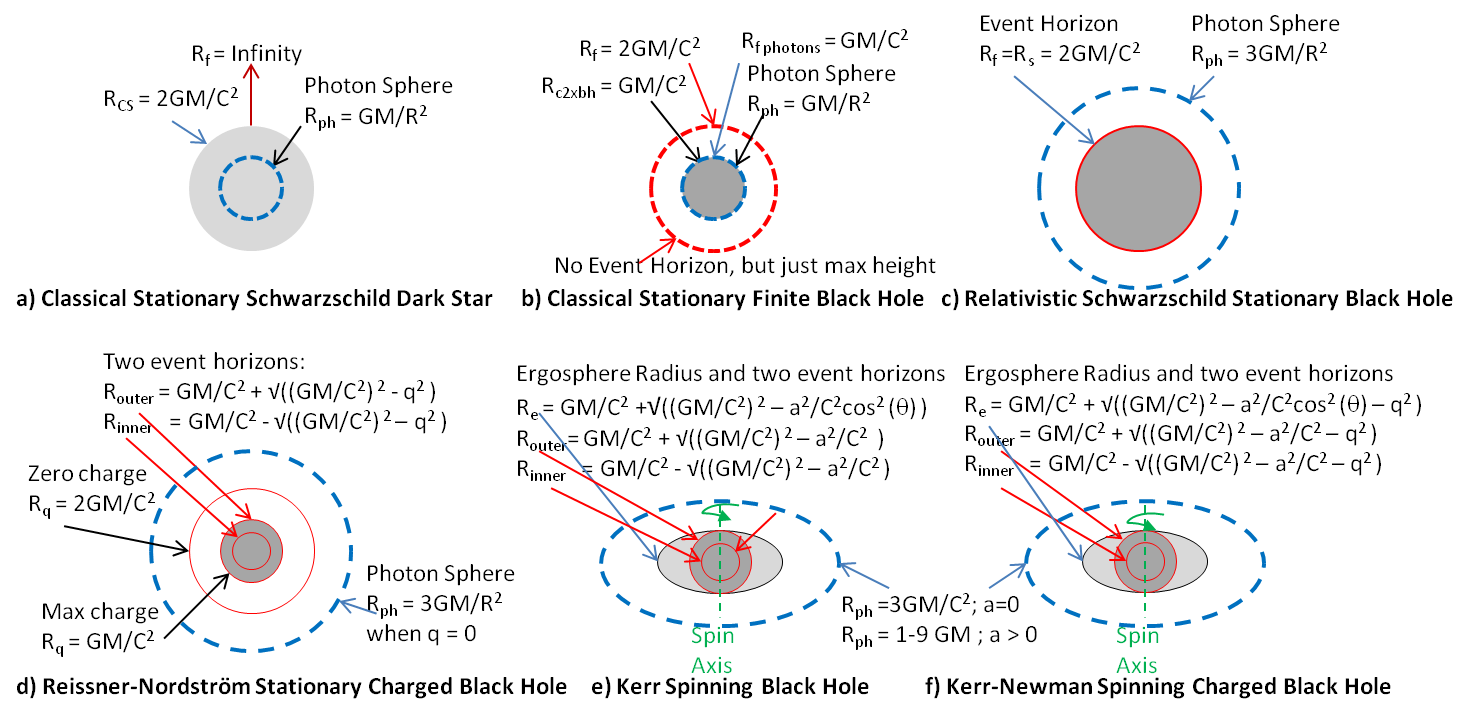}
\captionsetup{width=0.9\textwidth}
\caption{Black Hole Radii for Different Black Hole Types. All black holes have a radius between $GM/C^2$ and $2GM/C^2$\cite{kerr},\cite{KerrBH1},\cite{KerrBH2},\cite{kerrNewman},\cite{Zienikov},\cite{ChargedPH},\cite{Romero}, and \cite{QuantumBH}.  The classical Schwarzschild black hole most often derived in text books (was originally called a dark star by Michell/Laplace 1796) can keep light from escaping, but light can be seen at infinity ($R_f=\infty$).   A classical finite (twice) height black hole with final height $R_f=2GM/C^2$ has a radius $R_{c2xbh}=GM/C^2$.  All the remaining black holes would also not let light escape beyond a maximum height of $R_f = 2GM/C^2$.  All these black holes have photon sphere radii, $R_{ph}$ that bends light in circular orbits $R_{ph} = GM/C^2$ to $3GM/C^2$ (and larger) for spinning black holes.  All the stationary black holes (on the top) are expected to collapse into a singularity at the center. However, the spinning and charged black holes (on the bottom) can keep themselves from collapsing using spin or a repulsive charge force.  A spinning black hole is larger at its equator just like the spinning earth due to centrifugal forces (forming an elliptical shape called an ergosphere).  Spinning black holes also drags space around at its event horizon.   If one incorporates quantum mechanics into the black hole equations, one would get identical answers but some of the mass and radii values may need to come in discrete legal values, possibly at resonance frequencies with the gravitational waves.}
\label{BHRadii}
\end{center}
\end{figure}

\begin{table}
\centering
\caption{Black Hole Radii Equations}
\begin{center}
\bigskip
\noindent\makebox[\textwidth]{%
\begin{tabular}{| l | c | c | c | c | }
  \hline  
\pbox{10cm}{$Black$ Hole \\Description} & \pbox{10cm}{Event Horizon Radius \\(or Ergosphere Radius)} & \pbox{10cm}{Maximum Height \\$R_f$} & \pbox{10cm}{Photon Sphere\\$R_{ph}$} \\
\hline 
Classical Schwarzschild & $R_{cs} = 2GM/C^2$ &	$R_f=Infinity$ &	$R_{ph} = GM/C^2$\\ \hline 
Classical finite (twice) height & $R_{c2xbh} = GM/C^2$ & $R_{f} = 2GM/C^2$ & $R_{ph} = GM/C^2$\\ \hline 
Relativistic Schwarzschild &$R_{s} = 2GM/C^2$&$R_{f} = 2GM/C^2$&	$R_{ph} = 3GM/C^2$\\ \hline
Relativistic Charged &$R_{q} = GM/C^2 +/- \sqrt((GM/C^2)^2 -q^2)$&$R_{f} = 2GM/C^2$&	$R_{ph} = 3GM/C^2$\\ \hline
Relativistic Spinning Kerr &$R_{spin} = GM/C^2 + \sqrt((GM/C^2)^2 - a^2/C^2*cos^2(\theta))$&$R_{f} = 2GM/C^2$&	$R_{ph} = 1-9GM/C^2$\\ \hline
Relativistic Spinning Charged &$R_{spinq} = GM/C^2 + \sqrt((GM/C^2)^2 - a^2/C^2*cos^2(\theta) -q^2)$&$R_{f} = 2GM/C^2$&	$R_{ph} = 1-9GM/C^2$\\ \hline
\end{tabular}} 
\bigskip
\end{center}
\label{TableBlackHoleRadiiEquations1}
\end{table}

Although each black hole described above are computed differently under various conditions, their black hole radii are all between $GM/C^2$ and $2GM/C^2$; and their maximum final height $R_f$ is less than or equal to $2GM/C^2$.  They each have a photon sphere between $GM/C^2$ and $3GM/C^2$ to $9GM/C^2$ where they can bend light up to 360 degrees.   The low density black holes described throughout this paper will use this range of black hole radii and applies to both the classical and relativistic black hole definitions.  Throughout the remainder of the text, $R_s=2GM/C^2$ will refer to the larger Relativistic Schwarzschild radius and $R_{bh}=GM/C^2$ will used to represent all the black holes with the smaller radii, but mainly targeting a maximum spinning or charged black hole.  But reality may be somewhere between the two based on a lesser spin rate or charge of the black hole. 

\section{Low Density Black Hole Calculations}
The larger and more massive black holes have significantly smaller average densities.  All of the black hole radii discussed previously are between ($R_s = \frac{2GM}{C^2} \text{ and  }(R_{bh} =\frac{GM}{C^2}$).  The mass of each of these black holes is proportional to their Radius (since $2, G, C^2$ are all constants).   Thus, the mass of each black hole increases proportionally with its radius, but its volume ($Vol$) grows proportionally to its radius cubed (since $Vol= \frac{4\pi}{3}R^3$). Density is mass divided by volume. Thus the average mass density,  ($P_{avg}=\frac{M}{Vol}$) is inversely proportional to the radius squared (e.g. $1/R^2$).
 
The Relativistic Schwarzschild Black Hole ($R_s=2\frac{GM}{C^2}$) with assumed spherical Volume ($Vol = \frac{4\pi}{3}R_s^3$) has the following average mass density. 
\begin{align}
M &= \frac{R_sC^2}{2G} \\
P_{avg} &= \frac{M}{Vol} \\
P_{avg} &= \frac{R_sC^2/(2G)}{4/3\pi R_s^3} \\
P_{avg} &= \frac{3C^2}{8 \pi G R_s^2} \label{RsDenFun}
\end{align} 
Using  $C = 3*10^8 m/sec$ and $G = 6.67384*10^{-11}N(m/kg)^2$ the average mass density can be expressed in $kg/m^3$. 
\begin{align}
P_{avg} &= \frac{1.61*10^{26}}{R_s^2} kg/m^3 \qquad (R_s \text{ in meters}) \\
P_{avg}  &= \frac{1.797*10^{-6}}{R_s^2}  kg/m^3 \qquad (R_s \text{ in LY})
\end{align}
Solving for $R_s$ yields the following.
\begin{equation}
R_s = \frac{.0013401}{\sqrt{P_{avg}}} LY \qquad (P_{avg} \text{ in } kg/m^3) 
\end{equation}

Similarly, for the other black holes, ($R_{bh}=\frac{GM}{C^2}$) with assumed spherical Volume ($Vol = \frac{4\pi}{3}R_{bh}^3$) has the following average mass density. 
\begin{align}
M &= \frac{R_sC^2}{G} \\
P_{avg} &= \frac{M}{Vol} \\
P_{avg} &= \frac{R_{bh}C^2/G}{4/3\pi R_{bh}^3} \\
P_{avg} &= \frac{3C^2}{4 \pi G R_{bh}^2} \label{densityEquation}
\end{align} 
Using  $C = 3*10^8 m/sec$ and $G = 6.67384*10^{-11}N(m/kg)^2$ the average mass density can be expressed in $kg/m^3$. 
\begin{align}
P_{avg} &= \frac{3.22*10^{26}}{R_{bh}^2}  kg/m^3 \qquad   (R_{bh} \text{ in meters}) \\
P_{avg}  &= \frac{3.592*10^{-6}}{R_{bh}^2}  kg/m^3 \qquad (R_{bh} \text{ in LY})
\end{align}
Solving for $R_{bh}$ yields the following.
\begin{equation}
R_{bh} = \frac{.0018952}{\sqrt{P_{avg}}} LY \qquad (P_{avg} \text{ in } kg/m^3) 
\end{equation}

Black hole characteristics are tabulated in Tables \ref{TableSameDen} and \ref{TableSameMass} for various mass densities, $R_{bh}$ and $R_s$ radii, mass, $Gs$, and tidal force $dGs$.    The average densities are listed in $kg/m^3$ and in equivalent hydrogen atoms per cubic volume.  Table \ref{TableSameDen} contains black holes data with common densities.  Table \ref{TableSameMass} contains black hole data with common mass.  
Note that within each row of Table \ref{TableSameDen}, for a given common density, $R_s$ is .707 times the size of $R_{bh}$, and .707 times the Mass $M$.  Note that within each row of Table \ref{TableSameMass}, for a given common Mass, $R_s = 2R_{bh}$, and thus $P_{avg}$ decreases by 8 for the same mass.
\begin{table}
\centering
\caption{Black Hole Characteristics with Same Densities}
\begin{center}
\bigskip
\noindent\makebox[\textwidth]{%
\begin{tabular}{| l | c | c | c | c | c | c |c| c| }
  \hline  
\pbox{10cm}{Similar \\density}&\pbox{10cm}{Density \\Atoms/Vol} & \pbox{10cm}{Density \\($kg/m^3$)} & \pbox{20cm}{Radius \\$R_{bh}$} &\pbox{20cm}{Radius \\$R_s$ } &\pbox{20cm}{Mass $R_{bh}$ \\ (SM)}&\pbox{20cm}{Mass $R_s$\\ (SM)}&\pbox{20cm}{$G_s \text{ at }R_{bh}$ \\ $\frac{GM/R^2}{10}$}&\pbox{20cm}{Tidal \\ Force \\ $2GM/R^3$} \\
\hline 
1 SM black hole &	89 /$fm^3$&	1.5E+20&	1.48 Km&	1.04 Km&	1.0E+00&	0.707&	6.E+12&	5.E+16\\ \hline  
1000 SM black hole&	88530 /$pm^3$&	1.5E+14&	1.48 Mm&	1.04 Mm&	1.0E+03&	707&	6.E+09&	5.E+10\\ \hline
1 Million SM black hole&	0.09 /$pm^3$&	1.5E+08&	1.48 Bm&	1.04 Bm&	1.0E+06&	7.1E+05&	6.E+06&	5.E+04\\ \hline
Water (1000kg/$m^3$)\cite{zeilik56}&	598.8 /$nm^3$&	1.0E+03&	3.79 AU&	2.68 AU&	3.8E+08&	2.7E+08&	2.E+04&	4.E-01\\ \hline
1 B SM black hole&	88.53 /$nm^3$&	1.5E+02&	9.9 AU&	7 AU&	1.0E+09&	7.1E+08&	6.E+03&	5.E-02\\ \hline
Air (1.2 kg/$m^3$)&	0.74 /$nm^3$&	1.2E+00&	108 AU&	76 AU&	1.1E+10&	7.8E+09&	6.E+02&	4.E-04\\ \hline
1 T SM black hole&	88530 /$\mu m^3$&	1.5E-04&	0.16 LY&	0.11 LY&	1.0E+12&	7.1E+11&	6.E+00&	5.E-08\\ \hline
2000 atoms / $\mu m^3$&	2000 /$\mu m^3$&	3.3E-06&	1.04 LY&	0.73 LY&	6.7E+12&	4.7E+12&	9.E-01&	1.E-09\\ \hline
One atom / $\mu m^3$&	1 /$\mu m^3$&	1.7E-09&	46.4 LY&	32.8 LY&	3.0E+14&	2.1E+14&	2.E-02&	6.E-13\\ \hline
GC (70 M SM/$pc^3$)&	0.0217 /$\mu m^3$&	3.6E-11&	315 LY&	222.74 LY&	2.0E+15&	1.4E+15&	3.E-03&	1.E-14\\ \hline
MW Disk*2180&	850 /$mm^3$&	1.4E-15&	50 KLY&	36 KLY&	3.2E+17&	2.3E+17&	2.E-05&	5.E-19\\ \hline
MW Disk*128&	50 /$mm^3$&	8.4E-17&	207 KLY&	147 KLY&	1.3E+18&	9.4E+17&	5.E-06&	3.E-20\\ \hline
One atom / $mm^3$&	1 /$mm^3$&	1.7E-18&	1.47 MLY&	1.04 MLY&	9.4E+18&	6.7E+18&	6.E-07&	6.E-22\\ \hline
MW Disk*10&	0.39 /$mm^3$&	6.6E-19&	2.3 MLY&	1.65 MLY&	1.5E+19&	1.1E+19&	4.E-07&	2.E-22\\ \hline
MW Disk*2&	86.03 /$cm^3$&	1.4E-19&	5 MLY&	3.54 MLY&	3.2E+19&	2.3E+19&	2.E-07&	5.E-23\\ \hline
Solar region 1SM,2LY&	41.84 /$cm^3$&	7.0E-20&	7.2 MLY&	5.1 MLY&	4.6E+19&	3.3E+19&	1.E-07&	2.E-23\\ \hline
MW Disk&	40.08 /$cm^3$&	6.7E-20&	7.3 MLY&	5.2 MLY&	4.7E+19&	3.3E+19&	1.E-07&	2.E-23\\ \hline
MW Galaxy*10                &     	5.36 /$cm^3$&	9.0E-21&	20 MLY&	14.2 MLY&	1.3E+20&	9.1E+19&	5.E-08&	3.E-24\\ \hline
One atom / $cm^3$&	1 /$cm^3$&	1.7E-21&	46.4 MLY&	32.8 MLY&	3.0E+20&	2.1E+20&	2.E-08&	6.E-25\\ \hline
MW:1e12 SM,50 KPC \cite{zeilik280}&	0.62 /$cm^3$&	1.0E-21&	59 MLY&	42 MLY&	3.8E+20&	2.7E+20&	2.E-08&	4.E-25\\ \hline
MW;2e11 SM,50KLY\cite{lin}&	0.54 /$cm^3$&	9.0E-22&	63 MLY&	45 MLY&	4.1E+20&	2.9E+20&	2.E-08&	3.E-25\\ \hline
MW;4e11 SM,50KPC\cite{zeilik447}&	0.25 /$cm^3$&	4.1E-22&	93 MLY&	66 MLY&	6.0E+20&	4.2E+20&	1.E-08&	1.E-25\\ \hline
Intergalactic gas&	0.1 /$cm^3$&	1.7E-22&	147 MLY&	104 MLY&	9.4E+20&	6.6E+20&	6.E-09&	6.E-26\\ \hline
325 MLY Radius &	0.02 /$cm^3$&	3.4E-23&	325 MLY&	230 MLY&	2.1E+21&	1.5E+21&	3.E-09&	1.E-26\\ \hline
750 MLY Radius&	0.0038 /$cm^3$&	6.4E-24&	750 MLY&	530 MLY&	4.8E+21&	3.4E+21&	1.E-09&	2.E-27\\ \hline
LSubGrp3TSM,1.5Mly\cite{zeilik447}&	298 /$m^3$&	5.0E-25&	2.7 BLY&	2 BLY&	1.7E+22&	1.2E+22&	4.E-10&	2.E-28\\ \hline
Rs Critical Density\cite{CriticalDen}	&	5.67 /$m^3$&	9.5E-27&	19.5 BLY&	13.8 BLY&	1.2E+23&	8.8E+22&	5.E-11&	3.E-30\\ \hline
Rs Critical Density*2&	11.34 /$m^3$&	1.9E-26&	13.8 BLY&	9.7 BLY&	8.8E+22&	6.2E+22&	7.E-11&	7.E-30\\ \hline
Ra Critical Density *8&	45.36 /$m^3$&	7.6E-26&	6.9 BLY&	4.9 BLY&	4.4E+22&	3.1E+22&	1.0E-10&	3.E-29\\ \hline
LSubGrp.75TSM,Mpc \cite{zeilik447} &	58 /$m^3$&	9.7E-26&	6.1 BLY&	4.3 BLY&	3.9E+22&	2.8E+22&	2.E-10&	3.E-29\\ \hline
LocalGrp; 3TSM,5Mly\cite{atlas2}&	8.05 /$m^3$&	1.3E-26&	16.3 BLY&	12 BLY&	1.0E+23&	7.4E+22&	6.E-11&	5.E-30\\ \hline
LocalGrp.75TSM,5Mly\cite{atlas2}&	2 /$m^3$&	3.4E-27&	32.7 BLY&	23 BLY&	2.1E+23&	1.5E+23&	3.E-11&	1.E-30\\ \hline
LocalSuperCluster\cite{atlas1}&	2 /$m^3$&	3.4E-27&	32.7 BLY&	23 BLY&	2.1E+23&	1.5E+23&	3.E-11&	1.E-30\\ \hline
One atom / $m^3$&	1 /$m^3$&	1.7E-27&	46.4 BLY&	32.8 BLY&	3.0E+23&	2.1E+23&	2.E-11&	6.E-31\\ \hline
LocalSuperCluster&	0.34 /$m^3$&	5.6E-28&	80 BLY&	57 BLY&	5.1E+23&	3.6E+23&	1.E-11&	2.E-31\\ \hline
Visible Density of U.	\cite{VisibleDen}&	0.18 /$m^3$&	3.0E-28&	109 BLY&	77 BLY&	7.0E+23&	4.9E+23&	9.E-12&	1.E-31\\
\hline
\end{tabular}}
\bigskip
\end{center}
\label{TableSameDen}
\end{table}

\begin{table}
\centering
\caption{Black Hole Characteristics with Same Mass}
\begin{center}
\bigskip
\noindent\makebox[\textwidth]{%
\begin{tabular}{| l | c | c | c | c | c | c | c |}
  \hline  
\pbox{10cm}{Similar density \\ as $R_{bh}$}&\pbox{10cm}{Density $P_{bh}$ \\Atoms/Vol} & \pbox{10cm}{Density $P_s$\\Atoms/Vol} & \pbox{20cm}{Density $P_{bh}$\\ ($kg/m^3$)} &\pbox{20cm}{Density $P_s$ \\  ($kg/m^3$)} &\pbox{20cm}{Radius \\ $R_{bh}$}&\pbox{20cm}{Radius\\ $R_s$}&\pbox{20cm}{Mass\\(SM)} \\
\hline 
1 SM black hole &	89 /$fm^3$&	11 /$fm^3$&	1.5E+20&	1.8E+19&	1.48 Km&	2.95 Km&	1.0E+00\\ \hline
1000 SM black hole&	88530 /$pm^3$&	11066 /$pm^3$&	1.5E+14&	1.8E+13&	1.48 Mm&	2.95 Mm&	1.0E+03\\ \hline
1 Million SM black hole&	0.09 /$pm^3$&	0.01 /$pm^3$&	1.5E+08&	1.8E+07&	1.48 Bm&	2.95 Bm&	1.0E+06\\ \hline
Water (1000kg/$m^3$)\cite{zeilik56}&	598.8 /$nm^3$&	74.9 /$nm^3$&	1.0E+03&	1.3E+02&	3.79 AU&	7.59 AU&	3.8E+08\\ \hline
1 B SM black hole&	87.82 /$nm^3$&	11.07 /$nm^3$&	1.5E+02&	1.8E+01&	9.9 AU&	19.7 AU&	1.0E+09\\ \hline
Air (1.2 kg/$m^3$)&	0.74 /$nm^3$&	0.09 /$nm^3$&	1.2E+00&	1.5E-01&	108 AU&	216 AU&	1.1E+10\\ \hline
1 T SM black hole&	88530 /$\mu m^3$&	11066 /$\mu m^3$&	1.5E-04&	1.8E-05&	0.16 LY&	0.31 LY&	1.0E+12\\ \hline
2000 atoms / $\mu m^3$&	2000 /$\mu m^3$&	250 /$\mu m^3$&	3.3E-06&	4.2E-07&	1.04 LY&	2.07 LY&	6.7E+12\\ \hline
One atom / $\mu m^3$&	1 /$\mu m^3$&	0.13 /$\mu m^3$&	1.7E-09&	2.1E-10&	46.4 LY&	92.8 LY&	3.0E+14\\ \hline
GC (70 M SM/pc3)&	0.0217 /$\mu m^3$&	0.0027 /$\mu m^3$&	3.6E-11&	4.5E-12&	315 LY&	630.02 LY&	2.0E+15\\ \hline
MW Disk*2180&	864.55 /$mm^3$&	108.1 /$mm^3$&	1.4E-15&	1.8E-16&	50 KLY&	100 KLY&	3.2E+17\\ \hline
MW Disk*128&	50 /$mm^3$&	6.25 /$mm^3$&	8.4E-17&	1.0E-17&	207 KLY&	415 KLY&	1.3E+18\\ \hline
One atom / $mm^3$&	1 /$mm^3$&	0.13 /$mm^3$&	1.7E-18&	2.1E-19&	1.47 MLY&	2.93 MLY&	9.4E+18\\ \hline
MW Disk*10&	0.39 /$mm^3$&	0.05 /$mm^3$&	6.6E-19&	8.2E-20&	2.3 MLY&	4.68 MLY&	1.5E+19\\ \hline
MW Disk*2&	86.03 /$cm^3$&	10.75 /$cm^3$&	1.4E-19&	1.8E-20&	5 MLY&	10 MLY&	3.2E+19\\ \hline
Solar region 1SM,2LY&	41.84 /$cm^3$&	5.23 /$cm^3$&	7.0E-20&	8.7E-21&	7.2 MLY&	14.3 MLY&	4.6E+19\\ \hline
MW Disk&	40.08 /$cm^3$&	5.01 /$cm^3$&	6.7E-20&	8.4E-21&	7.3 MLY&	14.7 MLY&	4.7E+19\\ \hline
MW Galaxy*10&                     	5.36 /$cm^3$&	0.67 /$cm^3$&	9.0E-21&	1.1E-21&	20 MLY&	40 MLY&	1.3E+20\\ \hline
One atom / $cm^3$&	1 /$cm^3$&	0.13 /$cm^3$&	1.7E-21&	2.1E-22&	46.4 MLY&	92.8 MLY&	3.0E+20\\ \hline
MW:1e12 SM,50 KPC \cite{zeilik280}&	0.62 /$cm^3$&	0.08 /$cm^3$&	1.0E-21&	1.3E-22&	59 MLY&	118 MLY&	3.8E+20\\ \hline
MW;2e11 SM,50KLY\cite{lin}&	0.54 /$cm^3$&	0.07 /$cm^3$&	9.0E-22&	1.1E-22&	63 MLY&	127 MLY&	4.1E+20\\ \hline
MW;4e11 SM,50KPC\cite{zeilik447}&	0.25 /$cm^3$&	0.03 /$cm^3$&	4.1E-22&	5.2E-23&	93 MLY&	186 MLY&	6.0E+20\\ \hline
Intergalactic gas&	0.1 /$cm^3$&	0.01 /$cm^3$&	1.7E-22&	2.1E-23&	147 MLY&	293 MLY&	9.4E+20\\ \hline
325 MLY Radius &	0.02 /$cm^3$&	0.0025 /$cm^3$&	3.4E-23&	4.3E-24&	325 MLY&	650 MLY&	2.1E+21\\ \hline
750 MLY Radius&	0.0038 /$cm^3$&	0.0005 /$cm^3$&	6.4E-24&	8.0E-25&	750 MLY&	1500 MLY&	4.8E+21\\ \hline
LSubGrp3TSM,1.5MLY\cite{zeilik447}&	298 /$m^3$&	37.25 /$m^3$&	5.0E-25&	6.2E-26&	2.7 BLY&	5 BLY&	1.7E+22\\ \hline
Rs Critical Density\cite{CriticalDen}	&	45.36 /$m^3$ &	5.67 /$m^3$&	7.6E-26&	9.5E-27&	6.9 BLY&	13.8 BLY&	4.4E+22\\ \hline
LSubGrp.75TSM,Mpc\cite{zeilik447}&	58 /$m^3$&	7 /$m^3$&	9.7E-26&	1.2E-26&	6.1 BLY&	12.2 BLY&	3.9E+22\\ \hline
Rs Critical Density*2&	90.72 /$m^3$&	11.34 /$m^3$&	1.5E-25&	1.9E-26&	4.9 BLY&	9.7 BLY&	3.1E+22\\ \hline
LocalGrp; 3TSM,5MLY\cite{atlas2}&	8.05 /$m^3$&	1.01 /$m^3$&	1.3E-26&	1.7E-27&	16.3 BLY&	33 BLY&	1.0E+23\\ \hline
LocalGrp.75TSM,5MLY \cite{atlas2}&	2 /$m^3$&	0.25 /$m^3$&	3.4E-27&	4.2E-28&	32.7 BLY&	65 BLY&	2.1E+23\\ \hline
LocalSuperCluster&	2 /$m^3$&	0.25 /$m^3$&	3.4E-27&	4.2E-28&	32.7 BLY&	65 BLY&	2.1E+23\\ \hline
One atom / $m^3$&	0.98 /$m^3$&	0.12 /$m^3$&	1.6E-27&	2.1E-28&	46.8 BLY&	93.5 BLY&	3.0E+23\\ \hline
LocalSuperCluster\cite{atlas1}&	0.34 /$m^3$&	0.04 /$m^3$&	5.6E-28&	7.0E-29&	80 BLY&	160 BLY&	5.1E+23\\ \hline
Visible Density of U.	\cite{VisibleDen}&	0.18 /$m^3$&	0.02 /$m^3$&	3.0E-28&	3.8E-29&	109 BLY&	218 BLY&	7.0E+23\\
\hline
\end{tabular}}
\bigskip
\end{center}
\label{TableSameMass}
\end{table}
\vspace{3 mm}
The $Gs$ force was computed using the following equation \cite{zeilikP3}. 
\begin{equation} 
Gs(R)= \frac{GM}{9.8R^2} \qquad (\text{in earth } Gs )
\end{equation}

Any $G$ force, even the 3 trillion $Gs$ would not be felt in free fall or while in orbit around the black hole.  However, the tidal forces, the amount of $dGs$ pulling on the center of the planet (or body) versus the outer side could rip one apart. Let $\Delta r$ be the distance between the center and the outer side. Tidal forces across an earth size planet were computed using the following equation \cite{zeilik46},
\begin{equation}
\Delta Gs(R)=Gs(R)-Gs(R+R_{earth}) =\pm \frac{2GM \Delta r}{R^3}    
\end{equation}
where $R_{earth}=\Delta r= 6353000m$. 
This table does not represent a new fundamental equation, but is just deriving $R_s$ and $R_{bh}$ from density before computing the other parameters.  Identical results can be found computing the equivalent density directly from the traditional black hole equations $R_s = 2GM/C^2$ and $R_{bh}=GM/C^2$.  That is, given a mass $M$, compute $R_s$ and $R_{bh}$, and then compute the Volume using $Vol=4/3 \pi R^3$, then compute density from $P=M/Vol$  in $kg/m^3$.  In fact, Table \ref{TableSameMass} was computed in this fashion directly from the mass, and Table \ref{TableSameDen} was derived from the mass densities, with identical results.

\subsection{High Density Stellar Black Holes ($R=1.5*10^3$ - $3*10^9 m$; $M=1$ - $10M SM$)}

The data starts for a 1 SM stellar black hole (although in nature stellar black holes are believed to start at 2.8 SMs). The mass density of the stellar black holes are listed at the top of Tables \ref{TableSameDen} and \ref{TableSameMass} with masses of 1, 1000, and 1 million solar masses (SMs).  Their densities are huge, because their large mass (e.g. 1 solar mass) is crammed into very small radii (e.g 1.5 Km).   

Note, that as the Mass increases by 1000 between the first three rows in  Tables \ref{TableSameDen} and \ref{TableSameMass}, the Radii also increase by 1000, since both $R_s$ and $R_{bh}$ are proportional to $M$.  The density values drop very quickly between these three rows (1 million fold with each 1000 increase in mass).  Although their Mass increases 1000 fold, their radii also increase 1000 fold, and their volume (not shown) increase 1 billion fold since volume is proportional to $R^3$.  Since density = mass/volume; the 1000 fold increase in mass is over come by the 1 billion fold increase in volume, resulting in a million fold reduction in density.  Thus, although the mass is increasing, the density decreases due to its much larger volume.

The 1 SM black hole is very dense since it contains 1 solar mass  in a very small 1.5km radius.  It contains an equivalent mass of 11 to 89 hydrogen atoms in a cubic femtometer ($fm =1*10^{-15}$ meters). Since the volume of a neutron is about 1.76 $fm^3$, the 1 SM finite black hole is 48 times denser than a neutron (or neutron star).   The $R_{bh}$ radius of a 1 K SM black hole would be 1500 km, which is just slightly smaller than the radius of the moon.   The $R_{bh}$  radius of a 1 M SM black hole would be 1.5 million km, which is about twice the size of the sun’s radius, but contains the mass of a million suns.  Thus, stellar black holes are small and extremely dense. 

For the 1 solar mass black hole, the $G$ force is over 6 trillion Gs and its tidal force is $5*10^{16} dGs$ and would rip anything apart crossing the black hole boundary.  However, the tidal force drops off quickly and is only 10 M $dGs$ for the 1000 SM and 50,000 $dGs$ for the 1 million SM stellar black holes. These high tidal dG forces around stellar black holes are believed to be the source of X-Ray emissions due to the high energy release of ripping apart atoms from material crossing the boundary of stellar black holes.

Possible candidates for stellar sized high density black holes, within the Milky Way includes Sagittarius A*  as well as smaller black holes believed to be within the globular clusters.  Sagittarius A* is believed to be a 4 M SM black hole within the core of the Milky Way Galaxy. Other Million SM black holes are reported in the cores of other external galaxies.

\subsection{Solar System Size Black Holes ($R =3.79$ - $216 AU$; $M = 387M$ - $11B SM$)}
With a common density of water from Table \ref{TableSameDen}, a Schwarzschild black hole would have a radius of 2.68 AU and a mass of 270 BSM (billion solar masses).  A maximum spinning or charged black hole would have a radius of 3.79 AU and a mass of 387 B SM.     An Astronomical Unit (AU) is the orbital distance of the earth around the sun which is $1.496*10^{11}$ meters.  At 2.68 AU, the Schwarzschild radius would be out past twice the orbital radius of Mars around the sun and at 3.79 AU the spinning or charged black hole radius would be almost out to the orbital radius of Jupiter around the sun.   The G force drops to 20,000 $Gs$ for the spinning or charged black hole and the tidal force drops to 0.4 $dGs$.   

Using Table \ref{TableSameMass} the density of the 1 B SM finite black hole would be about $1/6^{th}$ the density of water and be about 10 AU in radius (just past the orbital radius of Saturn).  The density of the 1 B SM Schwarzschild black hole would be $1/50^{th}$ the density of water and have $R_s=20$ AU, and would extend out just past the orbital radius of Uranus.  The G force drops to 6000 $Gs$ for the spinning or charged black hole and the tidal force drops to $1/20^{th}$ of a $dG$.  

With the common density of air from Table \ref{TableSameDen},  black holes would have $R_s$  and $R_{bh}$ radii of 76 and 108 AU, which are about 2 to 2.5 times the average orbital radius of Pluto around the sun and have masses of 7.8 B and 11 B SM. The $G$ force drops to 600 $Gs$ and the tidal force is .0004 $dGs$.  

Since the tidal forces of these solar system size black holes are not unreasonable, they may not emit x-rays. Possible candidates for solar system sized moderate density black holes are dark nebula gas clouds, the very center of galactic cores, and quasars.
\subsection{Light Year Size Black Holes ($R = .16$ to $100 LY$; $M = 1T$ to $300T SM$)}
Using Table \ref{TableSameMass} the trillion SM black holes, (i.e. million million), have $R_{bh}$ and $R_s$ radii of .16 and .31 LY and have densities of $1.5*10^{-4}$ $kg/m^3$ to $1.8*10^{-5} kg/m^3$, which is 8,000-64,000 times less dense than air.  The $G$ force is just 6 $Gs$ and the tidal force has become $.5*10^{-9}$ (nano) $dGs$.  Thus, one could orbit very close to this black hole without being ripped apart. 

With the common density of 2000 atoms per cubic micro meter ($\mu m^3$) from Table \ref{TableSameDen},  black holes would have $R_s$ and $R_{bh}$ radii of .7 and 1 LY, and have masses of about 4.7 and 6.7 T SM (which represents more mass than current mass estimates for the entire Milky Way galaxy).  The $G$ force of this black hole would $G=.9Gs$, and the tidal force would be a billionth $dG$.

With the common density of one atom per cubic micro meter ($\mu m^3$) from Table \ref{TableSameDen},  black holes would have $R_s$ and $R_{bh}$ radii of 33 and 46 LY, and have masses of about 200 and 300 T SM (which represents 200 to 300 times the current mass estimates for the entire Milky Way galaxy).  This seems impossible at first, but it is only equivalent to 1 hydrogen atom per cubic $\mu$m.   One hydrogen atom per cubic micrometer would still be about 1 billionth of the density of air and thus would not be out of the question. The $G$ force of this black hole would be $.02Gs$, and the tidal force would be $6*10^{-13} dGs$.  

Possible candidates for light year sized low density black holes are dark nebula gas clouds, globular clusters (if they have additional dark matter), or the entire galactic core of a very large galaxy, (perhaps with additional dark matter thrown in).

\subsection{Galaxy Size Black Holes ($R=50K$ - $500KLY$; $M=3*10^{17}$ - $3*10^{18}SM$)}

If the average density is equivalent to about 125 to 2180 times the density of the Milky Way disk, from Table \ref{TableSameMass}, the $R_{bh}$ radii becomes 50K and 200K LY respectively.  This would be small enough to just encompass most galaxies.   From most galaxy speed curves, this seems improbable, unless the speed curves are only measuring the mass of the galactic disk, and most of the mass is in the outer edges of the galaxy. The average density would only be equivalent to  850 to 50 hydrogen atoms per cubic millimeters ($mm^3$)

Candidate galaxy sized black holes are entire elliptical galaxies (but would need additional dark matter) .

\subsection{Million Light Year Black Holes ($R=1M$ - $750MLY$; $M=9*10^{18}$ - $5*10^{21} SM$)}

Tables \ref{TableSameDen} and \ref{TableSameMass} continue with mass densities producing MLY size black holes of larger, along with common densities seen in nature, that would need to be extended outward to fill the total volume to achieve the large total mass.

Note that the average mass density within these low density black holes includes the total mass of all the stars, dust, interstellar gas, and dark matter divided by their volumes.  Consider our immediate solar region. The sun’s one solar mass divided by the volume of sphere with a radius of 2 LY would average about 40 hydrogen atoms per cubic cm, just counting the mass of the sun.  The mass of the interstellar gas and the planets would be added to this.  If this solar region density was extended outward it would create a low density black holes in 5-7 M LYs. For this larger black hole, the average density includes the mass of its internal galaxies, including the total sum of its stars, galactic disk, and globular clusters, interstellar and intra-galactic gases, plus any hidden dark matter. 

The mass density of the Milky Way Galactic disk is equivalent to about 40 hydrogen atoms/$cm^3$ but can be higher in denser galaxies.  Tables \ref{TableSameDen} and \ref{TableSameMass} include density entries corresponding to the mass density of the Milky Way Galactic disk, and 2, 10,100, and 2000 times these values along with their corresponding characteristics.  

The average density of the Milky Way Galaxy as a whole varies with estimates but is about .5 hydrogen atoms/$cm^3$.  If the average density of the Milky Way Galaxy is extended outward, this would make a low density black hole in 63 MLY.   This would just be large enough to include the Local Supercluster.  But its total mass would be $4*10^{20}$ SM, which would be over 10,000 times the current mass estimate of the local Supercluster.   This seems like an impossibly large mass, but it only averages out to be .5 hydrogen atoms per cubic cm.   

Possible million light year black holes are the Local Supercluster and the Abell 1689 Galaxy cluster. 

\subsection{Billion Light Year Black Holes ($R = 1B$ - $200BLY$; $M=4*10^{22}$ - $5*10^{23} SM$)}

The mass density of the Milky Way Local Subgroup varies with estimates ranging between 58 to 300 hydrogen atoms/$m^3$ depending on their total mass densities. If the local subgroup mass densities were extended outward, they would form a low density black hole in 2 to 6 BLYs.

The next rows in Tables \ref{TableSameDen} and \ref{TableSameMass} are related to the critical densities.  Many scientists believe the average density of the universe is near the critical density of $9.47*10^{-27}$ $kg/m^3$ which is equivalent to about 5.67 hydrogen atoms per cubic meter.  The critical density is the density that would prevent the universe from expanding to infinity, even if the mass was traveling outward at the speed of light.  This matches up precisely with the definition of a low density Schwarzschild black hole with radius $R_s$ and using our equation comes out exactly to the 13.8 B LY radius.  Thus, the $R_s$ equation driving Tables \ref{TableSameDen} and \ref{TableSameMass} appear to match other calculations of critical density.

If one believes that the density of the universe is exactly at this value, they should realize that this would constitute a large low density black hole, with the gravity trying to cancel the momentum from the big bang.  Using the Schwarzschild Classical black hole definition, it could still expand to infinity over all eternity before stopping and reversing inward.  And yes, if this were the case, by definition, we would be living in a Schwarzschild Classical low density black hole!  Using the  Schwarzschild Relativistic black hole definition, the edge would be an event horizon and the expansion would stop immediately.

If the universe had exactly twice the critical density, equivalent to about 11.34 atoms of hydrogen per cubic meter, and had the perceived 13.8 BLY radius, it would constitute a classical finite low density black hole.    If this were the case, the universe would begin to radically slow down and come to a stop within 200\% $R_{bh}$ height which would be 27.6BLY.   And yes, if this were the case, by definition, we would be living in a finite (relativistic) low density black hole.  Using the relativistic spinning or charged black hole models, the edge would be an event horizon and the expansion would stop immediately.

The remaining entries of Tables \ref{TableSameDen} and \ref{TableSameMass} are for values that are less than the critical densities.  Extending the estimated densities of the Local group results in $R_{bh}= 12$ to 23 BLYs  and $R_s= 16$ to 33 BLYs.  If the visible densities of the Local Super cluster was extended outward, one would get a $R_{bh} = 23$ to 57 BLYs and $R_s=32$ to 80 BLYs.   Using the density of 1 hydrogen atom per cubic meter results in $R_{bh} = 32$ BLYs and $R_s= 46$ BLYs.  Using just the visible density of the universe puts the radius at $R_{bh}=77$ BLYs  and $R_s=109$ BLYs.  These radii are all beyond the current estimated 13.8 BLY radius of the universe.  If the universe has any of these densities less than the critical density, then the universe does not have to be inside a low density black hole.

If the expanding universe is below its critical mass and is almost a black hole, most matter traveling at less than the speed of light, would not escape.  The light at the edge would get out, and not return, but the light from lower orbits inside the black hole could still be curved inwards.  Thus, we would still look like a low density black hole, even before the universe collapsed completely into a black hole.   If this were the case, we would not be living in a low density black hole, but it may still look like we are.

Even if the complete universe is below its average critical mass density and not a low density black hole, typical models of gravitational mass densities are denser in the center where we are presumed to be observing from.  Thus we could still be in a low density black hole, even if the entire universe is not a low density black hole.

A candidate billion light year sized black hole is the universe itself, or just portions including its center. 

\section{Candidate Localized Black Holes} 

Many astronomers believe black holes will be hard to detect or photograph because their internal light cannot escape the event horizon and all external light striking it would be sucked in and not reflected.  Although a black hole traps all internal light inside its event horizon, any interstellar gasses just outside the event horizon of a stellar black hole should be ripped apart by the high tidal forces and cause X-Ray emissions.  Thus, stellar black holes should be easy to detect.   Within the Milky Way Galaxy, X-Ray emissions have been detected inside globular clusters and at the galactic core.  A 4 million solar mass black hole called Sagittarius A* is believed to be at the core of the Milky Way Galaxy.  We can't quite see it since the Milky Way galactic disk obscures the view, but the orbits of objects have been plotted and the masses computed at $4 MSM$ based on the orbit equations.   

\begin{figure}
\begin{center}
\includegraphics[width=.53\textwidth]{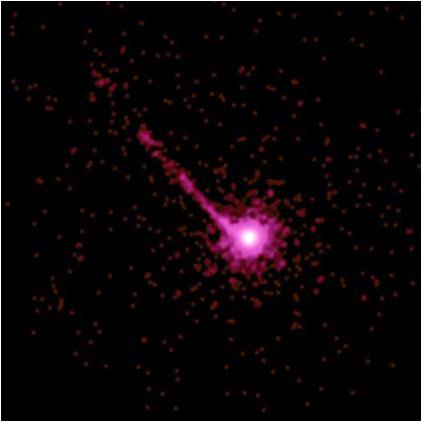}
\captionsetup{width=0.9\textwidth}
\caption{Chandra X-Ray Image of Quasar PKS 1127-145 \cite{ChandraPKS}.  Real Quasars are 600 MLY to 29 BLY away, and thus are hard to photograph.  All the really up close pictures of quasars are artistic renderings
This quasar is visible at an estimated 10 billion light years with an enormous X-ray jet extending at least a million light years from the quasar. Image credit NASA/CXC/SAO.}
\label{quasar}
\end{center}
\end{figure}

Additionally, if the black holes have an accretion disk just outside the event horizon, the influx of mass getting destroyed should glow extremely bright.   Quasars are the brightest objects in the universe and are believed to be the cores of distant galaxies each containing a giant black hole with several billion solar masses. A quasar can create more light than an entire galaxy.   A Hubble Space Telescope (HST) picture of a quasar about 10 BLY away is shown in Figure \ref{quasar} with a plume 1 MLYs long.   There are also prettier pictures of "artist rendering" of quasars.  Quasars are just too far away to get a good photo (nearest 600 MLYs) .  What would a quasar look like that is up to 4 trillions times the luminosity of the sun?  Perhaps it would look blindingly bright, but certainly not black.  

X-Rays and glowing accretion disks work really well for the stellar and solar system size black holes of 2.8, 1000, millions, and billions of solar masses, but the tidal forces and their dramatic effect will become negligible on the larger LY size black holes or larger.  Thus, we will need another mechanism to see the bigger ones because their direct viewing of internal light will be blocked by their event horizons.  External influx of matter also may not generate x-rays or give off enough light to be visible.  Thus, the primary mechanism for finding large black holes will need to change to gravitational lensing and the bending of external light. 
\begin{figure}
\begin{center}
\includegraphics[width=.53\textwidth]{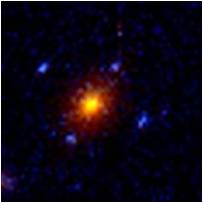}
\captionsetup{width=0.9\textwidth}
\caption{Galaxy Lens Quadrupling Quasar (HST 14176-5226) \cite{EinsteinCross}.  A light from a distant quasar (11 BLY) far behind but immediately behind a galaxy (7 BLYs) finds four pathways around the intermediate galaxy and shows up as four identical (but slightly distorted quasars) shown on the left.  The bending angle $\alpha=tan^{-1}(1MLY/4000MLY)=.014^o$. This shows that light bending is real and that even a small bending angle can duplicated objects. This image was taken by the Hubble Space Telescope (HST).}
\label{EinsteinCross}
\end{center}
\end{figure}

An Einstein's Cross, as shown in Figure \ref{EinsteinCross}, is created when an intermediate massive object (in this case a galaxy) is directly in front of a second more distant object (in this case a quasar).    The direct viewing of the quasar is blocked by the intermediate galaxy.  However, the light is bent around the left, right, top, and bottom so that four images are seen of the distant quasar.  Thus, although the primary light may be absorbed by the intermediate massive object (e.g. galaxy or black hole), gravitational lensing can create two and sometimes four duplicate objects of anything behind the intermediate object (including a hard to see black holes).   

\begin{figure}
\begin{center}
\includegraphics[width=.41\textwidth]{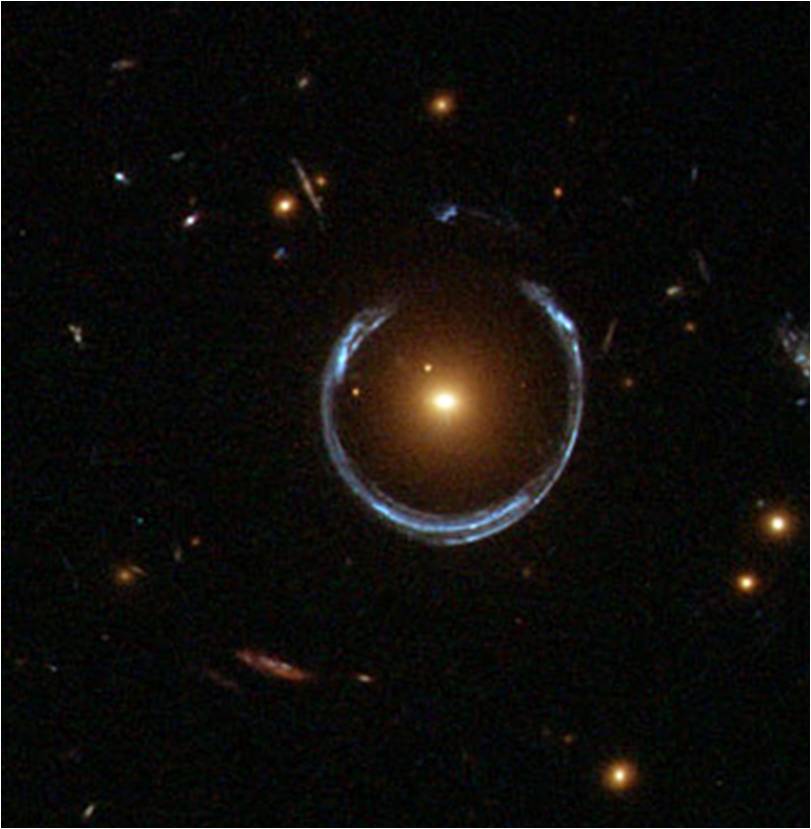}
\captionsetup{width=0.9\textwidth}
\caption{Einstein Ring Horseshoe \cite{EinsteinRing}.  When an object is exactly centered behind the intermediate object, light can flow around every angle of the center object and produce a continuous ring of light (called an Einstein Ring, or partial Einstein Ring Horseshoe in this case).   Also note that there appears to be duplicate pairs or triplets inside and outside the ring.  This image was taken by the HST. }
\label{EinsteinRing}
\end{center}
\end{figure}

\begin{figure}
\begin{center}
\includegraphics[width=.9\textwidth]{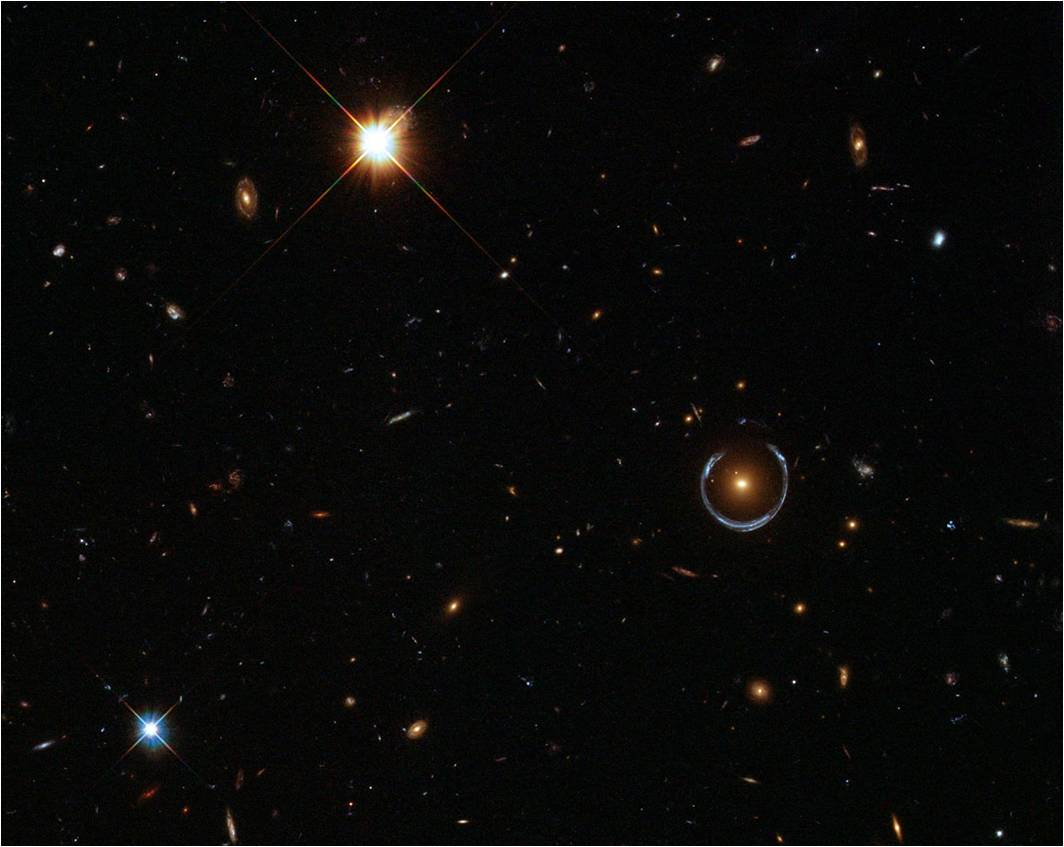}
\captionsetup{width=0.9\textwidth}
\caption{Einstein Ring Horseshoe Surrounding Space \cite{EinsteinRing}.  The duplicated pairs seem to extend outward in space.  This image was taken by the Hubble Space Telescope (HST). }
\label{EinsteinRing2}
\end{center}
\end{figure}

Similarly, when the object is exactly centered behind the intermediate object, light can flow around every angle of the center object and produce a continuous ring of light (called an Einstein’s Ring as shown in Figures \ref{EinsteinRing} and \ref{EinsteinRing2}.    Also note that there appears to be candidate duplicate pairs about the Einstein's Ring.   Thus, one can look for black holes by looking for duplicate objects and gravitational lensing rings, even if they cannot see the center object because of an unseen event horizon. 

\begin{figure}
\begin{center}
\includegraphics[width=\textwidth]{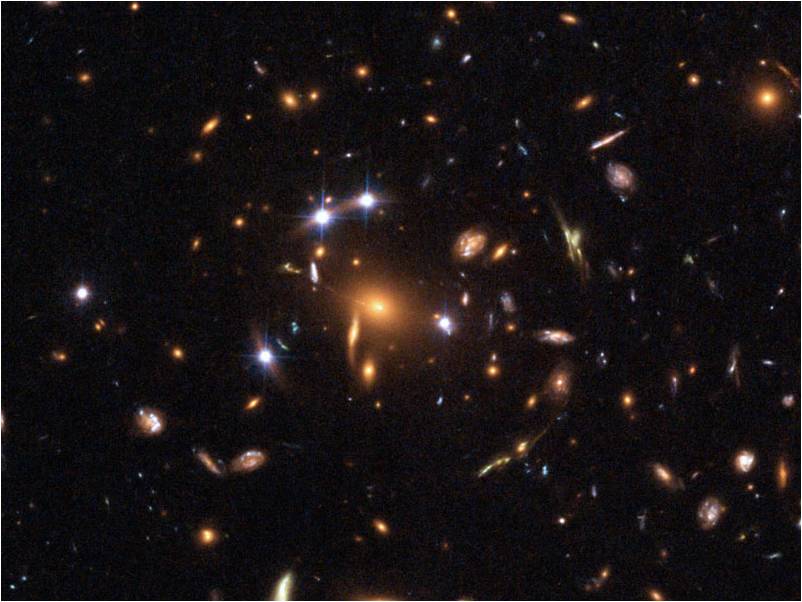}
\captionsetup{width=0.9\textwidth}
\caption{Galaxy Cluster SDSS J1004+4112 \cite{SDSS}.  This Hubble Space Telescope image of a cluster of galaxies that acts as the huge gravitational lens.  The five bright white points near the cluster center are actually images of a single distant quasar being lensed around a central host galaxy.  This galaxy cluster is at a distance of about 7 billion light years toward the constellation of Leo Minor.   Image Credit: ESA/Hubble.}
\label{SDSS}
\end{center}
\end{figure}

\begin{figure}
\begin{center}
\includegraphics[width=\textwidth]{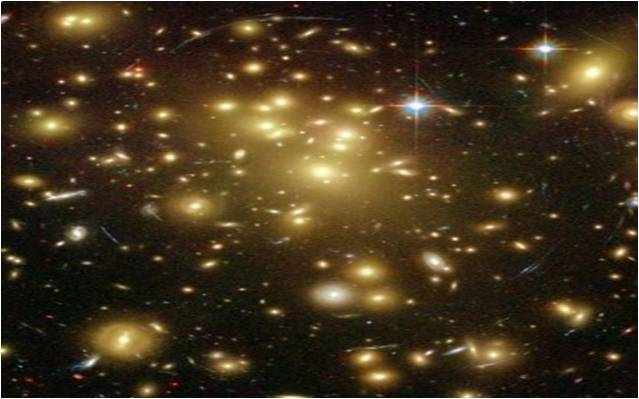}
\captionsetup{width=0.9\textwidth}
\caption{Abell 1689 Galaxy Cluster with Gravitational Lensing \cite{abell}. The Abell 1689 Galaxy Cluster is one of the densest galaxy clusters with an estimated mass of $2 * 10^{15}$ SM and radius of 1 MPC (or 1.63 MLY) which would have a density equivalent to 155,000 $atoms/mm^3$.  For this mass $R_{bh} = 100$ MLY and $R_s = 50$ MLY;  Thus the Abell Galaxy Cluster is only off by a factor of 30-60 from being a black hole.   Unseen dark matter could put this into a localized low density black hole.  The image also exhibits heavy gravitational lensing and a redshift  $z=0.183$.  Image Credit: NASA, ESA, L. Bradley (JHU), R. Bouwens (UCSC), H. Ford (JHU), and G. Illingworth (UCSC)}
\label{abell}
\end{center}
\end{figure}

\begin{figure}
\begin{center}
\includegraphics[width=\textwidth]{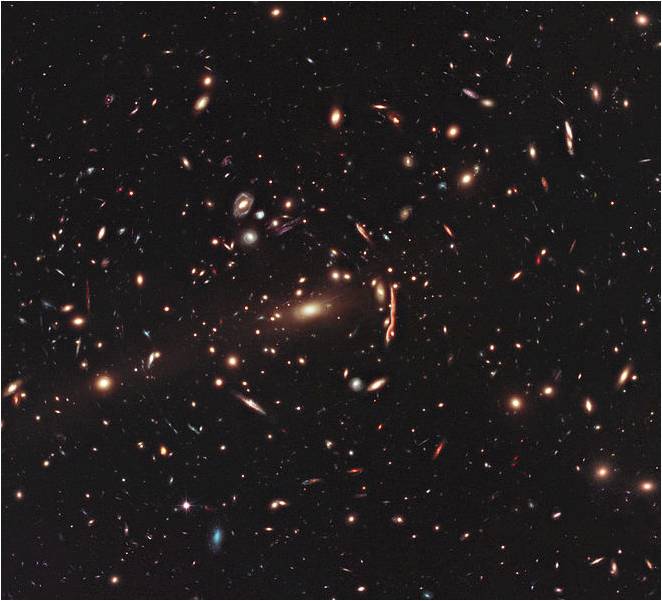}
\captionsetup{width=0.9\textwidth}
\caption{Galaxy Cluster MACS J1206 \cite{MACS}.  This image from the NASA/ESA Hubble Space Telescope shows the galaxy cluster MACS J1206.}
\label{MACS}
\end{center}
\end{figure}

\begin{figure}
\begin{center}
\includegraphics[width=\textwidth]{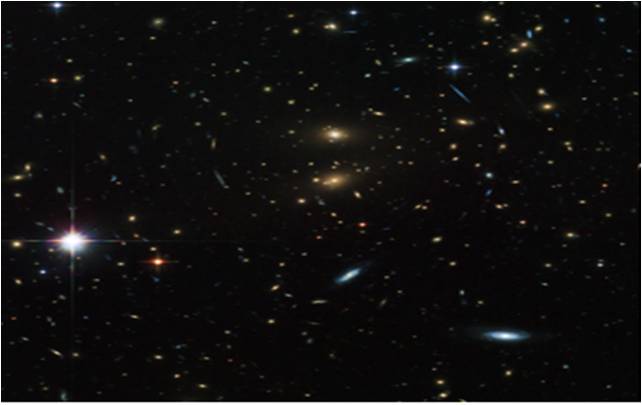}
\captionsetup{width=0.9\textwidth}
\caption{LCDCS-0829 Galaxy Cluster\cite{lcdcs}. The galaxy cluster shows signs of gravitational rings as well as multiple combinations of three galaxies.  Thus, a large part of this cluster could be caused by a smaller rotating low density black hole in the center, which is dragging space around in a spiral pattern, and causing multiplications of the base three galaxies.   Image Credit: ESA/Hubble.}
\label{LCDCS}
\end{center}
\end{figure}

Galaxy Cluster SDSS J1004+4112 , the Abell 1689 subgroup within the Virgo Super Cluster, Galaxy Cluster MACS J1206, and LCDCS-0829 Galaxy Cluster are good candidate localized low density black holes  (Figures \ref{SDSS}, \ref{abell}, \ref{MACS}, and \ref{LCDCS}).   These galaxy clusters show significant signs of gravitational lensing and possibly duplicated galaxy images that take different paths around or through the black holes.  

Einstein’s theory of general relativity predicts that gravity bends light near any massive object.  Einstein's theory predicts that our sun bends light 1.75 arcsec or .0005 degrees at its surface (measured during a solar eclipse).   
The example lensing in the Einstein's cross bent the light a fraction of a degree to create this four duplicate stars and this Einstein ring of Figure  \ref{EinsteinCross} $\alpha=tan^{-1}(1MLY/4000MLY)=.014^o$.  What level of light bending can be expected from a black hole?  
\newpage
Einstein's deflection angle, valid for small angles, is $D_a(R )= 4GM/(RC^2)$ radians \cite{lightBending}.  At $10R_s$, a Schwarzschild black hole can bend light roughly
\begin{subequations}
\begin{align}
D_a(10Rs)  &= \frac{4GM}{10\left(\frac{2GM}{C^2}\right)C^2}\\
&=0.2 \text{ radians} = 11.5 \text{ degrees.}
\end{align}
\end{subequations}

At $20R_{bh}$, a spinning or charged black hole can bend light roughly
\begin{subequations}
\begin{align}
D_a(20R_{bh})  &= \frac{4GM}{20\left(\frac{GM}{C^2}\right)C^2}\\
&=0.2 \text{ radians} = 11.5 \text{ degrees.}
\end{align}
\end{subequations}

Thus, light out to $10*R_s$ and $20*R_{bh}$ will bend light 11.5 degrees.  This means that all stars, within 11.5 degrees of $10*R_s$ and $20*R_{bh}$ will be duplicated.   The primary star image will be seen directly (a little farther away from the black hole), and a duplicate image will appear on the far side of the black hole.  For large black holes, with a radius of 1 LY, 1000 LY, 1 million light years, $10*R_s$ and $20*R_{bh}$ could be quite large.  In addition, light closer to the black hole just outside its photon sphere $R_{ph}$ would be bent up to 360 degrees as listed in Section 1 and computed in Appendix A.  Thus, all stars farther out than 11.5 degrees will be duplicated on the far side of the black hole between $R_{ph}$ and ${R_bh*20}$.  These duplicate images would appear like real stars or galaxy clusters.  Thus, even relativistic black holes with a fully formed event horizon would appear like star or galaxy clusters from the outside because light from each external star or galaxy would be bent around the black hole and seen by the observer as an optical illusion duplicate star, galaxy, or cluster.  
\vspace{3 mm}

The following thought experiment may help.  Hold up your thumb at arm’s length, and pick a distant object, say 5 to 10 degrees to the right.  Pretend this distant object is an external galaxy, and your thumb is a black hole.  You would see the external galaxy from the direct light on the right because the black hole is too far away to dramatically affect its light path on the right.  Now consider light from the object (external galaxy) heading to the far left of the black hole (say 5 to 10 degrees left of your thumb).  This light traveling to the left would also not be dramatically affected by the black hole and would head off towards the left unobserved.  But now consider what happens to the external galaxy’s light at decreasing angles from the left that is shining closer to the left of the black hole.  The light will start being bent towards the observer after passing by the black hole.  The light will be bent slightly at first, and then more and more, up to a point where it will be bent 360 degrees and then collapse into the black hole.  But between these extremes, there exist an angle where the light will be bent exactly towards you (the observer), and the observer should see a duplicate galaxy just to the left of the black hole (thumb).  The duplicated galaxy will appear more distant than the original galaxy since the light had to travel farther (distance to the black hole + distance from black hole to observer).  

\vspace{3 mm}
This same process will create a duplicate image illusion of every star or galaxy that is shining light on the black hole.  Since all external stars and galaxies shine light on the black hole, all stars and galaxies will be duplicated. All the duplicate images would appear around the black hole, as a cluster of stars or galaxies.  Perhaps this may look identical to the galaxy clusters of Figures \ref{SDSS}, \ref{abell}, \ref{MACS}, and \ref{LCDCS}.

\vspace{3 mm}
If the mass is just short of the black hole definition, then the edge would not be a fully formed event horizon and light from inside could also be seen from the middle.   Since each of these (nearly) black holes are probably spinning to keep from collapsing, the light coming from within might take multiple paths through the core or poles of the nearly black hole as it migrates outward.  Due to frame dragging of the spinning mass could add a spiral pattern to the duplicated objects.   

\vspace{3 mm}
So what would a picture of a black hole possibly look like?  Galactic cores like Sagittarius A*, Quasars, Globular clusters, galaxy clusters, and highly lensed galaxy clusters.  Note that none of these black holes are really black, but are some of the brightest objects in the universe.  

\vspace{3 mm}
This of course is the view from the outside of a black hole.   Section 8, 9, and 10 will discuss what it might look from inside of a black hole.

\newpage
\section{Expanding Universe and Disproof of the Big Bang Theory}
Is the expanding universe theory consistent with a universe which may be a black hole? 
\\[2ex] \textbf{Theorem: }\textsl{An expanding universe isn't consistent with Newton's law of gravity. }
\\[1ex] \textbf{Proof by Contradiction:} \\
Assume the following, 
\begin{enumerate}
\item The universe is expanding 
\item Newton's law of gravity is valid through this expansion (or at least the latter portion of this expansion)
\end{enumerate} 
Following Newton's law of gravity the black hole equations listed in Section 1 and derived in Appendix A are valid through this expansion (or at least the latter portion of this expansion). Recall, these derivations only required calculus and Newton's law of gravity and Newton's law of motion.
\\[1ex] Consider the following three cases: 
\\[1ex] \textbf{Case 1:} Assume the current density of the universe with the current radius ($R_{current}$) is at the critical density ($P_{critical}$) and critical mass ($M_{critical}$) such that it can be viewed as a Schwarzschild black hole. That is, 
\begin{align}
 P_{critical}  &= \frac{1.797*10^{-6}}{R_{current}^2}  kg/m^3 \\
M_{critical}&=P_{critical}*V=P_{critical}*\frac{4\pi}{3}R_{current}^3 \\
R_{current}&=\frac{2GM_{critical}}{C^2}
\end{align}
Then it could be viewed as a low density black hole with $R_s$ or $R_{cs} =R_{current}$.  $R_{current}$ is believed to be at 13.8 BLY which would make $P_{critical}=9.47*10^{-27} kg/m^3$.   
Using the  Schwarzschild \ul{Relativistic} black hole definition, the edge would be an event horizon and the expansion would stop immediately.  Using the weaker Schwarzschild \ul{Classical} black hole definition, it could still expand to infinity over all eternity before stopping and reversing inward.  
\\[1ex] Computing the \ul{finite} (classical or smaller relativistic) black hole radius for the same mass ($M=M_{critical}$) yields, 
\begin{align}
R_{bh} = \frac{GM}{C^2} = \frac{1}{2}\left(\frac{2GM}{C^2}\right) = \frac{1}{2}R_s = \frac{1}{2}R_{current} = 6.9 \text{ BLY.}
\end{align}
Note these black hole radii were also provided in Table $\ref{TableSameMass}$.  When the universe expanded through this $R_{bh}$ value of 6.9BLY in the past to get to $R_{current}$, it would have been half the size, eight times denser, and would have been a finite black hole.  If this was the case it would not have been able to double in size by definition of a finite classical black hole.  Even mass traveling at the speed of light at the boundary of a finite classical black hole only can increase just up to 100\% of its $R_{bh}$ radius to get back to the current radius.  But since no mass can travel at the speed of light (other than light), the hard mass could not have been able to expand back to the current radius.  Even light photons would be stopped at the photon sphere.  Therefore the finite classical black hole (in the past) could not double its size to get to the current radius.  This is a contradiction, so one of our assumptions must be invalid!   Thus, the universe cannot be exactly at the critical density in an expanding universe or the denser finite black hole in the past would have stopped the expansion.   

A relativistic finite black hole (in the past) with $R_{bh} = R_{current}/2$ would have had an event horizon at $R_{current}/2$ that would had stopped the expansion at $R_{current}/2$.
\\[1ex] \textbf{Case 2:} Assume the expanding universe is currently denser than the critical density $P_{critical}$, then this would be like the Schwarzschild black hole just discussed above at the exact critical mass density plus some extra mass added.  In the past, at half its radius, it would be eight times denser, and be like the finite black hole at eight times the critical density, but with some extra mass added.  Thus, the universe expansion would have been stopped by this finite black hole, plus extra mass. Thus, that universe could not have doubled in radius to get to $R_{current}$.  This is a contradiction, so one of our assumptions must be invalid!   Thus, the universe can't be denser than  $P_{critical}$ in an expanding universe.
\\[1ex] \textbf{Case 3:} If one assume that the universe is expanding but is currently at a lesser density than the critical density $P_{critical}$, that is $P_{current}=fP_{critical}$ where $f<1$. Then we would not currently be in a low density black hole.  However, in an expanding universe, the universe would have been much denser in the past and would still have been a finite black hole at a smaller radius.
\\[1ex]To show this 
\begin{subequations}
\begin{align}
M&=P_{current}\frac{4\pi}{3}R_{current}^3 \\
&=fP_{critical}\frac{4\pi}{3}R_{current}^3 \\
&=fM_{critical}
\end{align}
\end{subequations}

Let $M=fM_{critical}$, where $f<1$ is the lesser (fractional mass) such that $P=fP_s$ and $M_{critical}$ is the mass density of the critical universe above.  Then 
\begin{subequations}
\begin{align}
R_{bh} &= \frac{GM}{C^2} \\
             &= \frac{1}{2} \left(\frac{2GM}{C^2}\right) \\ 
            & = \frac{1}{2} \left(\frac{2GfM_{critical}}{C^2}\right)\\
           &= \frac{1}{2} f R_{current} \\
           &= 6.9 f \text{ BLY.}
\end{align}
\end{subequations}
Since f is less than 1, the universe would have been in a finite relativistic black hole prior to 6.9 BLY, which would have prevented it from more than doubling in size to 13.8 BLY. 

For example, if one assumes a density of one hydrogen atom per cubic meter ($1.67*10^{-27} kg/m^3$); $f =  1.678*10^{-27}/(9.47*10^{-27}) = 1/5.67$; $R_{bh} = 1/5.67*6.9 \text{ BLY} = 1.22 \text{ BLY}$.  But if the lesser dense universe was a finite black hole at a smaller size $f*6.9$ BLY (e.g. 1.22 BLY), it would  not have expanded greater than 100\% of this size (e.g. 2.44 BLY) to get to its current size of 13.8 BLY.  This is a contradiction, so one of our assumptions must be invalid!   Thus, the universe cannot be at less than the critical density in an expanding universe.  

Since all possible densities were covered in the above three cases, in an expanding universe and  encountered contradictions with each one, the invalid assumption must be that the universe is an expanding universe.  Therefore the expanding universe theory is inconsistent with Newton's law of gravity. \\ \qed \\

Thus, the expanding universe theory from the big bang is invalid because the smaller denser universe in the past would have constituted a finite (classical or relativistic) black hole that would have stopped its expansion. 

Thus, we are not in an expanding universe, and probably never were in an expanding universe, and the big bang probably did not happen.  The Doppler redshift originally associated with expanding universe will need to be replaced with another mechanism.  Possible candidates are translational redshift, gravitational redshift, and intrinsic redshift mechanisms, or a combination.  These will be covered in Section 7.  

The only possible way I see to preserve an expanding universe due to the big bang, is to suspend Newton's law of gravity in its earlier phases (perhaps the expansion phase), until the universe got past its finite black hole radius ($R_{bh}$) or fractional $R_{bh}$ (1.25-6.9 BLY). This would bypass the contradiction found in the analysis above.  That is, declare $F = GmM/R^2$ invalid for up to half of the expansion.  But ideally, physical theories should be self consistent and not require the suspension of the laws of physics.  

Physicists had to suspend the laws of physics during the initial seconds of the explosion until each fragment was less dense than a black hole. They also envisioned some fragments were micro black holes and perhaps stellar black holes, that now may be contributing to dark matter. However they failed to realize that the collection of all fragments also constituted a relativistic large low density black hole, out until it passes one half its critical density radius or fractional radius (e.g. 1.22BLY to 6.9 BLY).

Some physicists have also introduced the concept of dark energy to try to add a 10 to 25\% correction needed to match an expanding universe that is perceived as expanding even faster over time.  This basically adds an unknown repulsive force to override or counteract the gravitational force, $F= GmM/R^2$.   

This dark energy force would also have to be able to keep the finite black hole from collapsing when the expanding universe was a half, quarter, and eighth  its current radius, when it was 8, 64, and 512 times denser.  Thus, the dark energy repulsive force would also have to scale with size.  However, I believe one should first explore all the alternatives offered by the simpler gravity equation above before tweaking constants, adding terms, or adding new (unseen) physical concepts, especially if it is pretty clear that the universe is not expanding.  This exploration process of the remaining alternatives is the goal of the rest of this paper.

\newpage
\section{Non-Expanding Universe Options}
If we are not in an expanding universe, other options include: 1) an infinite universe 2) collapsing universe, 3) non-expanding or slowly collapsing universe, 4) a reflecting universe or 5) an oscillating universe.
\subsection{Infinite Universe}
If the universe is infinite, with finite density, however small, we would be in a low density black hole since the mass would extend to a finite black hole radius corresponding to that density.   Thus, infinite space would collapse into disjoint low density black holes.   If the universe is infinite we would probably be in one of the disjoint low density black holes.   Even if we were "lucky" and within a lagrange point exactly between two disjoint low density black holes, the collection of nearby disjoint low density black holes including ourselves would still constitute a larger low density black hole.  Thus, if the universe was infinite, we would be in a low density black hole.

An infinite universe made of disjoint low density black holes escapes Olber’s paradox \cite{olber} of being infinitely bright with its infinite lights since the light available within each black hole would be finite.  Thus, if we are in an infinite universe of disjoint low density black holes, the universe could look finite because the infinite external light is trapped in external black holes.  

If this is the case, the infinite universe would be lying to us by cloaking itself in an infinite number of invisible event horizons.   However, since we would be inside a localized black hole, the internal light from our localized black hole would be "reflected" backwards by our event horizon, and create the illusion (second lie) that we are in an infinite universe  (see reflective universe subsection below).   But, if this is the case, the second lie would actually cancel the first lie and accidentally convey the truth, that we would be actually within an infinite universe.   Unfortunately, if we are trapped within a localized black hole, we may never know.   If we are in an infinite universe of an infinite number of black holes, but no one can see their light, are they really there?

If we are in an infinite universe and other lagrange galaxies exist just outside our localized black hole, we may be able to still see them, but they may not be able to see us. That is, they may not be able to see internal light from our black hole.   External galaxies may only be able to see our gravitational lensing effect and we would appear like a galaxy cluster, perhaps of the lucky "galaxies" at the lagrange points.  Similarly, external black holes may only be visible as galaxy clusters of lagrange galaxies.  But these may be overwhelmed by the reflections of internal galaxies.

\subsection{Rapidly Collapsing Universe}
If we were in a finite collapsing universe denser than $P_{critical}$, we would be in a low density black hole.  Collapsing into a black hole would be the expected natural process.   However, if the universe was rapidly collapsing, one would see a blue shift which is currently not observed, (and we would be crushed by now).  Thus, the universe is not rapidly collapsing.

However, the light from the edge of the universe may be up to 13.8 BLY old, and thus, all that we know is that the universe was not rapidly collapsing 13.8 B years ago.   Since that time, it could have begun collapsing very rapidly.   For example, 10 billion years ago it could have started to collapse and we wont find out until about 4 billion years from now.  But as of this time, we have no evidence that we are rapidly collapsing.

There are also very bright quasars being detected at large distances that are heavily redshifted.  This redshift would argue against a collapsing universe.  However, as we will see in Section 7, the quasar redshift could be caused by gravitational redshift, which could override a smaller blueshift due to a collapsing universe.  Additionally, the collapsing mass at the edge of the universe into the quasar would explain why there are no nearby quasars.  Thus, the quasars redshift could also be supporting evidence for a collapsing universe as well.

\subsection{Non-expanding or Slowly Collapsing Universe}
If the universe was less dense than $P_{critical}$ and in a non-expanding or slowly collapsing universe, we would not be in a low density black hole, yet, until the universe collapses and its density increases to $P_{critical}$.  Note: light could still get out, but the mass would continue to collapse.   But after it collapses to $P_{critical}$, then even light would no longer be able to escape.  Thus, if we are not in a low density black hole, we will eventually be in one.  

Additionally, most conventional models of gravitational mass densities have higher densities in the center (where we would be observing from).  Thus, we could still be in a low density black hole, even if the entire universe has not yet collapsed into a low density black hole. Thus, we are probably living in a low density black hole, or eventually will be.  

Note that a non-expanding or slowly collapsing universe would be the closest model to the existing expanding universe model.  The universe could even be the same size, but just need to replace the Doppler redshift explanation with an alternate mechanism.   The universe could be finite and non-reflecting.

\newpage
\subsection{Reflecting Universe}
\begin{figure}[H]
\begin{center}
\includegraphics[width=.9\textwidth]{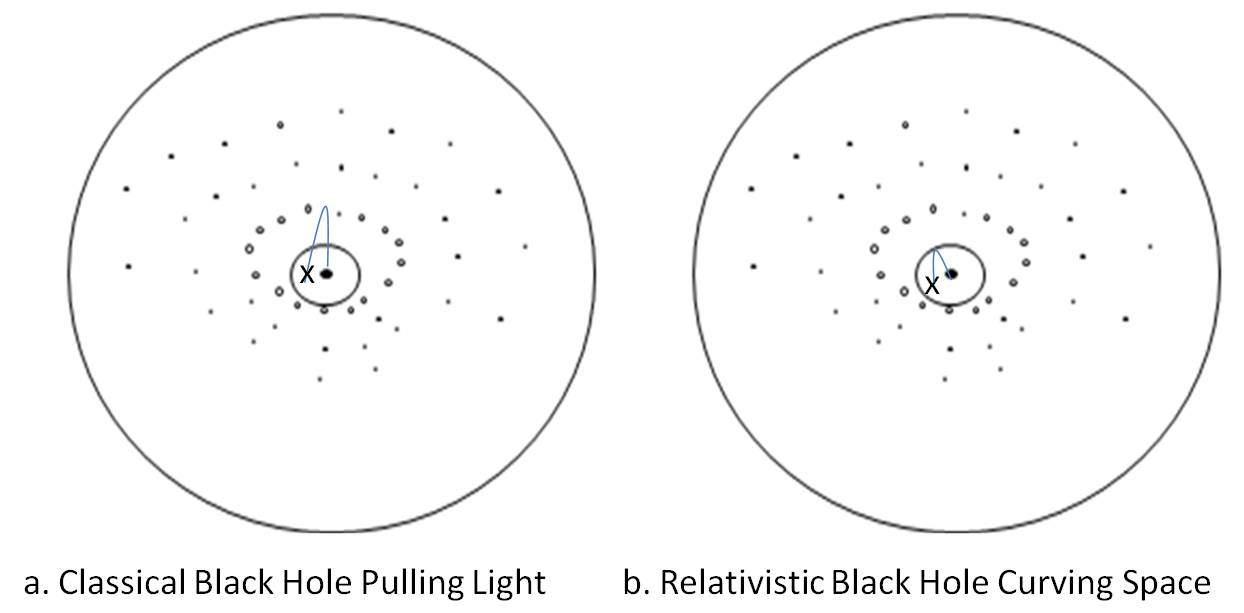}
\captionsetup{width=0.9\textwidth}
\caption{Light Sphere Illusion. If one assumes a single galaxy or galaxy cluster within a giant inner mirrored sphere, its own light would reflect off the mirror and create reflections in all directions, and appear like an infinite number of more distant galaxies.  Now, assuming that the center object is a low density black hole and the inner mirror sphere is the boundary of the black hole ($R_{bh}$).  Light traveling outward in a classical black hole (on the left) would be pulled backwards and be “reflected” inward by the gravitational pull (e.g. via the shown curved path to the observer at position x).  In a relativistic black hole (on the right), space would be curved at the event horizon and bend the light backwards.   Either of these low density black hole would create illusions similar to the reflecting mirror sphere. If there is only one galaxy inside the low density black hole, the "reflections" should come in periodic waves.  However, if the low density black hole contains galaxy clusters at varying depths, the reflections will appear less periodic.}
\label{lightsphereillusion}
\end{center}
\end{figure}

Consider, for example, galaxies or galaxy groups inside an inner mirrored sphere (small ellipse in the center) as depicted in Figure \ref{lightsphereillusion}. The light would ripple off the mirror in all directions, and create the illusion of an infinite number of galaxies in a much larger perceived sphere of Figure \ref{lightsphereillusion}.  A low density black hole could create a similar reflecting universe as the mirror as described in Figure \ref{lightsphereillusion}.   

The radius of different possible reflecting universes will be explored later, after showing that it is possible for a low density black hole to keep from collapsing.  And, if we are inside a reflecting universe, we would be living in a black hole. 

If the universe is finite as a whole, but not reflecting at its boundary, but we are in a localized low density black hole $R_{bh}$, this internal “mirror ball effect” reflection could still occur locally.   We would be able to see external galaxies as well, but they may not be able to see us.  That is, they may not be able to see the internal light from our black hole.   We would appear like a galaxy cluster due to our gravitational lensing of external galaxies.  

\newpage
\subsection{Oscillating Universe}
The oscillating universe, in the literature, generally comes from having the expanding universe eventually collapse, into a singularity, where there may be a subsequent explosion (subsequent big bang) and cause a subsequent expansion phase, thus repeating the process.   But this would run into the similar problems of the expanding universe, which requires breaking Newton's law of gravity.  Thus, oscillating universes in literature that depend on an expanding universe would not occur.

\subsubsection{Active Objects}
Another possible source for oscillation through temporary expansion is the active plumes of neutron stars, black holes, and quasars.  Recall, that the black hole escape velocity equations were for inert, ballistic projectiles.  Active objects with active plumes are more like relativistic rockets that could expand one outward against the forces of gravity.   If an active object is within an event horizon of a black hole with a negligible G force (e.g. less than a nano G), then an active object may be able to escape the event horizon.
\subsubsection{Diminished Gravity}
Another possible source for oscillation through temporary expansion would be if gravitational waves from inside the black hole were affected by the collapse of space around the black hole. That is, as mass collapses into a black hole, where nothing can escape or be seen externally, not even light nor other electromagnetic waves; its own internal gravitational waves might also become blocked and not escape, or at least  be diminished.  The light from external objects could still get in, thus gravity waves from external objects should still be felt, but the reverse may not be true.  However, this seems like it would break the principle of physics: for every action there is an equal and opposite reaction.  Thus, this seems implausible.  

\begin{figure}[H]
    \begin{center}
   \includegraphics[width=.69\textwidth]{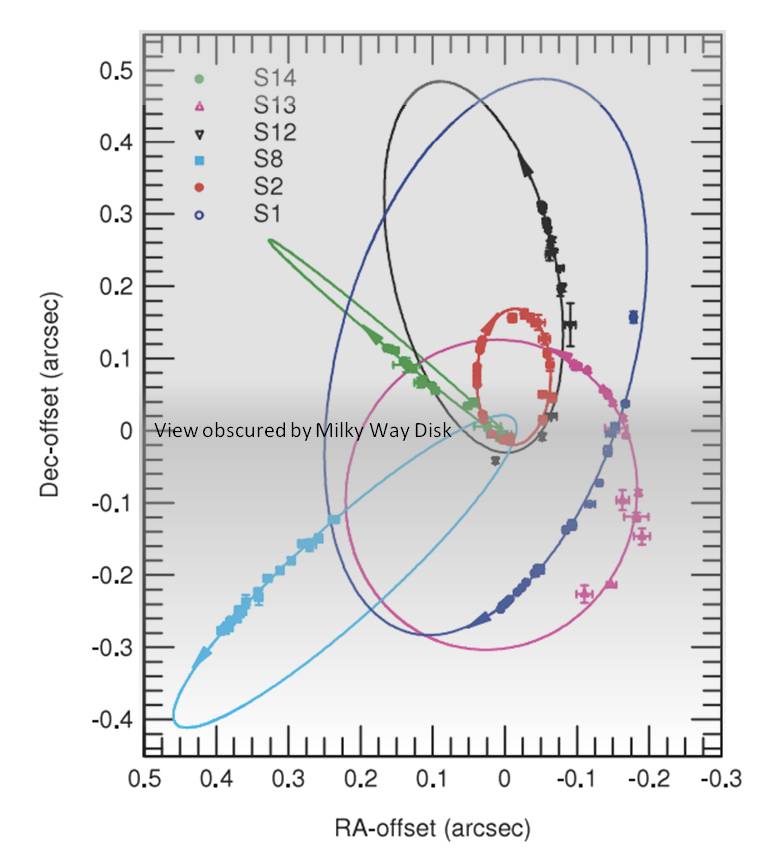}
\captionsetup{width=0.9\textwidth}
 \caption{Star Orbits around Sagittarius A* \cite{sagA}. Sagittarius A* is believed to be a  giant 4M SM black hole near the center of our Milky Way Galaxy.  The orbits of  6 stars around Sagittarius A* have been observed in elliptical orbits and tabulated. Using the periods of these orbits, the mass has been estimated at 4 M SM and the closest approach of the stars are between 110-2119 AU.   The radius of Sagittarius A* is not precisely known because it is obscured by the Milky Way disk, but it is exhibiting X-Rays and is presumed to be a black hole.  Since the closest approaches vary by a factor of 20, yet their computed mass for each are approximately 4.06 M SMs $\pm 1$\%, this would argue that the gravitational waves are not affected significantly by the curving of space due to the black hole's large gravitational field.  Note: Notional Milky Way obstruction added to figure from \cite{sagA}.}
\label{sagittarius}
\end{center}
  \end{figure}

\vspace{3 mm}
 \ul{Sagittarius A*:}  The “believed” black hole in the center of the Milky Way at Sagittarius A* is still attracting the stars in elliptical orbits at about a constant 4 million solar mass effect as shown in Figure \ref{sagittarius} and Table \ref{table2} \cite{sagA}.  If it is a finite black hole, the gravity does not seem to be dropping off!  That is, its gravity is still being felt.  But with a diminished gravity theory, shouldn’t the larger orbits feel diminished gravity at the various distances.  This would argue against the diminished gravity theory.  However, from the geometry above, Sagittarius A* does not necessarily need to be a black hole.   Based on its measured mass of 4 million solar masses, its radius would be $R_s=2GM/C^2 = 40$ light-seconds.  Its estimated location of 6.25 light-hours from the smallest orbit would put the smallest orbit at 563Rs. Thus, there is room for it being less dense than a black hole. That is, it could be just a nearly black hole with a radius of 60 light-seconds.

\begin{figure}[H]
\begin{center}
\includegraphics[width=0.9\textwidth]{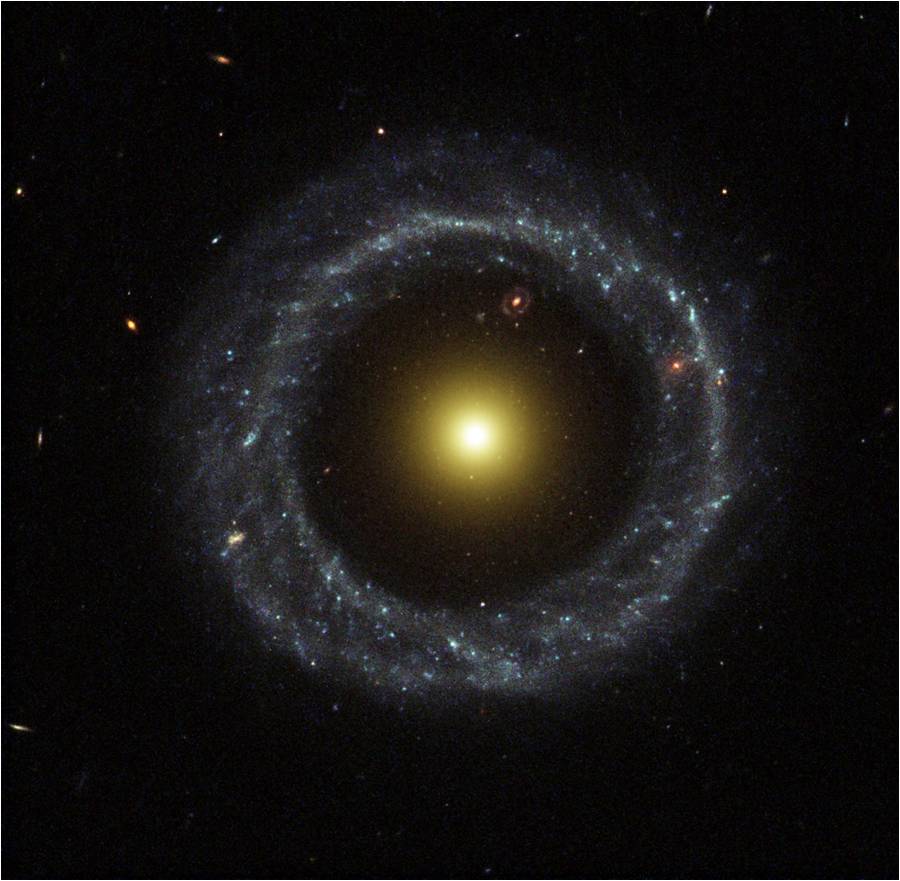}
\captionsetup{width=0.9\textwidth}
\caption{Hoag Ring Galaxy  \cite{hoag}. This Hoag Ring Galaxy appears to be missing its internal mass.  If the center has become a low density black hole, where no light or electromagnetic radiation can escape, maybe some of its gravitational waves are also affected (diminished) and are no longer able to hold in the galactic disk.   If this was the case, the centripetal force needed by stars orbiting the center would no longer be provided by the central gravitational force, and all stars would fly outward, possibly producing the observed Hoag Ring Galaxy.  Alternately, there could just be a strong intergalactic solar wind coming from the core.  Coincidentally, note the two baby Hoag like objects in the  background and possibly more in the distance.  Since the Hoag object is not a common phenomena, it is unlikely that there would be three (or more) in such a close proximity.  Thus, these miniature Hoag objects may be gravitational reflections from a low density black hole caused by the core of the Hoag Ring Galaxy. Image Credit: NASA and The Hubble Heritage Team (STScI/AURA) Acknowledgment: Ray A. Lucas (STScI/AURA)  }
\label{hoag}
\end{center}
\end{figure}

\ul{Hoag Object}: There is a rare and currently unexplained Hoag object believed to be a ring galaxy shown in Figure \ref{hoag} \cite{hoag}. This looks like a galaxy, with unexplained missing center matter.  This could be a galactic center that collapsed into a black hole in the middle, which possibly reduced its external gravitation effects, which then resulted in the entire collapsing disk to expand outward due to its gravity no longer sufficient to provide the needed centripetal force. However, it could also be just a strong solar wind pushing the outer disk outward.  

Coincidentally interesting, for this rare ring galaxy, there appears to be three identical baby galaxies of the exact same design in the picture.  What is the chance that that could happen?  My view is that independent of what caused the creation of the parent ring galaxy, the two (or three) smaller ones are possibly just two (or three) primary illusions of itself, with perhaps the smaller distant dots additional reflections or views.  If this were the case, this radical light bending supports the reflecting light theories.

If a black hole’s gravity is not felt outside the black hole, then a big bang universe may have a mechanism for legally suspending the laws of physics (or really honoring the new laws of diminished gravity).  A small dense big bang universe might be considered as a nested set of concentric black holes, each cutting off or reducing its gravity from the outer black holes, which could explode and expand without the gravity from the internal black holes slowing down the expanding outer shell.  If the speed of the exploding outer shell is near the speed of light, even after the inner black hole expanded above its critical density and no longer was a black hole, and would start putting out gravity waves traveling at the speed of light, it would take time for the inner gravitational waves to catch up with the outer moving shell.  

As one might suspect, I actually like the big bang theory and would like to preserve it.  It fits my current religious beliefs and gives us a definite start time for creation.  However, what I want and don’t want should not affect the cosmos or the laws of physics.  Using Occam’s razor, it may be simpler to possibly believe in an average density of 1 hydrogen atom per $m^3$, $cm^3$, or $mm^3$ than a huge explosion (along with its additional needed $\frac{1}{2}MC^2$ Kinetic Energy), plus the suspension or bending of the laws of physics.

The universe is not expanding (via Theorem 1) and all other cases result in a universe that is in a low density black hole or eventually will collapse into a black hole.   “Once you eliminate the impossible, whatever remains, no matter how improbable, must be the truth” (Sherlock Holmes and Mr. Spock).

Thus, for the rest of this paper, I will assume that $F=GMm/R^2$ is valid even after it becomes a black hole.  And, thus, \ul{We are probably living in a low density black hole, or eventually will be.}

\begin{table}
    \centering
      \captionsetup{width=0.9\textwidth}
      \caption{Star Orbits around Sagittarius A* \cite{sagA}.   Mass Calculations: Mass = $(4\pi^2 a^3) /( G P^2)$. Here $P$ is the period, $a$ is the aphelion ellipse radius and $R_{min}$ is the radius of nearest approach.  Note: data from  \cite{sagA} augmented with $R_{min}$ calculation.}
\begin{tabular}{| c | c | c | c | c | c| c|}
  \hline                        
Star & \pbox{20cm}{aphelion (ellipse radius) \\a(AU) }& \pbox{20cm}{Period\\ (years)} & \pbox{20cm}{Mass \\ M SM} & \pbox{20cm}{eccentricity \\ $\epsilon$} & \pbox{15cm}{ $R_{min}=a(1-\epsilon)$\\ \cite{zeilikP8}}\\
\hline
S2 & 980 & 15.24 & 4.05 & 0.876&122\\
S13 & 1750 & 15.24 & 4.13 & 0.395&1059 \\
S14 & 1800 & 36 & 4.04& 0.939&110\\
S12 & 2290 & 38 & 4.06 & 0.902&224\\
S8 & 2630 & 54.4 & 4.03 & 0.927&192 \\
S1 & 3300 & 94.1 & 4.06 & 0.358&2119\\
 \hline  
\end{tabular}
\label{table2}
\end{table}

The source of the initial hydrogen or dust is still an unanswered question.  It could have been created as a giant initial hydrogen or dust cloud or be continually formed incrementally (by a Creator or one of the steady state theory mechanisms or quantum mechanics creating something out of nothingness).  However, if it was always just there, it should have already collapsed earlier.  Thus, what has been presented so far has not ruled out the need for the dust (hydrogen), nor ruled out the beginning of time.  The exact date of the beginning, however, based on the big bang and Doppler redshift is no longer necessarily 13.8 billion years.  We will also need new mechanisms for the redshift and a mechanism to keep it from collapsing.

\section{Non-collapsing Black Hole via Rotation}
Black holes are assumed to collapse very quickly to a singularity, since no force seems available to counteract its strong gravitation forces.  However, simple centripetal force increases with the square of the velocity and can grow to balance the gravitational forces.  The conservation of angular momentum would also encourage this to happen.  The mass just needs to be orbiting fast enough to cancel the gravitational forces.
Setting the centripetal force ($F_{cent}$) equal to the gravitational force at any radius within the black hole, one can compute the orbiting velocity, $V$, to effectively cancel its gravitational force ($F_{grav}$) \cite{zeilikP10}.
\begin{align}
F_{grav} &= \frac{GmM}{R^2} = F_{cent} = \frac{mV^2}{R} \\
V^2&=\frac{GM}{R}
\label{OrbitalVequation}
\end{align}

\subsection{At the surface (with radii $R_s$ and $R_{bh}$)}
Let $V=V_s$ denote the velocity for an object orbiting at the \ul{surface}. 
\\[1ex] \ul{For a Schwarzschild black hole, using Equation $\ref{OrbitalVequation}$ becomes:}
\begin{align}
V_s^2 &= \frac{GM}{R_s} = \frac{GM}{(2GM/C^2)} = \frac{C^2}{2} \\
V_s&=\frac{C}{\sqrt2} = .707C
\end{align}
\ul{For an $R_{bh}$ black hole:} 
\begin{align}
V_s^2 &= \frac{GM}{R_{bh}} = \frac{GM}{(GM/C^2)} = C^2 \\
V_s &= C
\end{align}
One might think this is a mistake, since nothing can travel as fast as the speed of light C, and everything traveling less than the speed of light would be pulled in and even light would not be able to escape but be locked in a circular orbit.  But this is exactly the black hole effect we were expecting.  Thus, the outer velocities of $.707C$  and $C$ look OK.

It may also seem impossible to get mass traveling at the speed of light.  But even with the classical finite (twice) height black hole, any mass dropped at $2*R_{bh}$ would be traveling at or near the speed of light by time it hit $R_{bh}$.  It is just the return velocity profile of throwing up a rock at initial velocity C from the surface of the finite (twice) height black hole.  The rock would start out at velocity C, but slow down and stop at the top where $R=2*R_{bh}$, then it would fall backward and impact the surface back at its initial velocity C.  Thus, simply dropping a rock at $R=2*R_{bh}$ would also be traveling at velocity C when it reached the surface of the black hole. 
\subsection{Within the finite black hole at radius $R$}
Here we consider various mass density functions $P(R)$ and compute their mass function $M(R)$ and needed velocity functions $V(R)$ to keep the black hole from collapsing. 
\subsubsection{$P(R)$ that is inversely proportional to $R^2$}
First, assume a Kepler mass density function $P(R)$ that is inversely proportional to $R^2$. This is typically seen within solar systems. That is 
\begin{equation}
P(R) = \frac{K}{R^2}
\end{equation}
 where K is chosen in order to produce the same overall mass M of the black hole. 
Since density depends on radius, mass is also a function of the radius.  Consider $M(R)$, the mass of a sphere with radius $R$. This mass can be expressed using the spherical shell method with density $P(R)$ and surface area $4\pi R^2$, 
\begin{align}
M(R)&=\int_0^RP(r)4\pi r^2 dr \\
&=\int_0^R\frac{K}{r^2}4\pi r^2 dr \\
&=\int_0^RK4\pi dr \\
&=4\pi K R
\label{Mass1}
\end{align}
$K$ is a constant chosen so that total mass $M=M(R_{bh})$ is equal to the average density $P_{avg}$ times its total volume $Vol=\frac{4\pi}{3}R_{bh}^3$.  We use these in order to solve for an expression for $K$, 
\begin{align}
M = M(R_{bh})= 4\pi K R_{bh} &=P_{avg}*(\frac{4\pi}{3}R_{bh}^3) &\\
K &=\frac{1}{3} P_{avg}R_{bh}^2
\end{align}
Plugging $K$ into our expression for $M(R)$ from Eq. \ref{Mass1} yields
\begin{align}
M(R)&=4\pi*(\frac{1}{3} P_{avg}R_{bh}^2)* R\\
&=P_{avg}*\frac{4\pi}{3}R_{bh}^3\left(\frac{R}{R_{bh}}\right)\\
&=M*\left(\frac{R}{R_{bh}}\right)
\end{align}
Then the orbital velocity at radius R, $V(R)$ can be derived using Eq. \ref{OrbitalVequation} and $R_{bh} = \frac{GM}{C^2}$.
\begin{align}
V(R)^2 &=\frac{G*M(R)}{R}\\
&=\frac{G *(M*(\frac{R}{R_{bh}}))}{R}\\
&=\frac{GM}{R_{bh}}\\
&=C^2  \label{Csquared1}
\end{align}
therefore 
\begin{equation}
V(R)=C
\end{equation}
Thus the velocity $V$ is a constant independent of $R$.  Therefore, an object has to maintain the speed of light at all radii to keep from falling inward.  Thus, anything going less than the speed of light will collapse into the center.  Thus, any black hole with a Kepler distribution inversely proportional to $R^2$ will collapse, just like one sees in the movies.
\subsubsection{$P(R)$ that is inversely proportional to $R$}
Now assume a mass density function $P(R)$ that is inversely proportional to $R$. This is typically seen within Galactic disks. That is 
\begin{equation}
P(R) = \frac{K}{R}
\end{equation}
 where K is chosen in order to produce the same overall mass M of the black hole. 
Since density depends on radius, mass is also a function of the radius.  Consider $M(R)$, the mass of a sphere with radius $R$. This mass can be expressed using the spherical shell method with density $P(R)$ and surface area $4\pi R^2$, 
\begin{align}
M(R)&=\int_0^RP(r)4\pi r^2 dr \\
&=\int_0^R\frac{K}{r}4\pi r^2 dr \\
&=\int_0^RK4\pi r dr \\
&=2\pi K R^2
\label{Mass2}
\end{align}
$K$ is a constant chosen so that total mass $M=M(R_{bh})$ is equal to the average density $P_{avg}$ times its total volume $Vol=\frac{4\pi}{3}R_{bh}^3$.  We use these in order to solve for an expression for $K$,   
\begin{align}
M=M(R_{bh})= 2\pi K R_{bh}^2 &=P_{avg}*(\frac{4\pi}{3}*R_{bh}^3)\\
K &=\frac{2}{3}P_{avg}R_{bh}
\end{align}
Plugging $K$ into our expression for $M(R)$ (Eq. \ref{Mass2}) yields

\begin{align}
M(R)&=2\pi (\frac{2}{3} P_{avg}R_{bh}) R^2\\
&=P_{avg}*\frac{4\pi}{3}R_{bh}^3\left(\frac{R^2}{R_{bh}^2}\right)\\
&=M*\left(\frac{R}{R_{bh}}\right)^2
\end{align}

Then the orbital velocity at radius R, $V(R)$ can be derived using Eq. \ref{OrbitalVequation} and and $R_{bh} = \frac{GM}{C^2}$.
\begin{align}
V(R)^2 &=\frac{G*M(R)}{R}\\
V(R)^2 &=\frac{G*(M*\left(\frac{R}{R_{bh}}\right)^2)}{R}\\
&=\frac{GM}{R_{bh}}*\frac{ R}{R_{bh}}\\
&={C^2}*\frac{R}{R_{bh}}  \label{Csquared2}
\end{align}
therefore 
\begin{equation}
V(R)=\frac{C}{\sqrt{R_{bh}}}\sqrt R
\end{equation}
Thus $V(R)$ is proportional to $\sqrt R$. 
\begin{equation}
V(R)=C\sqrt \frac{R}{R_{bh}}
\end{equation}
At the surface, $R=R_{bh}$, velocity is the speed of light $V=C$. As one moves inwards the fractional radius  ($\frac{R}{R_{bh}}$) decreases and approaches zero.  The velocity is proportional to the square root of this fractional radius. 
Thus the black hole with mass distribution inverse proportional to $R$ does not need to collapse. 

\subsubsection{$P(R)$ that is a constant  }
Now assume a mass density function $P(R)$ that is constant.  That is 
\begin{equation}
P(R) = K
\end{equation}
where K is chosen in order to produce the same overall mass M of the black hole. Consider $M(R)$, the mass of a sphere with radius $R$. This mass can be expressed using the spherical shell method with density $P(R)$ and surface area $4\pi R^2$, 
\begin{align}
M(R)&=\int_0^RP(r)4\pi r^2 dr \\
&=\int_0^RK4\pi r^2 dr \\
&=\frac{4\pi}{3} K R^3
\label{Mass3}
\end{align}
$K$ is a constant chosen so that total mass $M=M(R_{bh})$ is equal to the average density $P_{avg}$ times its total volume $Vol=\frac{4\pi}{3}R_{bh}^3$.  We use these in order to solve for an expression for $K$, 
\begin{align}
M=M(R_{bh})= \frac{4\pi}{3} K R_{bh}^3&=P_{avg}*(\frac{4\pi}{3}R_{bh}^3)\\
K &= P_{avg}
\end{align}
Plugging $K$ into our expression for $M(R)$ (Eq. \ref{Mass3}) yields

\begin{align}
M(R)&=\frac{4\pi}{3}P_{avg} R^3\\
&=P_{avg}*\frac{4\pi}{3}R_{bh}^3\left(\frac{R^3}{R_{bh}^3}\right)\\
&=M*\left(\frac{R}{R_{bh}}\right)^3
\end{align}

\begin{figure}[H]
\begin{center}
\includegraphics[width=.8\textwidth]{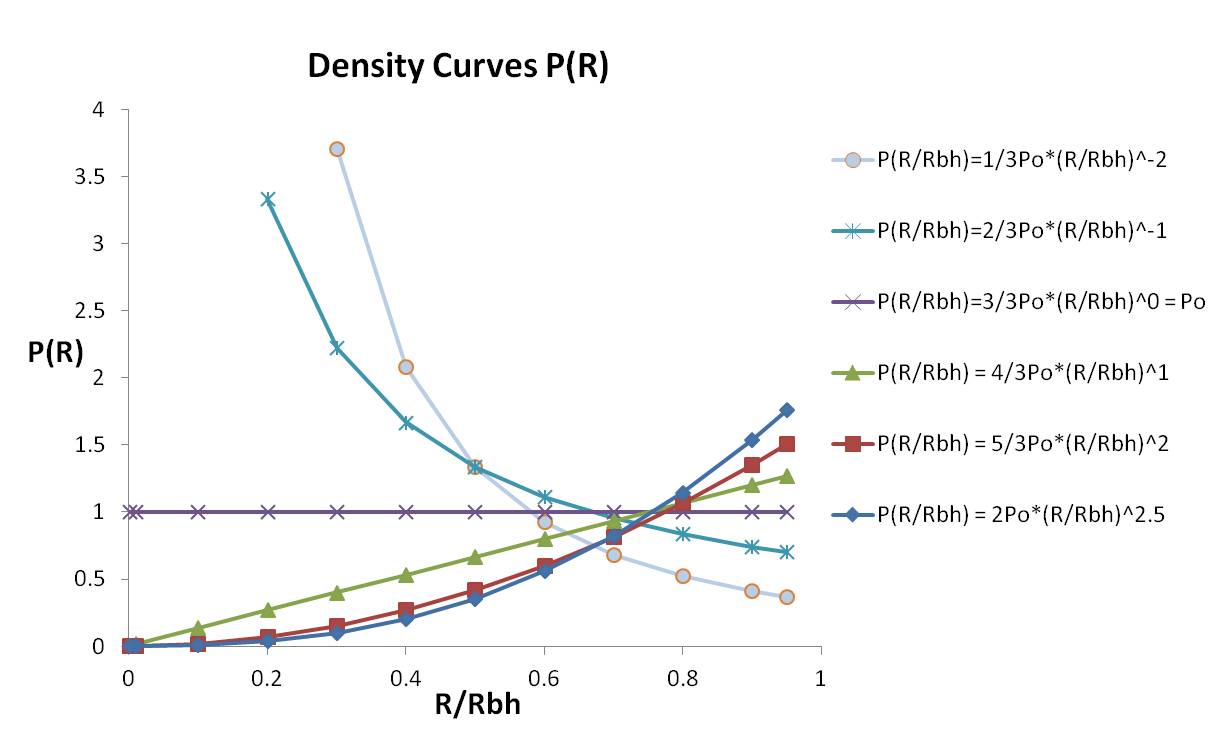}
\captionsetup{width=0.9\textwidth}
\caption{These are the density curves with $P(R)$ proportional to $1/R^2$, $1/R$, $K$, $R$, $R^2$, and $R^3$.  These are plotted at fractional radii $R/R_{bh}$.  The density continuously increases and approach infinity as $R$ approaches zero for the $1/R^2$ and $1/R$ functions;  $P$ is constant ($P_{avg}$) for the constant density distribution;  and $P$ decreases from the outside $R_{bh}$ radius and approaches zero for the $R$, $R^2$, and $R^3$ distributions.}
\label{DensityCurves}
\end{center}
\end{figure}

\begin{figure}[H]
\begin{center}
\includegraphics[width=.8\textwidth]{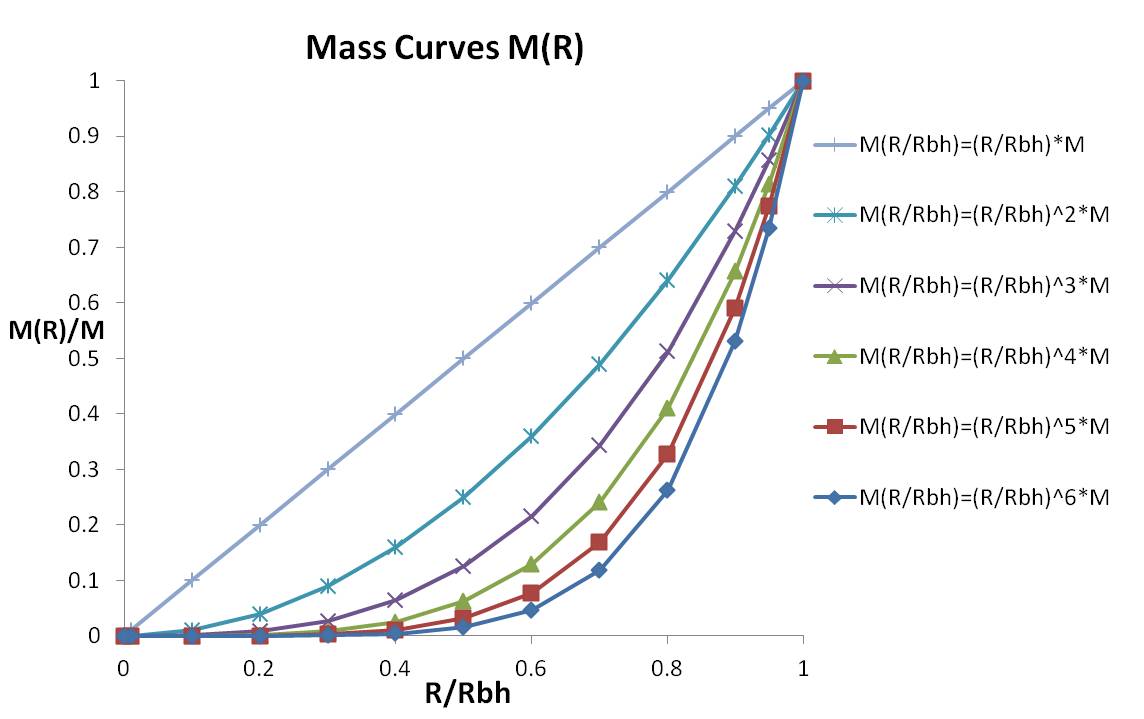}
\captionsetup{width=0.9\textwidth}
\caption{Internal Binding $M(r)$ for $R_{bh}$. These mass functions corresponding to the mass density functions $P(r)$ proportional to $1/r^2$, $1/r$, $K$, $r$, $r^2$, and $r^3$.  These are plotted at fractional radii $R/R_{bh}$.  These are the binding Mass functions, $M(r)$, and represent the sum of all mass with radius $r$.   They each have the same total mass M at $r=R_{bh}$, and drops off at various rates towards zero in the center.   The mass drops off linearly for the $P(r)$ proportional to the $1/R^2$ mass density function; thus half its mass is within $R=1/2 R_{bh}$.   The other functions have the majority of their mass outside, away from the enter. }
\label{InternalBindingMass}
\end{center}
\end{figure}

\begin{figure}[H]
\begin{center}
\includegraphics[width=.8\textwidth]{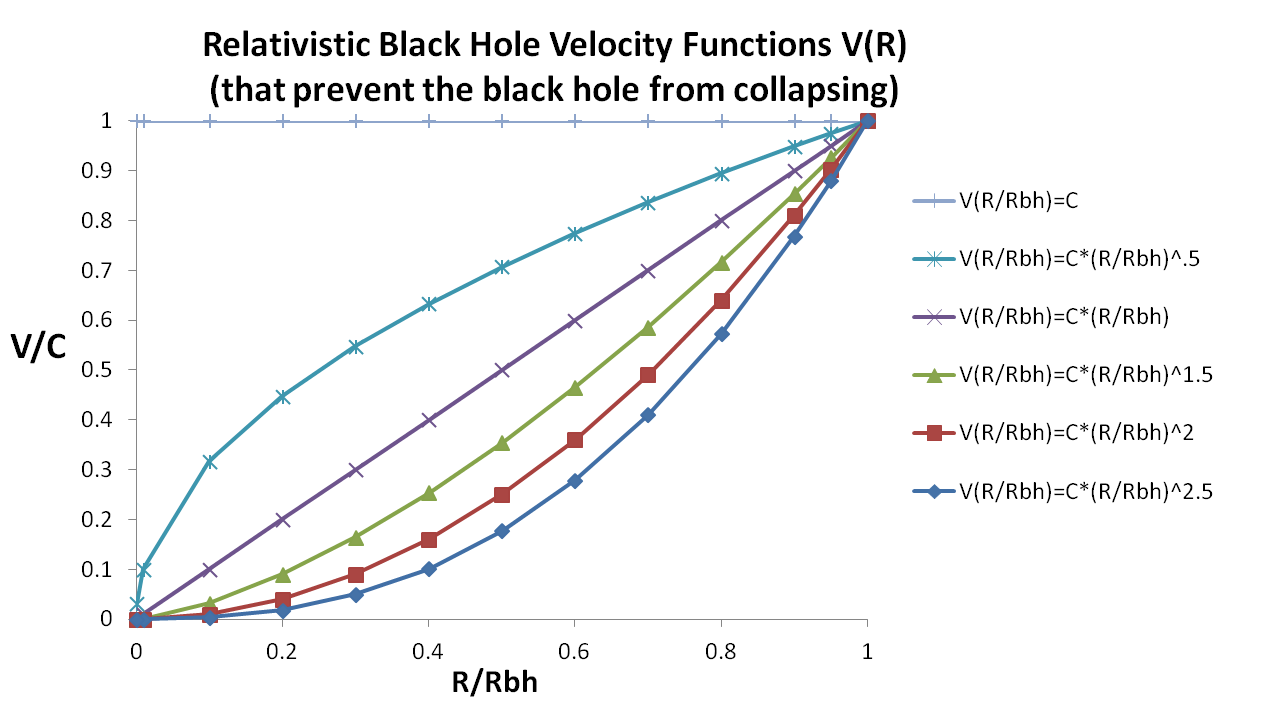}
\captionsetup{width=0.9\textwidth}
\caption{Velocity Curves for $R_{bh}$ These velocity functions corresponding to the binding mass functions, and  $P(r)$ functions proportional to $1/r^2$, $1/r$, $K$, $r$, $r^2$, and $r^3$.  These are plotted at fractional radii $R/R_{bh}$.  These are the orbital velocity functions, $V(r)$ and represents the orbital velocity at radius $r$, to exactly balance the gravitational binding mass inside radius $r$.  The mass outside the $r$ radius actually pulls outward, but the outward force is exactly cancelled by the outward pulling force on the far side the sphere outside of radius $r$ (due to the equivalence principle \cite{zeilikP9} ).   They each must be traveling at the speed of light, $C$, at the black hole boundary ($R_{bh}$).  The top function, that corresponds to the $P(R)$ proportional to $1/R^2$ function, must maintain the speed of light at all radii to keep from collapsing.  Since this is impractical in nature, this black hole will collapse.  The other velocity functions drops off  from the boundary at $V=C$ decreasingly at various rates towards $V=0$ in the center.   Thus, any mass density curve that falls off slower than $1/R^2$, can balance the huge gravitational forces with the appropriate velocity functions.}
\label{VelocityCurves}
\end{center}
\end{figure}

\begin{figure}[H]
\begin{center}
\includegraphics[width=.8\textwidth]{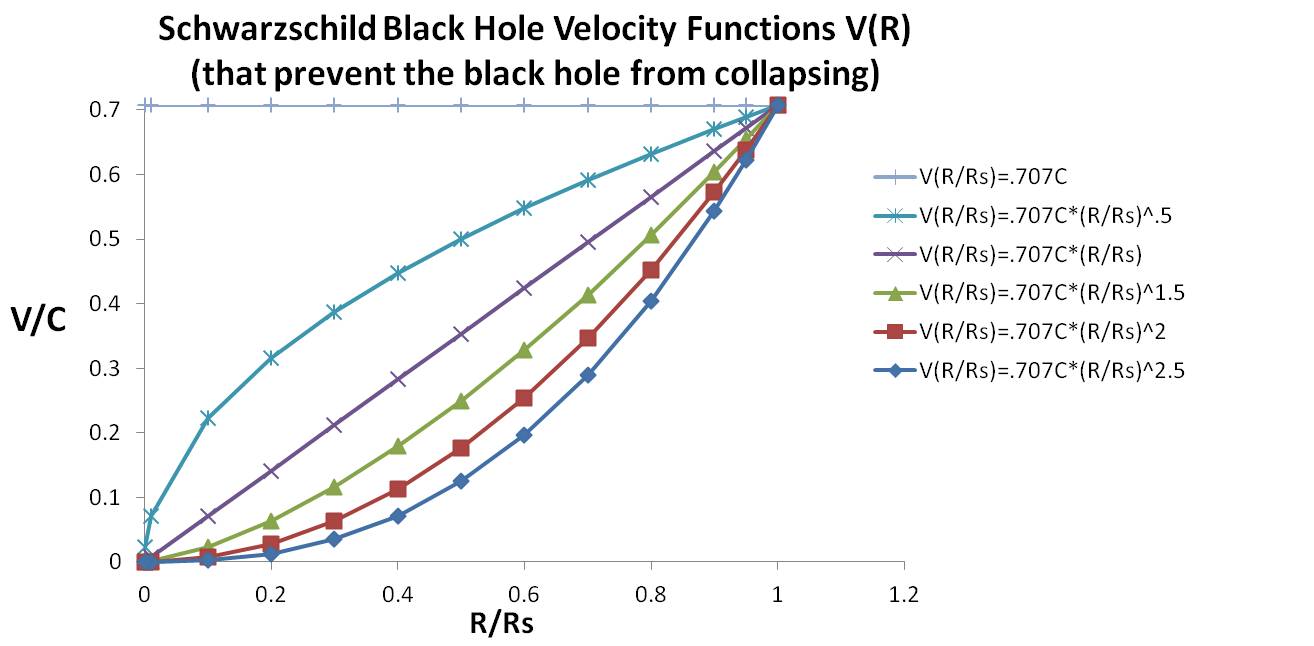}
\captionsetup{width=0.9\textwidth}
\caption{Velocity Curves for $R_{s}$ These velocity functions corresponding to the binding mass functions, and  $P(r)$ functions proportional to $1/r^2$, $1/r$, $K$, $r$, $r^2$, and $r^3$.  These are plotted at fractional radii $R/R_{s}$.  These are the Schwarzschild Black Hole orbital velocity functions, $V(R)$ and represents the orbital velocity at radius $R$, to exactly balance the binding mass inside this $R$ radius.  These are similar to the Relativistic Black Hole orbital velocity functions, but they go up to $.707C$ instead of to $C$.}
\label{RsVelocityCurves}
\end{center}
\end{figure}

Then the orbital velocity at radius R, $V(R)$ can be derived using Eq. \ref{OrbitalVequation} and Eq. \ref{densityEquation}
\begin{align}
V(R)^2 &=\frac{G*M(R)}{R}\\
V(R)^2 &=\frac{G*(M*\left(\frac{R}{R_{bh}}\right)^3)}{R}\\
&=\frac{GM}{R_{bh}}*\frac{R^2}{R_{bh}^2}\\ 
&=\frac{C^2}{R_{bh}^2}R^2 \label{Csquared3}
\end{align}
therefore 
\begin{equation}
V(R)=\frac{C}{R_{bh}}R
\end{equation}
Thus $V(R)$ is proportional to $R$. 
\begin{equation}
V(R)=C \left(\frac{R}{R_{bh}}\right)
\end{equation}
At the surface, $R=R_{bh}$, velocity is the speed of light $V=C$. As one moves inwards the fractional radius  ($\frac{R}{R_{bh}}$) decreases and approaches zero.  The velocity is proportional to this fractional radius. 
Thus the black hole with constant mass distribution does not need to collapse. 
\\[1ex]These calculations were continued for density functions proportional to $R$, $R^2$ and $R^3.$  These and the earlier black hole density, mass, velocity functions are shown in Table \ref{table3} and in Figures \ref{DensityCurves} through \ref{VelocityCurves}. 
\\[1ex] These calculations were also continued for the Schwarzschild black holes with density functions proportional to $1/R^2$, $1/R$, constant, $R$, $R^2$ and $R^3$.   The black hole density and mass functions do not change but the velocity functions are multiplied by an additional .707 as shown in Table \ref{table3b} and the new Schwarzschild velocity curves are plotted in Figure \ref{RsVelocityCurves}.  The extra .707 comes from the square root of 1/2 introduced by the final substitution of $\frac{C^2}{2}$ for $\frac{GM}{R}$ in Equations \ref{Csquared1}, \ref{Csquared2}, and \ref{Csquared3} instead of $C^2$; since $R_s = \frac{2GM}{C^2}$.
\begin{table}
\caption{Relativistic $R_{bh}=GM/C^2$ Density, Mass, Velocity Curves}
\begin{center}
\begin{tabular}{| c | c | c | c | }
  \hline                        
$P(R)$& $M(R)$& $V(R)$& $V(.5)$\\
\hline
$\frac{1}{3}P_{avg}R_{bh}^2/R^2$	& $M*(R/R_{bh})$ &$V=C$ & $C$ \\
$\frac{2}{3}P_{avg}R_{bh}/R$	 & $M*(R/R_{bh})^2$ &$V=C(R/R_{bh})^\frac{1}{2}$ & $0.707C$ \\
$P_{avg}$ constant & $P_{avg}$  &$V=C(R/R_{bh})$ & $0.5C$ \\
$\frac{4}{3}P_{avg}R/R_{bh}$	& $M*(R/R_{bh})^4$ &$V=C(R/R_{bh})^\frac{3}{2}$ & $0.35C$ \\
$\frac{5}{3}P_{avg}R^2/R_{bh}^2$ & $M*(R/R_{bh})^5$ &$V=C(R/R_{bh})^2$ & $0.25C$ \\
$2P_{avg}R^3/R_{bh}^3$ & $M*(R/R_{bh})^6$ &$V=C(R/R_{bh})^\frac{5}{2}$ & $0.177C$ \\
 \hline  
\end{tabular}
\end{center}
\label{table3}
\end{table}

\begin{table}
\caption{Schwarzschild $R_{s}=2GM/C^2$ Density, Mass, Velocity Curves}
\begin{center}
\begin{tabular}{| c | c | c | c | }
  \hline                        
$P(R)$& $M(R)$& $V(R)$& $V(.5)$\\
\hline
$\frac{1}{3}P_{avg}R_{bh}^2/R^2$	& $M*(R/R_{bh})$ &$V=.707C$ & $.707C$ \\
$\frac{2}{3}P_{avg}R_{bh}/R$	 & $M*(R/R_{bh})^2$ &$V=.707C(R/R_{bh})^\frac{1}{2}$ & $0.5C$ \\
$P_{avg}$ constant & $P_{avg}$  &$V=.707C(R/R_{bh})$ & $0.35C$ \\
$\frac{4}{3}P_{avg}R/R_{bh}$	& $M*(R/R_{bh})^4$ &$V=.707C(R/R_{bh})^\frac{3}{2}$ & $0.25C$ \\
$\frac{5}{3}P_{avg}R^2/R_{bh}^2$ & $M*(R/R_{bh})^5$ &$V=.707C(R/R_{bh})^2$ & $0.175C$ \\
$2P_{avg}R^3/R_{bh}^3$ & $M*(R/R_{bh})^6$ &$V=.707C(R/R_{bh})^\frac{5}{2}$ & $0.125C$ \\
 \hline  
\end{tabular}
\end{center}
\label{table3b}
\end{table}
As $R$ decreases to zero the density of the first two density curves of Figure 6 increase towards infinity at rate $R$ or $R^2$.  However their central mass, which is equal to density times volume, goes to zero since volume decreases to zero at a faster rate of $R^3$ (Figure \ref{InternalBindingMass}). Thus there does not need to be a singularity in the center, if the black hole does not collapse. 

As discussed earlier, the Kepler's $1/R^2$ density function will collapse as soon as its velocity drops below the speed of light (Figure \ref{VelocityCurves}). Any density curve that falls off slower than the  Kepler's $1/R^2$ density function, including all of the remaining densities curves will keep the black hole from collapsing at speeds less than $C$.   For a Schwarzschild black hole, the Kepler's $1/R^2$ density function will collapse as soon as its velocity drops below .707C (Figure \ref{RsVelocityCurves}) since the velocity to maintain orbit is .707C. Traveling at speeds greater than .707C using the classical Schwarzschild model would travel outwards but eventually fall back inwards. Using the relativistic Schwarzschild model the extra curvature of space would keep one from escaping the event horizon.

Any density curve that falls off slower than the  Kepler's $1/R^2$ density function, including all of the remaining densities curves will keep the black hole from collapsing at speeds less than $C$ and $.707C$.   The density proportional to $1/R$ or constant density $P_{avg}$ seems more likely.  The densities proportional to $R$, $R^2$, and $R^3$ require smaller velocities at lower orbits because most of their mass actually resides in higher orbits traveling at very high speeds.  

\vspace{3 mm}
Thus, black holes do not need to collapse if they are rotating and have a mass density function that drops off less than $1/R^2$.  Furthermore there does not need to be a singularity in the center.

\section{Sources of Redshifts }
Redshift ($z$) is the measured relative difference between the observed wavelength ($\lambda_{obsv}$) and emitted wavelength ($\lambda_{emit}$) of  light or other electromagnetic radiation from an object.
\begin{equation}
z=\frac{\lambda_{obsv}-\lambda_{emit}}{\lambda_{emit}}
\label{zwave}
\end{equation}
Note when $\lambda_{obsv}>\lambda_{emit}$ the shift $z$ is positive and the observed light will appear ``redder" than the emitted light. When $\lambda_{obsv}<\lambda_{emit}$ the shift $z$ is negative and the observed light will appear ``bluer" than the emitted light. These are denoted as redshifts and blueshifts.  
Redshifts are found in almost all distant galaxies and objects, and redshift increases with distance.    The actual observed data indicate redshift values for objects at different distances ($d$) given below.
\begin{equation}
z = .000076*d \qquad (\text{with d in MLY})
\label{redshift}
\end{equation}

\subsection{Relativistic Doppler Redshift}
Hubble found that distant objects largely had redshifts proportional to distance and he assumed the redshift to be associated with a Doppler velocity shift from an expanding universe.  
The classical linear Doppler redshift, $z$ for small values of $v$ is given \cite{zeilik477}, 
\begin{subequations}
\begin{align}
z  = \frac{V}{C}  &= .000076*d \qquad (\text{with d in MLY}) \\
V &= .000076* d* 300,000 \text{ km/sec} \\
V &= 23\text{ km/sec  * d} \qquad (\text{with d in MLY}) 
\end{align}
\label{dopplerZ}
\end{subequations}
Thus, assuming the redshift was a Doppler shift, this mapped into a velocity constant, called the Hubble constant $H_o$ for expansion velocity ($V$) of the universe \cite{zeilik434}. 
\begin{align}
V &= H_od \\
 &= 23\text{ km/sec } * d  \qquad (\text{with d in MLY}) 
\end{align}
The relativistic Doppler equations are given below \cite{zeilik477}:
\begin{align}
z +1&= (\frac{1+V/C}{1-V/C})^\frac{1}{2} \\    
\frac{V}{C} &= \frac{(z+1)^2 -1}{(z+1)^2+1}
\end{align}

The Relativistic Doppler and $z=\frac{v}{C}$ approximation values are tabulated in Table \ref{TableDoppler} and plotted below in Figure \ref{ZvsDistance}.  From the data and plots, the linear $d=z/.000076$ estimate is good out to 1.3 BLY with $z <0.1$ and reasonable out to 5 BLY with $z < 0.4$.  

\begin{table}
\caption{Relativistic Shifts and Errors}
\begin{center}
\begin{tabular}{| c | c | c | c | c | }
  \hline  
$\frac{V}{C}$& \pbox{20cm}{Relativistic\\Doppler Redshift\\$z= (\frac{1+V/C}{1-V/C})^\frac{1}{2} -1$} & \pbox{20cm}{Classical Linear\\Doppler Redshift \\$z=\frac{V}{C}$ for $V$ \textless\textless $C$} & \pbox{20cm}{$z=\frac{V}{C}$ \\ \%Err} &	\pbox{20cm}{$d=\frac{V}{H_o}$  \\$d=\frac{z}{.000076}$ \\(MLY)} \\
\hline
0.99&	13.107&	0.99&	-92\%&	13026 \\
0.95&	5.245&	0.95&	-82\%&	12500\\ 
0.9&	3.3589&	0.9&	-73\%&	11842\\
0.8&	2&	0.8&	-60\%&	10526\\
0.7&	1.3805&	0.7&	-49\%&	9211\\
0.6&	1&	0.6&	-40\%&	7895\\
0.5&	0.7321&	0.5&	-32\%&	6579\\
0.4&	0.5275&	0.4&	-24\%&	5263\\
0.3&	0.3628&	0.3&	-17\%&	3947\\
0.2&	0.2247&	0.2&	-11\%&	2632\\
0.1&	0.1055&	0.1&	-5\%&	     1316\\
0.08&	0.0834&	0.08&	-4\%&	1053\\
0.06&	0.0619&	0.06&	-3\%&	789\\
0.04&	0.0408&	0.04&	-2\%&	526\\
0.02&	0.0202&	0.02&	-1\%& 	263\\
0.001& 	0.001&	0.001& 	0\%&	13.16\\
\hline
\end{tabular}
\end{center}
\label{TableDoppler}
\end{table}

\begin{center}
\begin{figure}[H]
\includegraphics[width=\textwidth]{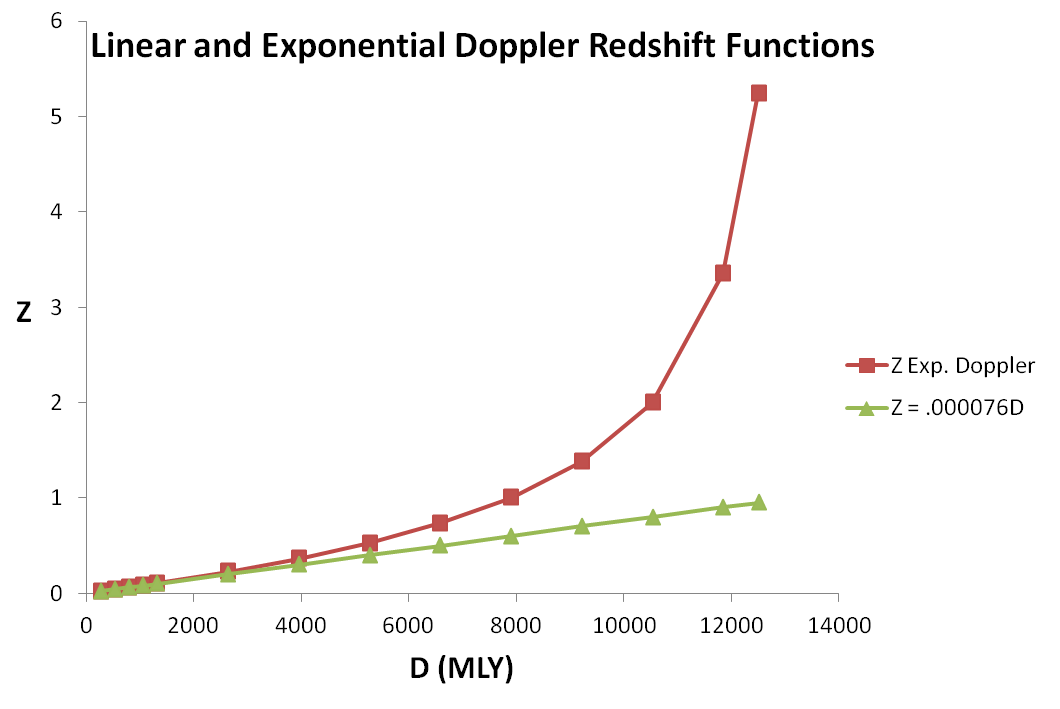}
\captionsetup{width=0.9\textwidth}
\caption{z v.s. Distance (MLY). These are plots of the linear Doppler redshift equation and the exponential (relativistic) Doppler equation.  The plots shows them almost identical (within $2\%$) out to 500 MLY, similar out to 1000 MLY, but radically different afterwards.  The linear Doppler redshift end at the edge of the expanding universe with a $z = 1$, corresponding to a velocity of $C$.  However, in nature, redshift values of distant quasars and galaxies have been measured with $z >1$. The relativistic Doppler equation permits redshift values greater than 1 as an object approaches the speed of light. }
\label{ZvsDistance}
\end{figure}
\end{center}

The equations depart after 1.3 BLY with $z$ values growing exponentially, where the linear approximation is limited to $z=1$ at the perceived radius of the universe.  Astronomers use these exponentially increasing $z$ values to explain the most distant quasars (with $z=7$) without requiring them to be at 92 BLY using the linear approximation equation (e.g. $z=7=.000076d MLY$; $d =7/.000076 = 92000 MLY$)

Before the year 2000, astronomers only had accurate distance measures based on redshifts to supernovae out to about 0.5 BLY.  However, in 2003, accurate super nova measurements have been extended out to about 1.3 BLY and the light appears to be following the theoretical non-linear relativistic curve \cite{hartnett}.  That is, it matches the relativistic Doppler equation.  But as indicated by Table  \ref{TableDoppler}, at this point, the linear equation is only off by less than 5\%.

However these researchers then plugged these same data into their best expanding universe models and couldn't find a fit unless they tweaked terms or included an accelerated expansion.   They chose to introduce a new energy source called Dark Energy to preserve their expanding universe model.  Normally, when one's best model does not fit the data, they should consider abandoning their model.   Hopefully, after reviewing Theorem 1 of Section 4 which disproves the expanding universe, cosmologist might want to reconsider their model.   

Even if the expanding universe and Doppler shift velocity interpretation are invalid, the observed redshift values of $z = .000076*d$ (with $d$ in MLY) still need to be explained.  

Without  a relativistic Doppler shift equation cosmologist might have concluded that the shift was linear out to their best data measurements, but perhaps with some slight deviations of $z$ starting to show up at .5 to 1.3 BLYs.   They would then look for a physical cause for these slightly increased $z$ values in the past.  Or they might find a different “best” fit that better matches the more distant data, and have the closer (more recent) points showing a slightly lower redshift value than expected (perhaps due to a contracting universe).    This paper will pursue both: investigating the linear redshift $z =.000076*d$ MLY function and also curves that depart starting around .5 BLY.  This paper will also pursue different explanation for the large redshifts of quasars.

\subsection{Cosmological Redshift}
Some scientists replaced the Doppler based redshift explanation with a cosmological redshift explanation where the velocities increases because space itself is expanding.  This allows for velocity values higher than C, without violating the speed of light.

Since the Cosmological redshift is based on an expanding universe, the cosmological shift would actually become zero in a non-expanding universe and a blueshift in a rapidly collapsing universe.  In any event, both the Doppler based redshift and Cosmological based redshifts would go away in a non-expanding universe. 

\subsection{Relativistic Transverse Redshift}
\begin{figure}
        \centering
        \begin{subfigure}[b]{0.48\textwidth}
	\includegraphics[width=\textwidth]{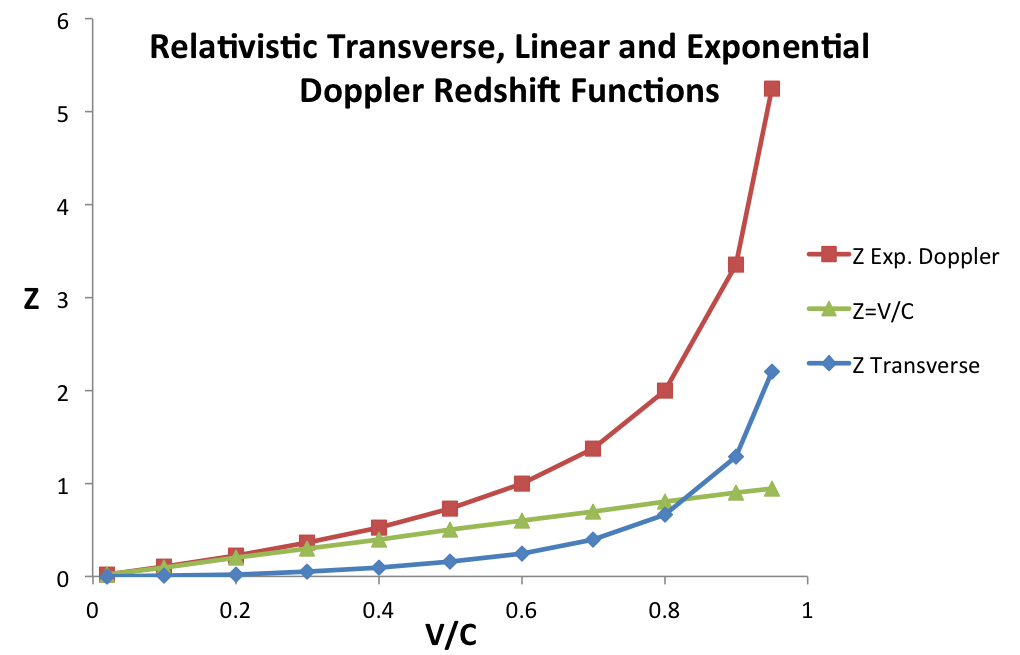}
\captionsetup{width=0.9\textwidth}
	\caption{$z$ v.s. Velocity ($\frac{V}{C}$) values. These are the relativistic Doppler, linear Doppler, and relativistic transverse redshift v.s. velocity functions.
	Note that the relativistic transverse redshift is significantly less than the relativistic Doppler redshift. However, both Relativistic function eventually go up exponentially near $V=C$.}
	\label{ZvsVelocity}
        \end{subfigure}
	\qquad 
        \begin{subfigure}[b]{0.45\textwidth}
	\includegraphics[width=\textwidth]{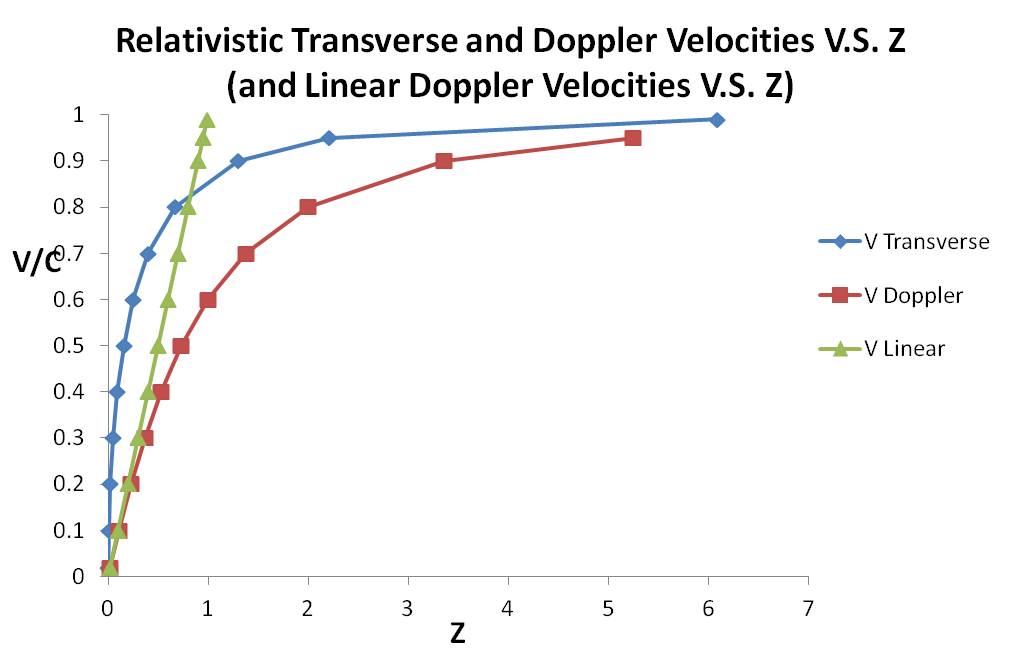}
\captionsetup{width=0.9\textwidth}
	\caption{Velocity ($\frac{V}{C}$) v.s. $z$ Value. These are velocities  plots of the relativistic transverse redshift and relativistic Doppler redshift functions v.s. the redshift. Note that the relativistic transverse velocity is significantly higher than the relativistic Doppler velocity to produce the same redshift.}
	\label{VelocityvsZ}
        \end{subfigure}
\caption{Relativistic Transverse Redshift Compared to the Relativistic Doppler Shifts}
\end{figure}

In addition to relativistic Doppler shift due to the speed of light moving away from the observer, a relativistic transverse redshift would also be observed due to time dilation (time slowing down) if the universe was rapidly rotating at relativistic speeds to keep from collapsing.  Einstein's relativistic transverse redshift ($z_t$) equations are \cite{gordon}:
\begin{align}
z_t &= \frac{1}{(1-V^2/C^2)^\frac{1}{2}} - 1\\
 \frac{V}{C} &= \left(1-\frac{1}{(z_t+1)^2}\right)^\frac{1}{2} \label{transverse}
\end{align}
These are shown compared to the relativistic Doppler shifts Figures \ref{ZvsVelocity} and \ref{VelocityvsZ} and Table \ref{TableTransverseDoppler}.  Note that the Transverse shift is considerably smaller than the Doppler shift. Thus, to get the observed redshift $z$ seen in Doppler, the transverse velocity would need to be greater (.e.g. $z_t=.25$ at $V=.6C$ for transverse instead of $V=.22C$ for Doppler).  

\begin{center}
\begin{figure}[H]
\includegraphics[width=.9\textwidth]{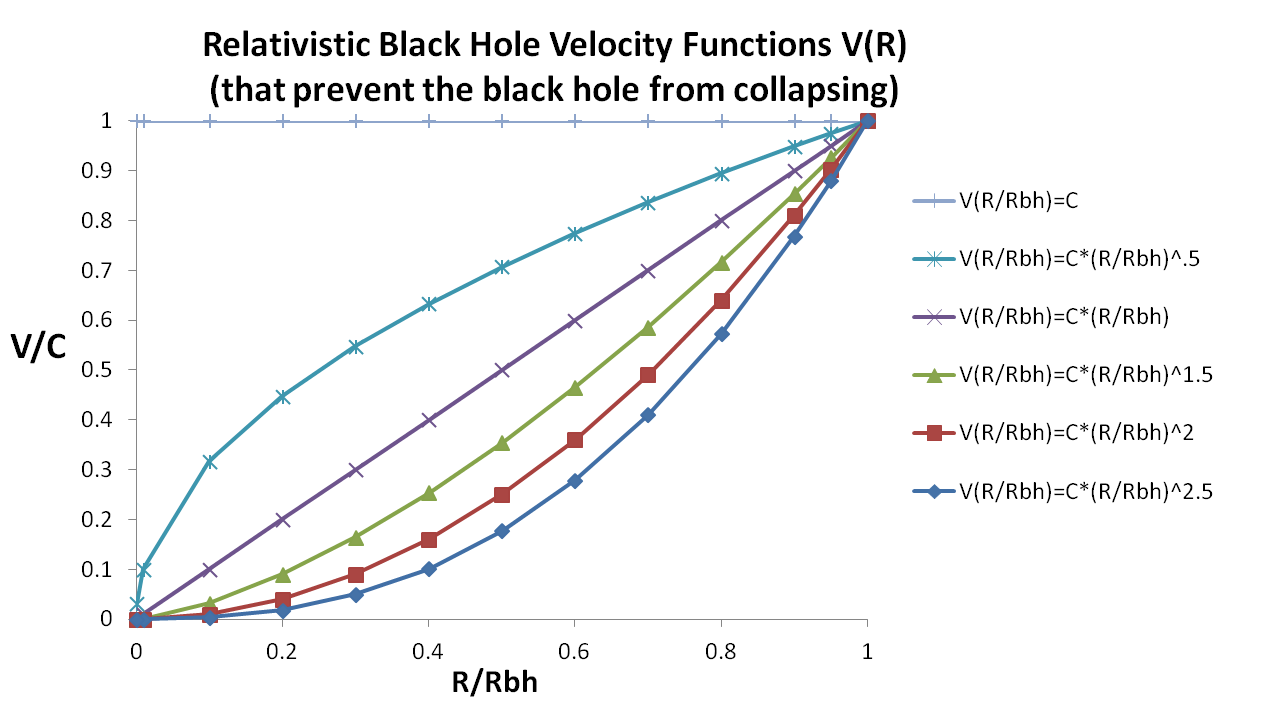}
\captionsetup{width=0.9\textwidth}
\caption{Velocity Curves for $R_{bh}$. These are plots of the velocity for the six mass density distributions (repeated from Figure \ref{VelocityCurves}). This will enable comparisons with the transverse redshift of Figure \ref{Redshift}.}
\label{VelocityCurves2}
\end{figure}
\end{center}

\begin{center} 
\begin{figure}[H]
\includegraphics[width=.9\textwidth]{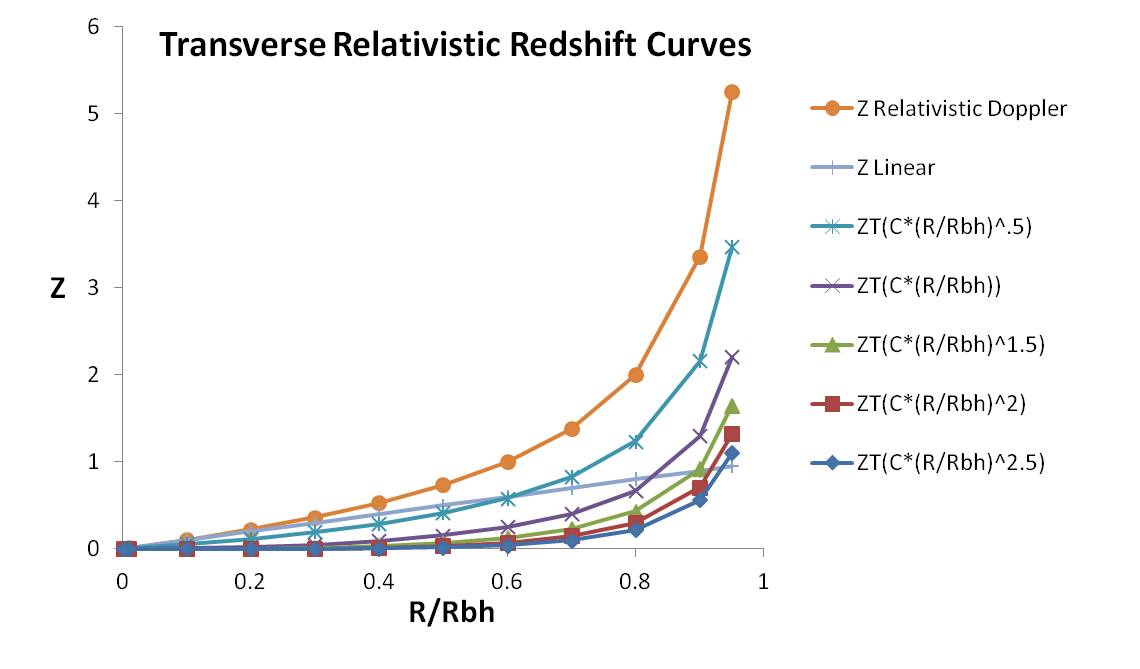}
\captionsetup{width=0.9\textwidth}
\caption{Transverse Relativistic Redshift. These are plots of the transverse redshift that would be produced from the velocity functions (of  Figure \ref{VelocityCurves2}) for the six mass density distributions.
Also included are the original relativistic Doppler and linear Doppler redshifts.  Note, no curves are a perfect match, but the  $V=C*(R/R_{bh})^{.5}$ is the closest and lies between the two and corresponds to the P(R) = 1/R mass density distribution. }
\label{Redshift}
\end{figure}
\end{center}

The rotating velocity curves (from Figure \ref{VelocityCurves}) are repeated in Figure \ref{VelocityCurves2} along with their matching translational redshifts curves in Figure \ref{Redshift}. All the velocity curves create relativistic transverse redshifts as their speeds approach the speed of light C.  The redshift for the $V=C$ curve was not plotted since this would always have infinite redshift.  Values are plotted up to .95C.  Higher values of $z_t$ could be achieved by approaching closer to the speed of light (e.g. .99C).  Thus, any of these speed curves can cause high $z_t$ values, near the edge of the Black hole.  

\begin{table}
\caption{Relativistic Shifts and Errors}
\begin{center}
\begin{tabular}{| c | c | c | c | c | c | }
  \hline  
$\frac{V}{C}$& \pbox{20cm}{Relativistic\\Transverse Redshift\\$z_t = \frac{1}{(1-V^2/C^2)^\frac{1}{2}} - 1$  }& \pbox{20cm}{Relativistic\\Doppler Redshift\\$z= (\frac{1+V/C}{1-V/C})^\frac{1}{2} -1$}& \pbox{20cm}{Classical Linear \\Doppler Redshift\\$z=\frac{V}{C}$} & \pbox{20cm}{$z=\frac{V}{C}$ for $V$ \textless\textless $C$\\ \%Err} &	\pbox{20cm}{$d=\frac{V}{H_o}$  \\$d=\frac{z}{.000076}$ \\ (MLY)} \\
\hline 
0.99&	6.0888&	13.107&	0.99&	-92\%&	13026\\
0.95&	2.2026&	5.245&	0.95&	-82\%&	12500\\
0.9&	1.2941&	3.3589&	0.9&	-73\%&	11842\\
0.8&	0.6667&	2&	0.8&	-60\%&	10526\\
0.7&	0.4003&	1.3805&	0.7&	-49\%&	9211\\
0.6&	0.25&	1&	0.6&	-40\%&	7895\\
0.5&	0.1547&	0.7321&	0.5&	-32\%&	6579\\
0.4&	0.0911&	0.5275&	0.4&	-24\%&	5263\\
0.3&	0.0483&	0.3628&	0.3&	-17\%&	3947\\
0.2&	0.0206&	0.2247&	0.2&	-11\%&	2632\\
0.1&	0.0050&	0.1055&	0.1&	-5\%&	1316\\
0.08&	0.0032&	0.0834&	0.08&	-4\%&	1052\\
0.06&	0.0018&	0.0619&	0.06&	-3\%&	789\\
0.04&	0.0008&	0.0408&	0.04&	-2\%&	526\\
0.02&	0.0002&	0.0202&	0.02&	-1\%&	263\\
0.001&	5E-07&	0.001&	0.001&	0\%&	13.16\\
\hline
\end{tabular}
\end{center}
\label{TableTransverseDoppler}
\end{table}

\begin{center} 
\begin{figure}[H]
\includegraphics[width=.8\textwidth]{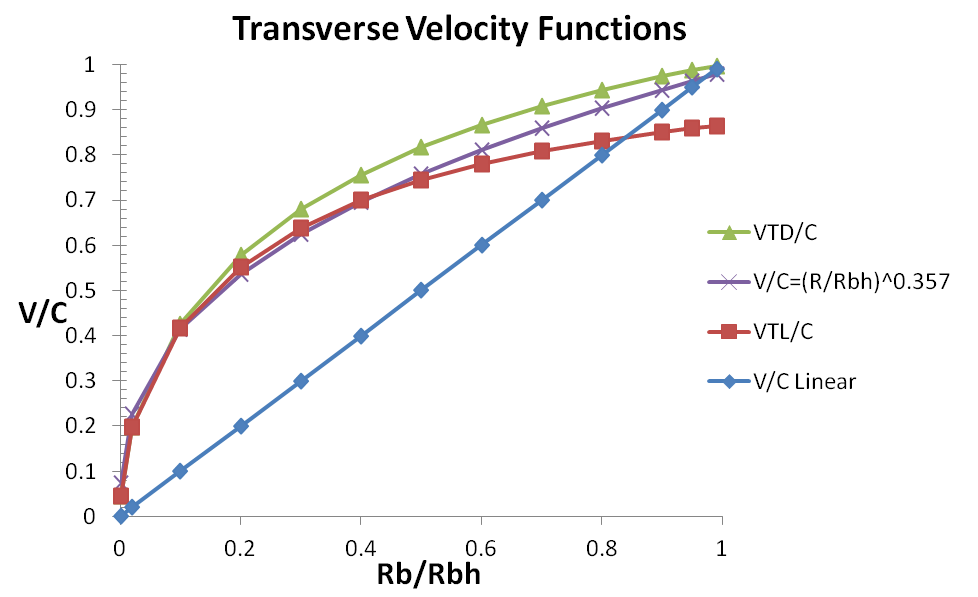}
\captionsetup{width=0.9\textwidth}
\caption{Transverse Velocity Functions. If one plugs the linear Doppler shift into the transverse velocity equation, on can get the velocity function that matches the linear Doppler shift (plotted as red squares)
If one plugs the relativistic Doppler shift into the transverse velocity equation, on can get the velocity function that matches the linear Doppler shift (plotted as green triangles)
A polynomial function (found by trial and error) between these two is $V/C = (R/R_{bh})^{0.375}$. Also, provided is the original linear Doppler redshift plot.}
\label{TransverseVeolocity}
\end{figure}
\end{center}

One could calculate the \ul{Transverse Linear} velocity ($V_{TL}$) curve that exactly matches the Hubble linear redshift curve ($z_l = .000076*d$ with d in MLY Eq. \ref{redshift}) by inserting $z_l$ in the transverse equation (Eq. \ref{transverse}):
\begin{equation}
\frac{V_{TL}}{C} = \left(1 - \frac{1}{(.000076d+1)^2}\right)^\frac{1}{2} \qquad (\text{with d in MLY}) 
\end{equation}
One could calculate the \ul{Transverse Doppler} velocity ($V_{TD}$) curve that exactly matches the relativistic Doppler redshift curve by replacing $z_t$ in the transverse equation (Eq. \ref{transverse}):

\begin{equation}
\frac{V_{TD}}{C} = \left(1 -\frac{1}{(z_d +1)^2}\right)^\frac{1}{2}
\end{equation}
where
\begin{equation}
z_d = \left(\frac{1+V/C}{1- V/C}\right)^\frac{1}{2}-1
\end{equation}
and this doppler $V/C=0.000076*d$ from Eq. \ref{dopplerZ} that is,
\begin{equation}
z_d = \left(\frac{1+0.000076*d}{1- 0.000076*d}\right)^\frac{1}{2}-1
\end{equation}
\\[1ex] Figure \ref{TransverseVeolocity} plots these two transverse velocity curves, along with the original Hubble relationship for Doppler velocity, and a polynomial function $\frac{V}{C} = \left(\frac{R}{R_{bh}}\right)^{0.375}$ curve between the two.  Note that even the $V_{TL}$ Red curve based on constant linear increase in $z$ versus distance ($z=.000076d$) curves up very rapidly in speed, to produce a linear observed $z$ shift.  The green curve includes additional velocities to match the non-linear distances based on the original Doppler equation which raises it partially higher.  

\begin{center} 
\begin{figure}[H]
\includegraphics[width=.8\textwidth]{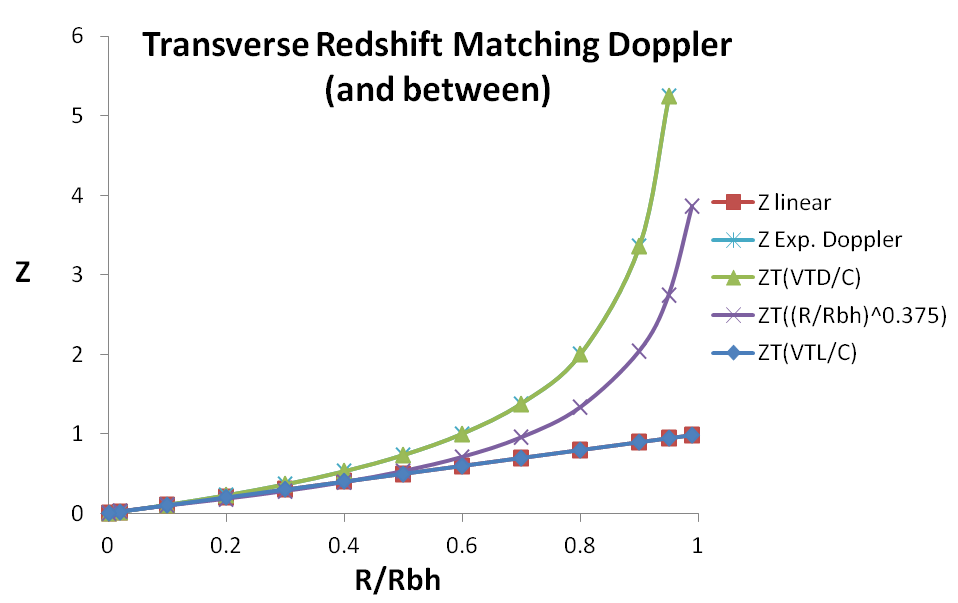}
\captionsetup{width=0.9\textwidth}
\caption{Transverse Z Functions. Running these transverse velocity values back through the transverse redshift function verifies that the transverse redshift function can produce identical redshift as either the original relativistic Doppler redshift (green triangle plot), or the linear Doppler (red squares plot).  The $z_t(R/R_{bh})^{0.375}$ comes between these values.}
\label{TransverseZ}
\end{figure}
\end{center}

Figure \ref{TransverseZ} plots the Transverse $z$ functions given the velocities from Figure \ref{TransverseVeolocity}.  These match the original Doppler shift and linear shifts, thus the transverse velocity curves would produce the observed redshift versus distance.  The intermediate $\left( \frac{R}{R_{bh}}\right)^{0.375}$ curve produces redshifts between these two.  The transverse redshift due to linear velocity function is plotted in blue. 

The $\frac{V}{C} = \left( \frac{R}{R_{bh}}\right)^{0.375}$ is a polynomial distribution like those computed earlier in Section 6 that closely fits between these two functions. The mass and mass density functions that correspond to this would be:
\begin{align}
M\left(\frac{R}{R_{bh}}\right) &= M* \left( \frac{R}{R_{bh}}\right)^{1.75} \\
P\left( \frac{R}{R_{bh}}\right)&= \frac{7}{12}P_{avg}\left( \frac{R}{R_{bh}}\right)^{-1.25}
\end{align}

The ($R/R_{bh})^{0.375}$ function was found by trial and error and then worked backwards to $P(R)$ proportional to $1/R^{1.25}$.
Following the procedure presented in Section 6 and assuming that the mass in not uniform, but the density is proportional to $1/R^{1.25}$. That is 
\begin{equation}
P(R)=\frac{K}{R^{1.25}}
\end{equation}
where $K$ is chosen in order to produce the same overall mass M of the black hole.  Since density depends on radius, mass is also a function of the radius.  Consider $M(R)$, the mass of a sphere with radius $R$. This mass can be expressed using the spherical shell method with density $P(R)$ and surface area $4\pi R^2$, 
\begin{align}
M(R)&=\int_0^RP(r)4\pi r^2 dr \\
&=\int_0^R\frac{K}{r^{1.25}}4\pi r^2 dr \\
&=\int_0^RK4\pi r^{0.75} dr \\
&=\frac{4\pi}{1.75} K R^{1.75}
\label{Mass4}
\end{align}
$K$ was chosen so that total mass $M =M(R_{bh})$ is equal to the average density $P_{avg}$ times its total volume $Vol=\frac{4\pi}{3}R_{bh}^3$.  We use this in order to solve for an expression for $K$, 
\begin{align}
M = M(R_{bh}) =\frac{4\pi}{1.75} K R_{bh}^{1.75} &= P_{avg}*(\frac{4\pi}{3}R_{bh}^3)\\
K &=\frac{7}{12} P_{avg}R_{bh}^{1.25}
\end{align}
Plugging $K$ into our expression for $M(R)$ (Eq. \ref{Mass4}) yields

\begin{align}
M(R) &= \frac{4\pi}{1.75} (\frac{7}{12} P_{avg}R_{bh}^{1.25}) R^{1.75} \\
&=P_{avg}*(\frac{4\pi}{3} R_{bh}^3)*\left(\frac{R^{1.75}}{R_{bh}^{1.75}}\right)\\
&=M*\left(\frac{R}{R_{bh}}\right)^{1.75}
\end{align}

Then the orbital velocity at radius R, $V(R)$ can be derived using Eq. \ref{OrbitalVequation} and $R_{bh}=\frac{GM}{C^2}$
\begin{align}
V(R)^2 &=\frac{G*M(R)}{R}\\
V(R)^2 &=\frac{G(M*\left(\frac{R}{R_{bh}}\right)^{1.75})}{R}\\
V(R)^2 &=\frac{GM}{R_bh}*\left(\frac{R}{R_{bh}}\right)^{.75}\\
&=C^2\left(\frac{R}{R_{bh}}\right)^{0.75}
\end{align}
Therefore 
\begin{equation}
V(R)=C \left(\frac{R}{R_{bh}}\right)^{0.375} \label{vfunr}
\end{equation}

\begin{center} 
\begin{figure}[H]
\includegraphics[width=.9\textwidth]{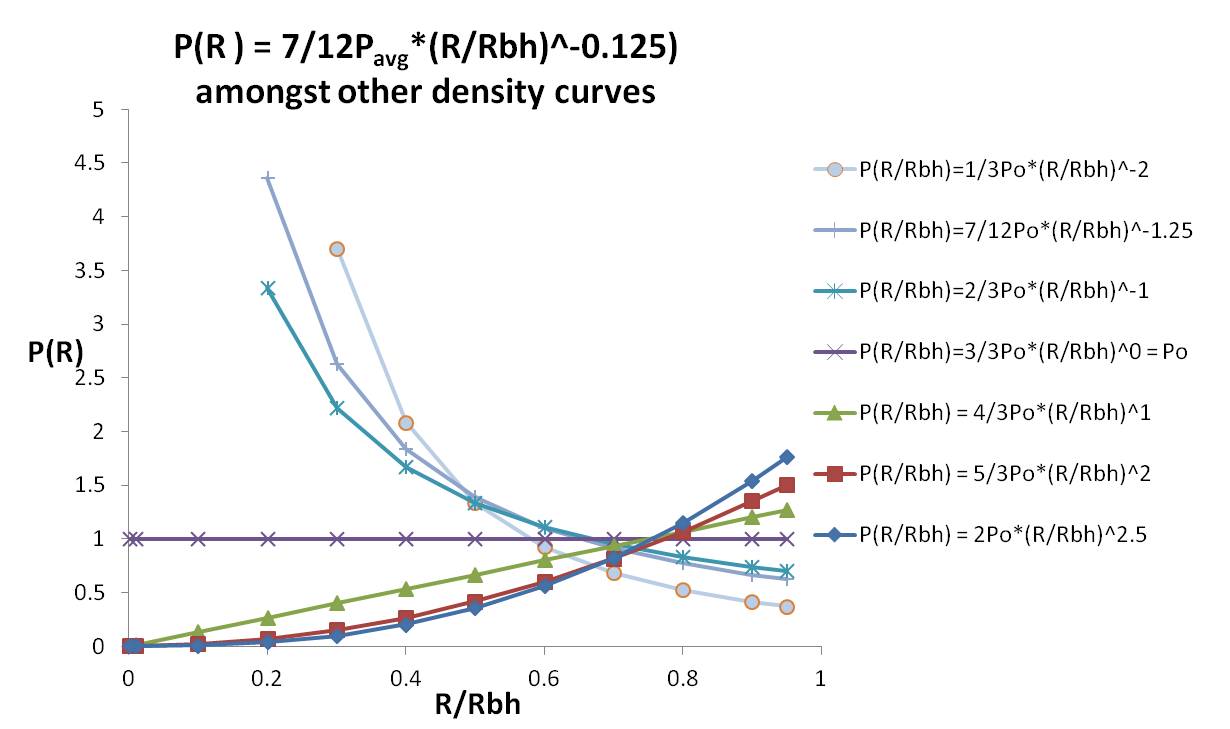}
\captionsetup{width=0.9\textwidth}
\caption{Transverse Density Function. This is the density curve that produces the matching transverse velocity curve (between the Doppler) redshift, plotted in relation to the other density curves.}
\label{TransverseDensity}
\end{figure}
\end{center}

\begin{center} 
\begin{figure}[H]
\includegraphics[width=.9\textwidth]{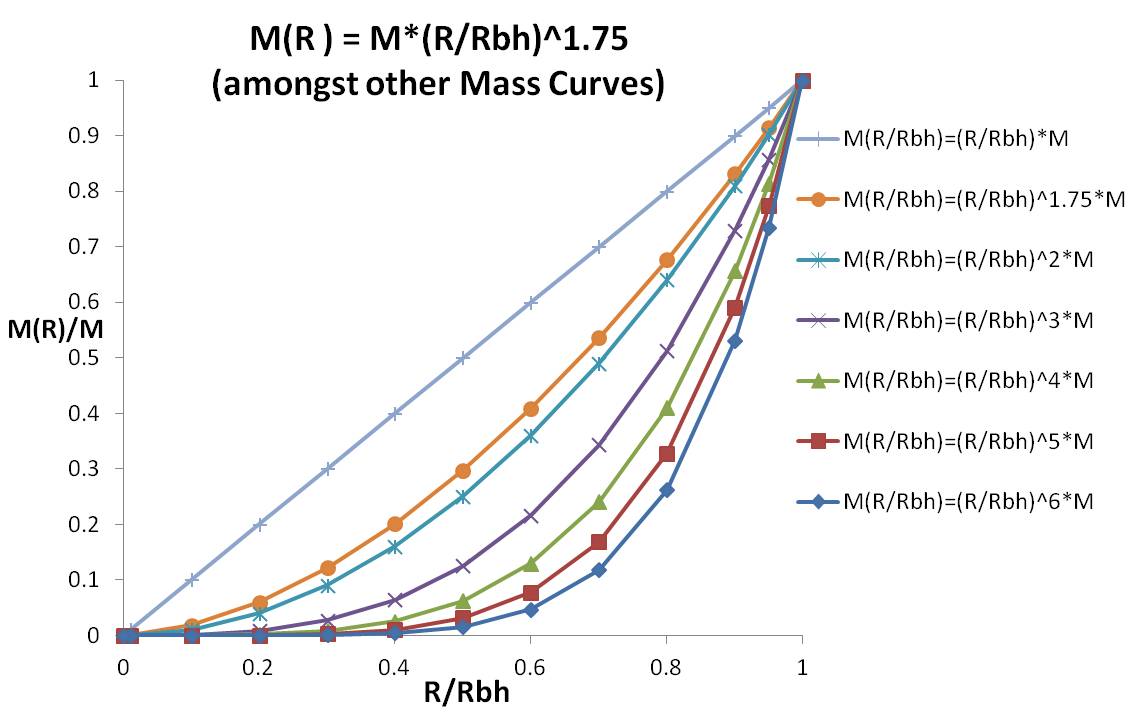}
\captionsetup{width=0.9\textwidth}
\caption{Transverse Mass Function. This is the binding mass curve that produces the matching transverse velocity curve (between the Doppler) redshift, plotted in relation to the other mass curves.}
\label{TransverseMass}
\end{figure}
\end{center}

\begin{center} 
\begin{figure}[H]
\includegraphics[width=\textwidth]{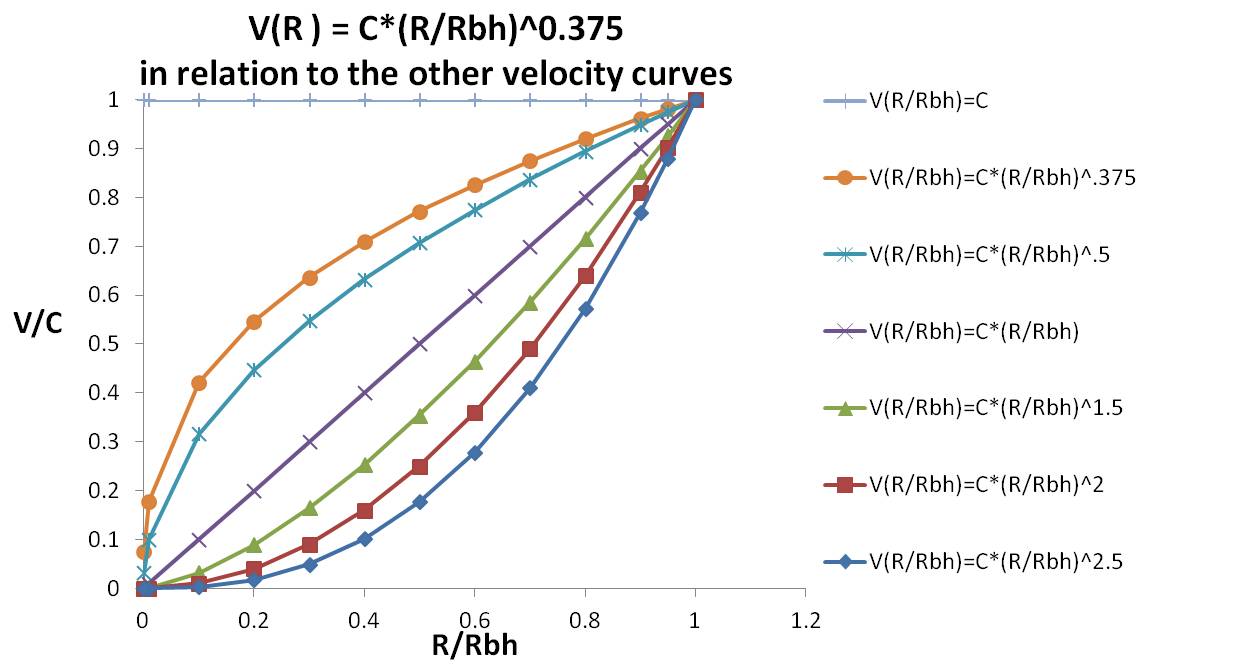}
\captionsetup{width=0.9\textwidth}
\caption{Transverse Velocity Function. This the matching transverse velocity curve (between the Doppler) redshift, plotted in relation to the other velocity curves.}
\label{TransverseVelocity}
\end{figure}
\end{center}

These density, mass, and velocity functions for the Transverse redshift are plotted in Figures \ref{TransverseDensity}, \ref{TransverseMass}, and \ref{TransverseVelocity}.  Most of its mass is on the outside, but the interior is still very dense due to its decreasing radius.   No singularity is needed since the mass is zero in the center.

Thus, using only transverse velocity redshift we have found the velocity, mass, and density distribution functions that fall in between the linear and Doppler redshift curves.  These will also keep a low density black hole from collapsing.  Note that the velocity and redshift curves apply to any of the $R_{bh}$ black holes.

\begin{center}
\begin{figure}[H]
\includegraphics[width=.9\textwidth]{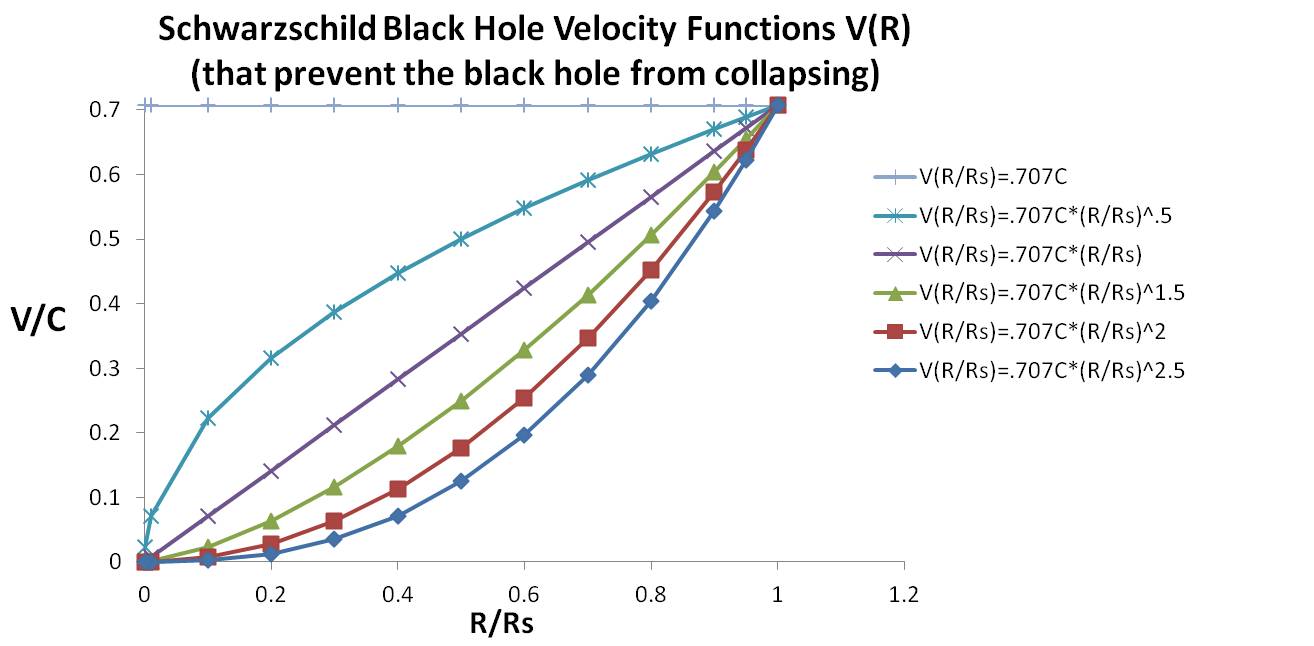}
\captionsetup{width=0.9\textwidth}
\caption{Velocity Curves for $R_{bh}$. These are plots of the Schwarzschild black hole velocity functions for the six mass density distributions (repeated from Figure \ref{RsVelocityCurves}). This will enable comparisons with the transverse redshift of \ref{RsRedshift}.}
\label{RsVelocityCurves2}
\end{figure}
\end{center}

\begin{center} 
\begin{figure}[H]
\includegraphics[width=.9\textwidth]{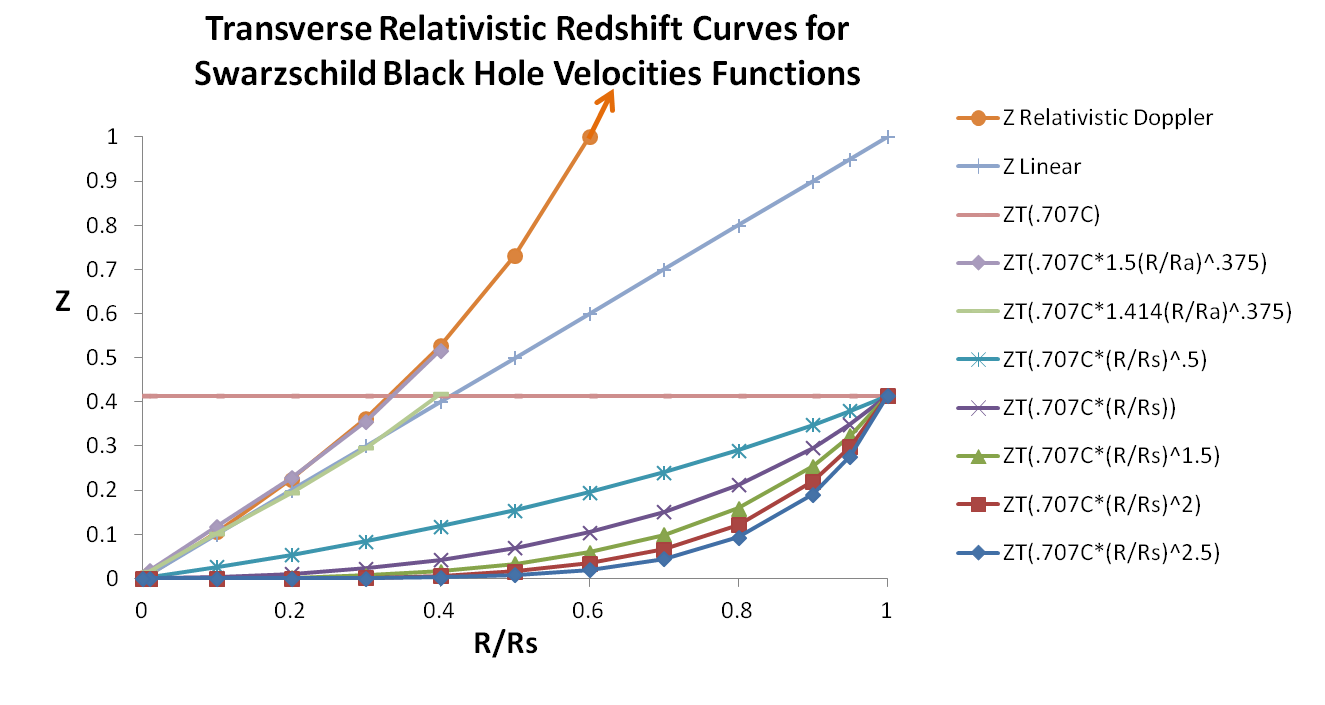}
\captionsetup{width=0.9\textwidth}
\caption{Transverse Relativistic Redshift for Schwarzschild Black Hole Velocity Curves. These are plots of the transverse redshift that would be produced from the velocity functions (of  Figure \ref{RsVelocityCurves2}) for the six mass density distributions.
Also included are the original relativistic Doppler and linear Doppler redshifts.  Note that the transverse redshift for Schwarzschild black holes are much smaller than even the linear Doppler shift, and have a max redshift of .414 corresponding to .707C.
Thus, the closest matches for the relativistic Doppler and linear Doppler redshift curves can only match out to .3 and .4 R/Rs.  The two curves that approximates these are  $V=.707*C*1.5*(R/R_{s})^{.375}$ and  $V=.707*C*1.414*(R/R_{s})^{.375}$. }
\label{RsRedshift}
\end{figure}
\end{center}

The transverse redshift functions could also be redone for the stationary Schwarzschild black holes with radius $R_s$.  The Velocity curves for the Schwarzschild black hole from Figure \ref{RsVelocityCurves} was repeated in Figure \ref{RsVelocityCurves2}.  The transverse redshift for these velocity curves are shown in Figure \ref{RsRedshift}.  Recall that the velocity functions are similar in shape to the $R_{bh}$ black holes, but the maximum velocity V is limited to .707C.  This translates into a maximum redshift of .414 in  Figure \ref{RsRedshift}.

The Relativistic Doppler redshift and Linear redshift functions are also plotted in  Figure \ref{RsRedshift}.  However, these exceed the .414 redshift at about $R/R_s = 0.3$ and $0.4$.  Thus, a transverse speed curve for Schwarzschild black holes will only be able to match the relativistic curve up to $R/R_s=0.3$ and match the linear Doppler redshift up to $R/R_s =0.4$.  The polynomial solutions for these two curves are $V(R)=.707*C*1.5*(R/R_s)^{0.375}$ and  $V(R)=.707*C*1.4*(R/R_s)^{0.375}$ respectively.  These are plotted in Figure \ref{RsRedshift}, but are hard to see since they are on top of the Relativistic and Linear Doppler functions.

Thus, using only transverse velocity redshift, we have found the velocity, mass, and density distribution functions that fall in between the linear and Doppler redshift curves (at least out to $R/R_s = 0.3$ and $0.4$).  A spinning charged black hole could have a maximum velocity somewhere between .707C and C.  As V(R) approaches C, and the Z values can grow to very large values with the polynomial function between the Relativistic and Linear Doppler redshift.   

\subsection{Gravitational Redshift}

\begin{figure}[H]
\begin{center} 
\includegraphics[width=0.75\textwidth]{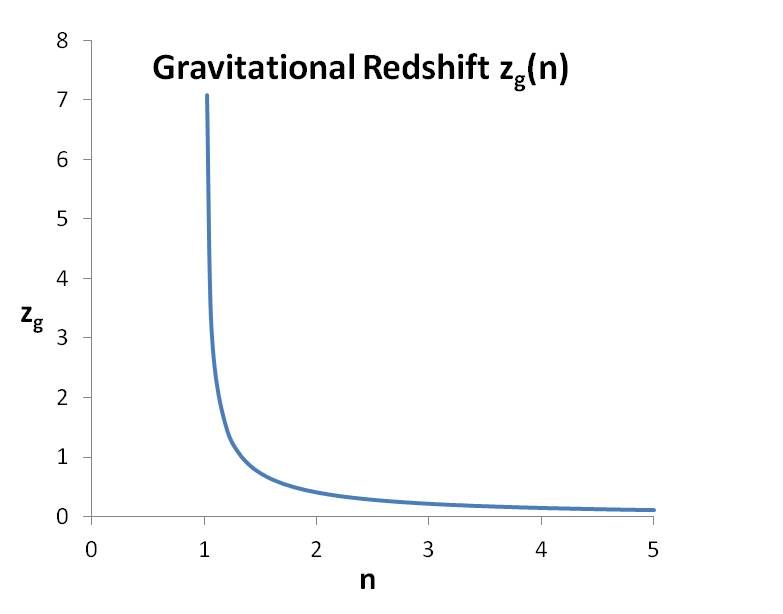}
\captionsetup{width=0.9\textwidth}
\caption{$z_g(n)$ Plot. This is the gravitational redshift function of an object outside a black hole, but within $nR_s$ radii of the black holes center. Note that $z_g$ approaches infinity, as n approaches 1 where the light is just outside the black hole. The gravitational redshift is caused by time dilation of the light within a gravity well. }
\label{zplot}
\end{center}
\end{figure}

In the theory of general relativity, gravity causes a time dilation within a gravitational well.  This causes time to slow down and lengthens wavelengths, which is perceived as a redshift.
The relativistic Gravitational redshift ($z_g$) equation for black holes on the emitting source is given below \cite{zucker}, 
\begin{equation}
z_g+1= \frac{1}{\left(1-\frac{2GM}{C^2R_{source}}\right)^\frac{1}{2}}
\end{equation}
where $R_{source}$ is the distance between the emitting source and the center of the black hole.  If the emitting source is inside a relativistic black hole, the light will be redshifted to infinity and nothing will be seen.  If the emitting source is just outside the black hole’s maximal height ($R_f$) the equation can be used to determine the gravitational shift. Consider  $R_{source}=nR_s=n\frac{2GM}{C^2}$ for $n>1$.
\begin{equation}
z_g+1 = \left(\frac{1}{1-1/n}\right)^\frac{1}{2}
\end{equation}
Therefore $z_g$ can be expressed as a function of $n$:
\begin{equation}
z_g(n) = \left(\frac{1}{1-1/n}\right)^\frac{1}{2}-1 \label{zfunn}
\end{equation}
See Figure \ref{zplot}.

If the source of the light is just outside the black hole boundary, it could have a significant shift.  Thus, gas emitted due to tidal force destruction of matter entering just outside the black hole, could be significantly redshifted.  For stellar black holes, where $R_s= 3000$ m, this region would be very close, and very small.  However, for the low density black holes, now $R_s$ could be AUs, light years, or MLYs or BLYs.

Also, being inside a small rotating low density black hole does not preclude adjacent low density black holes from contributing additional redshift. An additional larger encapsulating container low density black hole at larger radii surrounding the original black hole and its neighbors could provide even more redshift.  Additional adjacent container black holes and larger nested container low density back holes can provide additional gravitational redshifts, and also contribute to the observable linear redshift that increases with distance.   That is, the universe could be considered a nested set of low density black holes.

\subsection{Intrinsic Redshift of Interstellar Space}
Independent of this paper, there has been a growing debate about replacing the Doppler shift interpretation of redshift with an intrinsic redshift of space itself (or its interstellar medium) that causes light to redshift linearly with distance, roughly matching the linear Hubble redshift approximation equation.
Marmet et al. \cite{marmet} contains a survey of 59 of these theorized intrinsic redshift mechanism, including:
\begin{enumerate}
\item Gravitational Drag
\item Interaction of a massive Photon with Vacuum Particles
\item Interaction Between the Photon Energy and Vacuum Space
\item Light interaction with Microwaves and Radio Waves
\item Photon Decay
\item Finite Conductivity of Space
\item Photon-Graviton Interaction
\item Electrogravitational Coupling
\item Gravitomagnetic Effect
\item Interaction with the Intergalactic Gas
\end{enumerate}

These redshift concepts are attempting to find a real causal mechanism within the laws of physics to describe the linearly increasing redshift with distance.

To date, the Doppler shift and the expanding universe has been the leading redshift mechanism by most cosmologists.  However, after disproving the expanding universe model in this paper, along with its Doppler shift interpretation, one of the intrinsic redshift mechanism may get a closer reading.

If the linear redshift is just a function of space itself, then the redshift largely matches the observed phenomena.  However, at present, we currently don’t know the true source of the intrinsic redshift, assuming that intrisic redshift is real.  For the remainder of this paper, this paper will consider the following five possible candidates for intrinsic redshift: Bremsstrahlung radiation, Induced Electric Dipole redshift (IEDRS), simple charge based intrinsic redshift, gravitational drag, and bending redshifts.

\subsubsection{Bremsstrahlung Intrinsic Redshift}
In the literature, there is a new Non Doppler effect redshift based on Bremsstrahlung radiation.  Low density particles in space scatter light, in addition to the Compton scattering effect, the light can accelerate electrons.  According to electromagnetic theory, any accelerated charge must emit Bremsstrahlung radiation.   Although the energy emitted due to such acceleration is extremely small, the energy lost by Bremsstrahlung causes a slight redshift.  

Bremsstrahlung radiation is reported to impact all wavelengths equally and thus will appear similar to Doppler redshift.  An average density of the order of 0.01 atoms per cubic centimeter is sufficient to produce, on the Planck spectrum, an effect equivalent to that of a Doppler shift in agreement with the Hubble constant\cite{marmet1988}. 

If the Bremsstrahlung redshift is valid, no other mechanism is needed to match the perceived linear Doppler shift as long as the interstellar density is approximately .01 atoms per cubic centimeter.
\subsubsection{Induced Electric Dipole Redshift}
Also in the literature, there is another Non Doppler effect redshift based on electric charge effects on light \cite{mc}.  To quote that paper,    \textsl{``Small induced dipole forces acting on photons can account for the red-shift.  In atomic physics, gamma rays are known to split (electron positron pair production) while in the intense central electric fields of atomic nuclei and charged subatomic particles. The photon energy is converted to mass and kinetic energy, but due to the law of conservation of charge, the (-,+) charge pair must have been contained within the photon prior to pair production. It must be the induced electric dipole force that forces the charge pair to separate.  Less energetic photons do not have sufficient energy for pair production (1.02 MeV is the minimum energy required), but they must similarly contain a (-,+) charge pair. This pair will separate slightly while in any non-uniform electric field; causing an attractive force (the induced electric dipole force is always attractive). This small force acting on photons as they travel over astronomical distances will reduce photon energy (the red-shift)"}.

If Induced Electric Dipole redshift (IEDRS) theory is valid, no other redshift is needed to match the linear Doppler redshift.

\subsubsection{Charge Based redshift }
A charged particle in a changing electro-magnetic field will be accelerated.  This energy needs to come from the light, and will change the electromagnetic energy by redshifting the light.  Thus, when light passes through charged particles, the light will be redshifted.  Thus, light passing through ions in space should slow down and produce a slight redshift.

Even a neutral hydrogen atom consists of a charged electron and a charged proton, when considered separately.   The electromagnetic field (e.g. light) would accelerate each of these slightly, but would consume light energy and produce a slight redshift.

\subsubsection{Gravitational Drag (Shapiro Effect)}
Dr. Irwin I. Shapiro of the Lincoln Labs of the Massachusetts Institute of Technology, stated ``\textsl{...according to the general theory, the speed of a light wave depends on the strength of the gravitational potential along its path.}"  Light will experience a small, cumulative deceleration from the forces of this gravitational field acting at long range.  This does not smear the light nor provide a visible signature, but only a gravitationally induced time delay.  The photon’s interaction with gravity does not alter its path, nor change its characteristic, except by gradually decreasing its velocity and energy, showing up as a redshift \cite{marmet, shapiro}.

\subsubsection{Light Bending}
Since light also has momentum, the simple process of bending of light, independent of what actually caused the bending, may cause the momentum to change, and thus may cause its energy to change and be redshifted. 

\subsection{Other Sources of redshift}
\ul{Light reddening}: occurs when light hits particles in space.   They cause the light to redden, but not shift.  Astronomers are able to account for this phenomenon and these are not included in the redshift values. 
\newline \newline \ul{Index of Refraction}: $n = \frac{C}{V}$:  Light traveling through a medium (e.g. for air, n= 1.000277) will slow down and thus be redshifted. This would put $n = \frac{C}{V}$; Interstellar gas is a medium, but could cause a slow down.  However, the index of refraction slow down is constant, not accumulative, unless the gas gets increasing dense farther away.  Thus these are not believed to cause the linear redshift.

\subsection{Quasars Redshift}
The most distant objects visible (as measured by redshift $z$ values) are Quasi-stellar radio sources (``quasars"), which are very energetic distant active galactic cores.  A quasar is believed to be a compact region in the center of a massive galaxy, that surrounds its central region with a dust cloud the size that is 10–10,000 times the Schwarzschild radius of the black hole.  Since some quasars display changes in luminosity very rapidly from side to side, quasars are estimated to be not much larger than the Solar System (50 au = 7 light hours) \cite{zeilik480}. Quasars are extremely luminous objects with redshifts between 0.056  and 7.085 \cite{hewitt,mortlock}.  There is much debate as to whether quasars are large, luminous, distant objects or smaller, less luminous, closer objects.  

In either event, their redshifts will need to be created by transverse redshifts or from gravitational redshifts from localized black holes, or intrinsic redshifts, or a combination of these redshifts.

\subsubsection{Intrinsic Redshift}

Using only the linear distance redshift equation (Eq. \ref{redshift}) applicable to all the intrinsic redshift mechanisms, we get an equation for distance as a function of $z$:
\begin{align}
d(z) &=\frac{z}{.000076} \qquad (\text{with d in MLY}) \\
d(.056) &= \frac{0.056}{.000076} = 737 \text{ MLY} \\
d(7.085) &= \frac{7.085}{.000076} = 92\text{ BLY}
\end{align}
The linear equation generates extremely large values of $d$ well beyond the current big bang estimates of 13.8 BLY for the size of the universe.  However, now that the big bang is disproven, the universe could go out to 92 BLY since the big bang date of 13.8 is not necessarily valid.  The universe would also almost match the visible density of Table \ref{TableSameDen}.  However, a quasar would have to be extremely bright to be seen at a 92 BLY distance.

\subsubsection{Transverse Redshift for Quasars}
Using the transverse redshift equation (Eq. \ref{transverse}) we get an equation for $V/C$ as a function of $z$: 
\begin{subequations}
\begin{align}
\frac{V}{C}(z) &= \left(1-\frac{1}{(z+1)^2}\right)^\frac{1}{2} \\
\frac{V}{C}(0.056) &= \left(1-\frac{1}{(0.056+1)^2}\right)^\frac{1}{2} = .321 
\end{align}
Therefore a quasar must be traveling horizontally around us at a velocity of $V=.321C$ to produce the redshift of $z=0.056$. If was assume we are somewhere near the center of a black hole with radius 13.8 BLY we can solve for its distance using Eq. \ref{vfunr}: 
\begin{align}
\frac{V}{C}  &= \left(\frac{R}{R_{bh}}\right)^{.375} = .321\\
          \left(\frac{R}{R_{bh}}\right) &= .321^{8/3} = .0487\\
          d &= 13.8\left(\frac{R}{R_{bh}}\right) = 672 \text{ MLY}
\end{align}
\end{subequations}
Similarly for the largest $z$ value:

\begin{subequations}
\begin{align}
\frac{V}{C}(7.085) &= \left(1-\frac{1}{(7.085+1)^2}\right)^\frac{1}{2} = .992\\
\frac{V}{C}  &=  \left(\frac{R}{R_{bh}}\right) ^{.375} = .992\\
           \left(\frac{R}{R_{bh}}\right) &= .992^{8/3} = .978\\
          d &= 13.8 \left(\frac{R}{R_{bh}}\right) = 13.5 \text{ BLY}
\end{align}
\end{subequations}
This quasar would be traveling with velocity of $V=0.992C$ and at a distance of 13.5 BLY. Thus, just the transverse redshift could explain the high redshifts and bring them in closer.

\subsubsection{Gravitational Redshift for Quasars}
Using the gravitation redshift equation (Eq. \ref{zfunn}), one can derive an equation for $n$ as function of $z$: 
\begin{align}
z_g(n) &= \left(\frac{1}{1-1/n}\right)^\frac{1}{2}-1 \\
n(z) &=\left( 1-\frac{1}{(z+1)^2}\right) ^{-1}\\
n(0.056) &=\left( 1-\frac{1}{(z\ 0.056+1)^2}\right) ^{-1} = 9.7
\end{align}
Then $R=nR_s=9.7R_s$. Assuming the the quasar radius is $R=7$ Light Hours then $R_s=7/9.7=0.72$ Light Hours and $M=\frac{R_sC^2}{2G} = 0.5$ B SM. This is just enough to cover the ``closest" quasars with the lowest redshifts. 
\\[1ex] Similarly for the largest values, $z=7.085$, 
\begin{align}
n(7.085) &=\left( 1-\frac{1}{(z\ 7.085+1)^2}\right) ^{-1} = 1.016
\end{align}
Assuming the the quasar radius is $R=7$ Light Hours then $R_s=7/1.016 = 6.89$ Light Hours and $M=\frac{R_sC^2}{2G}=5$ B SM. Thus, the entire quasar redshift can come from gravitational redshift.  

Quasars are only perceived as far away due to their high redshift. They would have to be very large to be seen at that great a distance.  However, if the source of the redshift is gravitational redshift, then quasars can be much closer and smaller.  That said, since there are examples of gravitational lens duplication of quasars by intervening galaxies, Quasars are at least outside our galaxy, and beyond the next, and possibly beyond many more.  

\begin{figure}
        \centering
        \begin{subfigure}[b]{0.3\textwidth}
	\includegraphics[width=\textwidth]{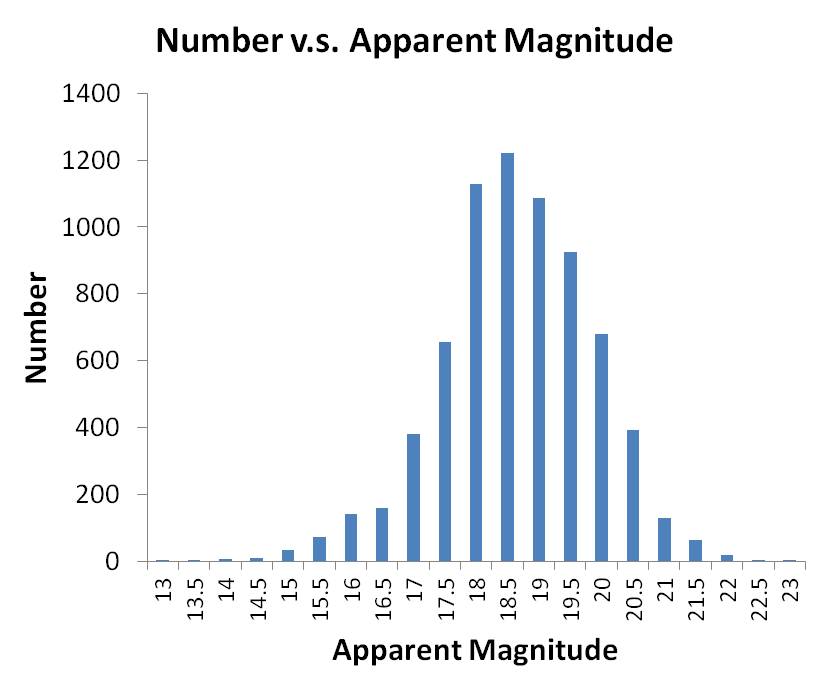}
	\caption{Number v.s. App. Magnitude }
	\label{QuasarNumberVsRedshift}
        \end{subfigure}
		\qquad 
        \begin{subfigure}[b]{0.3\textwidth}
	\includegraphics[width=\textwidth]{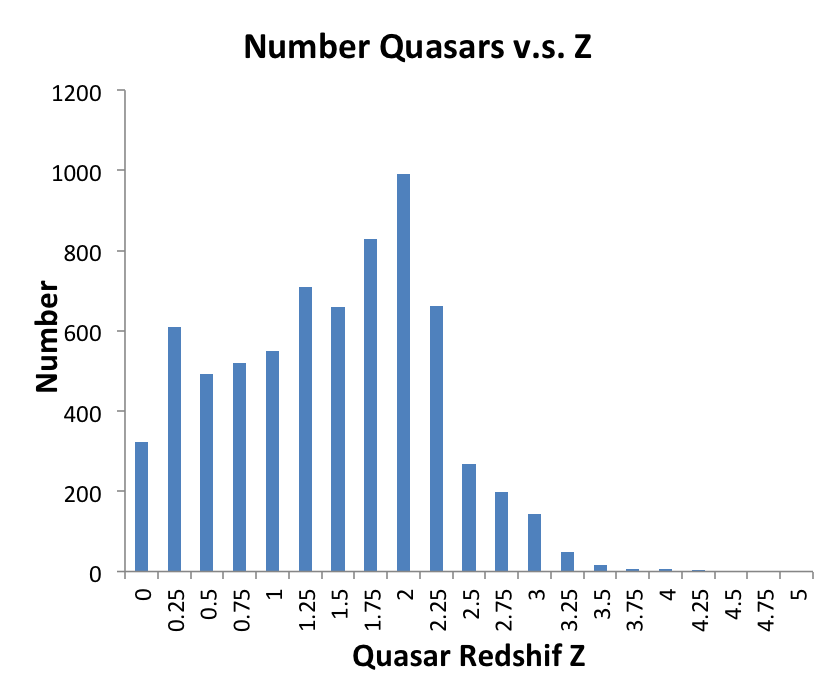}
	\caption{Number  v.s. redshift}
	\label{QuasarMagVsRedshift}
        \end{subfigure}
		\qquad
	\begin{subfigure}[b]{0.3\textwidth}
	\includegraphics[width=\textwidth]{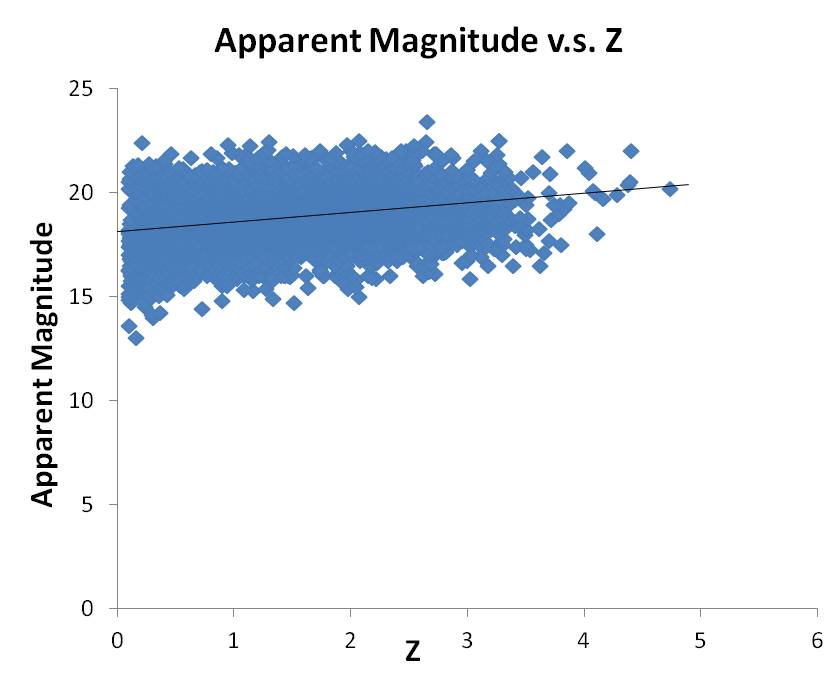}
	\caption{App. Magnitudes v.s. redshift}
	\label{AppMagVsRedshift}
        \end{subfigure}
\captionsetup{width=0.9\textwidth}
\caption{Plots of Quasar Data. (a) Shows that the apparent magnitude of quasars are about  $19 \pm 4$. (b) Shows that the average quasars range between .062 and 4, with a average about $z=2$.  Also, there appears to be groupings of preferred values (e.g. z=1.5, 2). (c) Shows that the apparent magnitude gets larger with $z$; however, stellar magnitudes are actually dimmer as they get larger, so the quasars are actually getting dimmer with increasing redshift (distance)\cite{hewitt}. Note: data plotted from \cite{hewitt}.}
\end{figure}

The apparent magnitude of quasars are about $19 \pm 4$ (\cite{hewitt}) shown in Figure \ref{QuasarNumberVsRedshift}.  Since we don’t know if they are close or far away, we can’t plot their $z$ shift versus distance.   Plots of number and apparent magnitudes are shown in Figure \ref{QuasarNumberVsRedshift} and \ref{QuasarMagVsRedshift} (from \cite{hewitt}).  Their apparent magnitudes increase with redshift.  Stellar magnitudes are actually dimmer as they get larger, so the quasars are actually getting dimmer with increasing redshift (distance).   

\begin{figure}[H] 
\begin{center} 
\includegraphics[width=.9\textwidth]{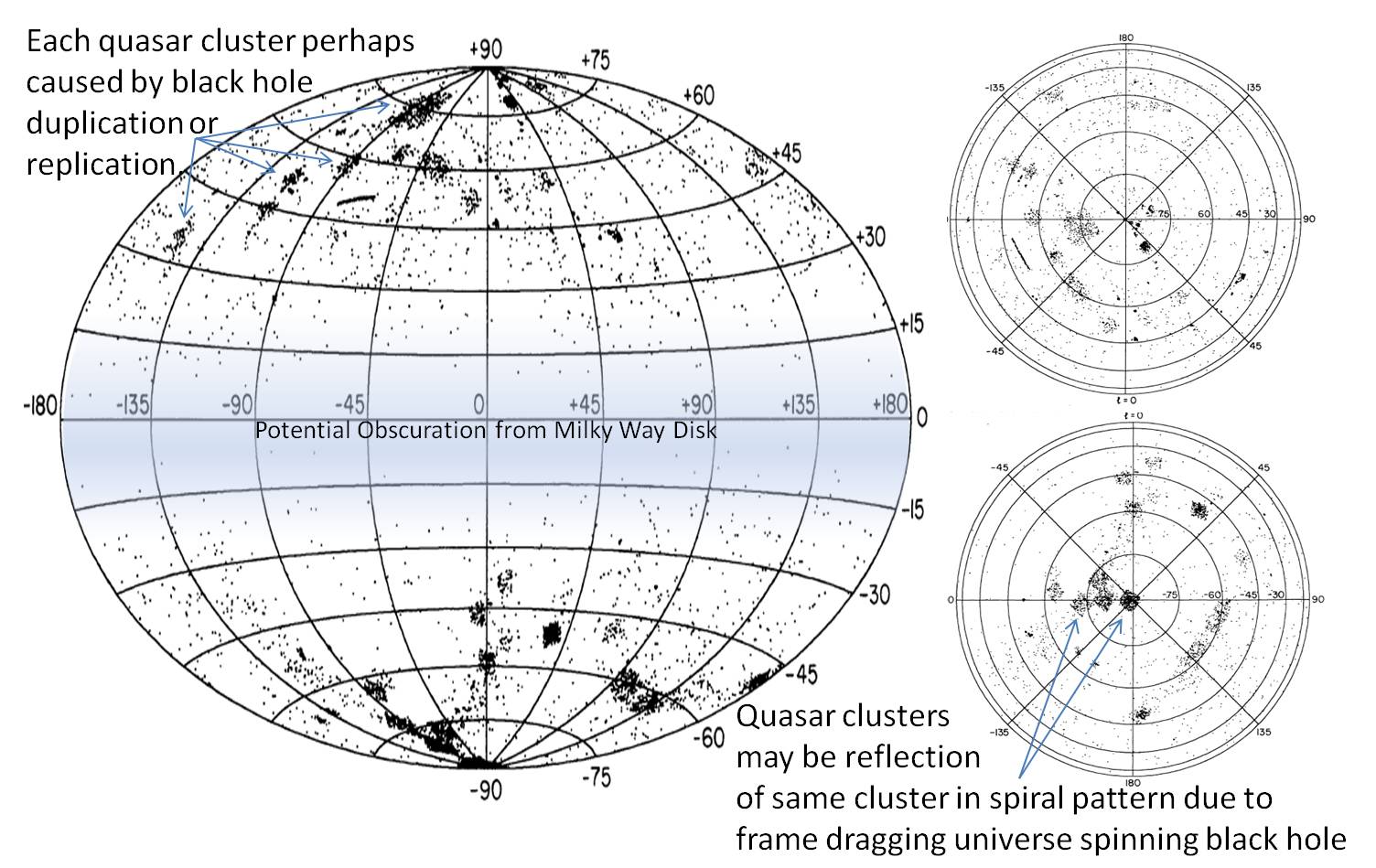}
\captionsetup{width=0.9\textwidth}
\caption{Distribution of 7315 QSOs on the sky in Galactic Coordinates A REVISED AND UPDATED CATALOG OF QUASI-STELLAR OBJECT A. Hewitt and G. Burbidge. This is the quasar data plotted in the sky in Galactic Coordinates.  Note, that although they appear across the sky, there are clusters or groupings of quasars, possibly in a rotating spiral pattern (see top view).  Although little quasars are shown along the Galactic Plane, their view is probably obscured by the Milky Way Disk \cite{hewitt}.  Note: annotations and Milky Way obsuration added to base figure 3D plots from \cite{hewitt}.}
\label{3Dquasar}
\end{center}
\end{figure}

The spatial distribution of quasars are shown in Figure \ref{3Dquasar}.  Although quasars appear at all angles, there are dense groups of quasars and the dense groups appear to be spiraling about the galactic poles.  Although there are fewer quasars visible near the galactic plane, their view is probably obscured by the Milky Way Galactic disk.  


The equation relating apparent magnitude $(m)$ to Absolute Magnitude $(M)$ as a function of distance $d$ is \cite{ZeilikP232}:
\begin{equation}
M = m + 5 - 5*log(d) \text{  (with d in parsecs)}
\end{equation}

An example quasars with $m=12.9$ and $d=2.44 BLY$ will have $M=12.9+5 -5*log(2.44*10^9/3.26) =-26.5$.  A one magnitude decrease corresponds to a 2.512 times increase in brightness \cite{ZeilikP226}.  This equation applies to two objects with either apparent magnitudes (m and n) or two objects with absolute magnitudes (M and N): 
\vspace{3 mm}
\begin{subequations}
\begin{align}
f_m/f_n &= 2.512^{(n-m)}   \text{  where f is the apparent flux. } \\
L_M/L_N &=2.512^{(N-M)}   \text{  where L is the absolute Luminosity}. \\
L/L_{sun} &= 2.512^{(4.83-M)}   \text{  using $4.83$ for the absolute magnitude of the sun} 
\end{align}
\end{subequations}
For $M=-26.5$, $2.512^{(4.83-(-26.5))}=2.7*10^{12}$, almost three trillion times brighter than the sun.  The Stephan-Boltzmann equation relates a star's Luminosity to its radius (R) and surface temperature (T)\cite{ZeilikP232}: 
\begin{subequations}
\begin{align}
L &= 4\pi R^2 \sigma *T^4                   \\
L/L_{sun} &= \frac{R^2T^4}{R_{sun}^2T_{sun}^4}  \\
L/L_{sun} &=  (\frac{R}{R_{sun}})^2*(\frac{T}{T_{sun}})^4 \\
T &= T_{sun}*(L/L_{sun} *R_{sun}^2/R^2)^{\frac{1}{4}}\\
T &= 6500K*(L/L_{sun} *R_{sun}^2/R^2)^{\frac{1}{4}}
\end{align}
\end{subequations}

A $1*10^{12}$ luminosity ratio could be explained if  $T=1000*T_{sun}$ since $1000^4=1*10^{12}$. A quasar could have $T=1000*6500K = 6.5$ Million degrees K. 
Alternately, the quasar could have the same temperature as the sun and a radius $R= 1000000*R_{sun}$ since $1000000^2=1*10^{12}$.  A radius $R=R_{sun}*10^6 = 1$ light month.

One can estimate the distance to the quasars using the linear redshift equation to compute absolute magnitudes:
\begin{subequations}
\begin{align}
z &=0.000076*d   \text {     with d in MLYs}\\
d &= z/0.000076   \text {     with d in MLYs}\\
M &= m + 5 - 5*log( z/0.000076 * 1000000LY*1 parsec/3.26LY)
\end{align}
\end{subequations}

Table \ref{tableMagnitudes} uses the average magnitudes at different z values from Figure \ref{AppMagVsRedshift} to compute $d, M, L/L_{sun}, L/L_{sun}*R_{sun}^2/R^2$, and temperature. The table assumes a $R_{bh}$ of 5 Billion solar mass black hole with radius of about 7 light hours, surrounded by a gas at $1, 10, 100, 1000$, and $10000*R_{bh}$.    At 92 BLY, the quasar would need to be about 3 trillion times as luminous as the sun, but the luminosity per unit area is only 23,000 times as luminous, requiring a temperature about 80,000K.  However, if a gas cloud surounds the black hole increasing the radius to $R=10, 100, 1000, and 10000*R_{bh}$, then the luminosity per unit area drops to  230, 2.3, .023, and ,.00023 that of the sun, and the temperature could drop to 25000, 8000, 2500, and 800 degrees K respectively.  Thus, the seemingly impossible to see quasar at 92 BLY, could just be a big cloud of hot gas surrounding a black hole.  Even a $10000*R_{bh}$ gas cloud can be seen at 658 MLY even at only 180 degrees K.  These clouds might be the diffuse objects in Figures \ref{EinsteinRing}, \ref{EinsteinRing2}, \ref{SDSS}, \ref{abell}, \ref{MACS}, and \ref{LCDCS}.  Why are there no quasars closer than about 700MLYs?  Perhaps quasars only exist in the past, as giant dust clouds, before they collapsed into current nearby denser objects like galaxies.

\begin{table}
\caption{Quasar Magnitudes, Distances, Luminosities, and Temperature}
\begin{center}
\noindent\makebox[\textwidth]{
\begin{tabular}{| c | c | c | c | c | c | c | c | c | c | c | c |c | c | c | }
  \hline  
$z$& \pbox{20cm}{$m$}& \pbox{20cm}{$d Mly$}& \pbox{20cm}{$M$}& \pbox{20cm}{$\frac{L}{L_{sun}}$}&\pbox{20cm}{$L_1R_{bh}$}&\pbox{20cm}{$T_1K$} &\pbox{20cm}{$L_{10}R_{bh}$}&	\pbox{20cm}{$T_{10}K$}& \pbox{20cm}{$L_{100}R_{bh}$}&\pbox{20cm}{$T_{100}K$}&\pbox{20cm}{$L_{1k}R_{bh}$}&\pbox{20cm}{$T_{1k}K$}& \pbox{20cm}{$L_{10k}R_{bh}$}&	\pbox{20cm}{$T_{10k}K$}\\

\hline 
7&	21&	92105&	-26.3&	2.7E+12&	23030&	80073&	230&	25321&	2.3&	8007&	0.023&	2532&	2.3E-04&	801\\
\hline 
6&	20.5&	78947&	-26.4&	3.2E+12&	26817&	83179&	268&	26304&	2.7&	8318&	0.027&	 2630&	2.7E-04&	832\\
\hline 
5&	20&	65789&	-26.5&	3.5E+12&	29515&	85197&	295&	26942&	3.0&	8520&	0.030&	2694&	3.0E-04&	852\\
\hline 
4&	19.5&	52632&	-26.5&	3.5E+12&	29938&	85501&	299&	27038&	3.0&	8550&	0.030&	2704&	3.0E-04&	855\\
\hline 
3&	19&	39474&	-26.4&	3.2E+12&	26690&	83081&	267&	26272&	2.7&	8308&	0.027&	2627&	2.7E-04&	831\\
\hline 
2&	18.5&	26316&	-26.0&	2.2E+12&	18800&	76112&	188&	24069&	1.9&	7611&	0.019&	2407&	1.9E-04&	761\\
\hline 
1&	     18&	13158&	-25.0&	8.8E+11&	7449&	60385&	74&	19096&	0.7&	6039&	0.007&	1910&	7.4E-05&	604\\
\hline 
0.5&	17.5&	6579&	-24.0&	3.5E+11&	2951&	47908&	30&	15150&	0.3&	4791&	0.003&	1515&	3.0E-05&	479\\
\hline 
0.05&	17&	658&	-19.5&	5.5E+09&	47&	16998&	0.47&	5375&	0.005&	1700&	4.7E-05&	538&	4.7E-07&	170\\
\hline
\end{tabular}}
\end{center}
\label{tableMagnitudes}
\end{table}

\newpage
\section{Non-Reflecting Universe}
\subsection{Near Black Hole Universe}
\begin{figure}[H] 
\begin{center} 
\includegraphics[width=.5\textwidth]{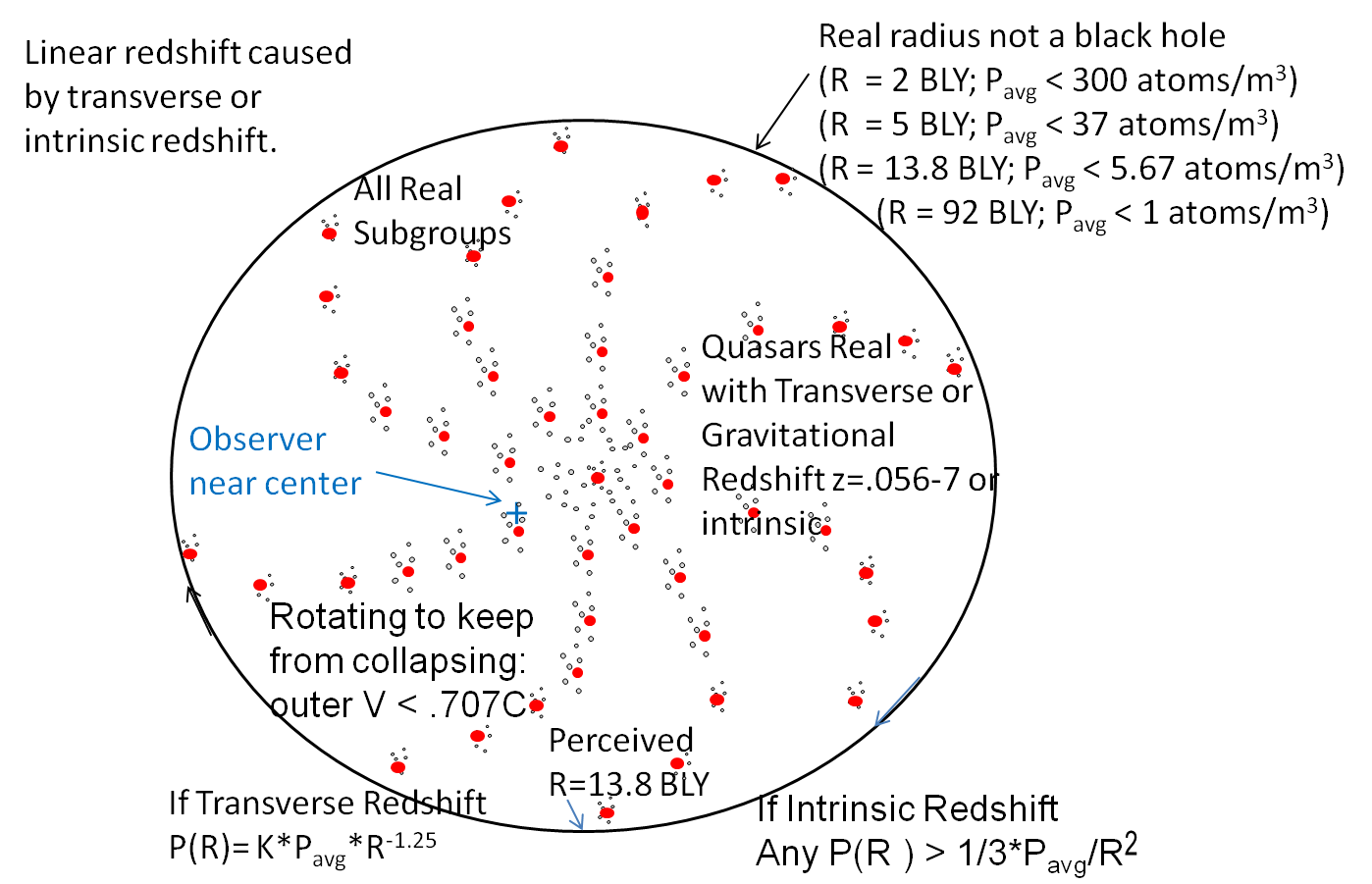}
\captionsetup{width=0.9\textwidth}
\caption{Non-Reflecting Universe.  If the universe is not a low density black hole then the universe would be non-reflecting.  Without the expanding universe model, the radius of the universe is no longer tied to $13.8 BLY$.  Since scientist have confirmed linear redshift from supernovae out to $1.3$ BLY, the universe is probably at least $2 BLY$, but could be much larger (e.g. $92BLY$).  However, its average density would have to decrease at larger radii to keep from becoming a black hole.  It would be rotating about its center to keep from collapsing with velocity less than $.707C$.
If the linear redshift is transverse, the redshift would be limited to $z=0.4$ with $V$ \textless $.707C$; and the radius of the universe would be limited to about $5 BLY$ and the density function would need to be proportional to $1/R^{1.25}$.   If the linear redshift is intrinsic, then larger radii are possible and any density function that falls off slower than $1/3*P_{avg}/R^2$ could keep the universe from collapsing.
The quasar redshift could also be caused by gravitational redshift with $z=.056-7$.
If the quasar redshift is based on linear redshift, the universe radii would need to be $92 BLY$ to match the max $z=7$ value, and have to be extremely bright.}
\label{NonReflectingUqg}
\end{center}
\end{figure}

If the universe is below its critical density of $5.67 atoms/m^3$, then it does not need to be a low density black hole.   It would not have the gravitation force density necessary to capture light, nor an event horizon to curve space sufficiently to keep light from escaping.  Thus, this nearly Black Hole universe would largely be a Non-Reflecting Universe as described in Figure \ref{NonReflectingUqg}. 

If the universe was radically collapsing, one would see a Doppler blueshift.  Thus, the universe is probably not rapidly collapsing.  The universe would need to be rotating and/or charged to keep it from collapsing. It will become a low density black hole in the future as it collapses inward, and becomes denser. Until that time, there could still be localized low density black holes within the universe of all sizes and we could be inside one of these localized black holes.  And if this is the case, our portion of the universe would be reflecting.    There could also be a lot of simple gravitational lensing.    But the universe, as a whole, would not need to be reflecting at its boundary.

\subsection{Black Hole Non-Reflecting Universe?}
Many cosmologist believe that the universe is so close to its critical density, that it probably is at its critical density.   If this is the case, then the universe would be a low density black hole.
The minimal low density black hole is a \ul{Stationary Classical} Schwarzschild, $R_{cs}$ black hole.  Light (or particles) going out from the edge of an $R_{cs}$ black hole at the edge of the universe, will return, but only after an infinite time.  Thus, this universe will largely not be a reflecting universe and would look almost identical to the nearly Black hole universe described above.  I believe this description matches what a lot of cosmologists are describing as their model of the universe, if they formalized their descriptions of the universe.  However, if the universe met the Schwarzschild criteria, relativistic mathematics, rather than classical mathematics, would probably be needed.  The relativist Schwarzschild would introduce an event horizon at the boundary due to the gravitational curving of space.  If this is the case, the universe would become a reflecting universe (and will be described further in the next section).  


For the many scientists who believed in intrinsic (and gravitational) redshift already, they may not have to change a thing.  Well perhaps one thing, instead of ending their papers with the statement “Since intrinsic redshift explains the observed redshift, there is no compelling evidence to prefer an expanding universe and the big bang theory.” to “Since the expanding universe and Big Bang theory has been undermined, there is compelling evidence to prefer a non-expanding universe theory based on intrinsic redshift.”  A second change will be the possibilities of a smaller reflecting universe or infinite universe.  

For those who embrace the expanding universe and Big Bang theories, realize that they are just theories.  That is, theories are our best models until new data or a better argument comes along to displace it.  At that point, one should just move on towards the next likely theory.  Well, this may be that time, or at least the time to reduce the expanding universe theory to the not compelling category and be a little more open to the alternatives.

This would be a good place to stop for this paper, before things really start getting radical.   Most Scientist may already be feeling a little unsettled at this point, and may need time to contemplate the ramifications of all that is covered thus far, before continuing down this rabbit hole.

\section{Reflecting Universes}
The remainder of this paper will be exploring potential reflecting universes based on a gravitational “mirror” balls.
Another word of caution is probably well overdue.  After losing the expanding universe, Doppler Shift, and the Big Bang, what we currently know is less than we think.  The reflecting universes, if they exist at all, will be hard to understand, and harder yet to prove, especially since by their very nature are lying to us.  My hope here is to find at least one plausible approach.  If the reader finds any one of these conjectures impractical, realize that at most only one of them can be ultimately true. 

\subsection{Current Size Universe 13.8 BLY}
\begin{figure}[H]
        \centering
        \begin{subfigure}[b]{0.48\textwidth}
	\includegraphics[width=\textwidth]{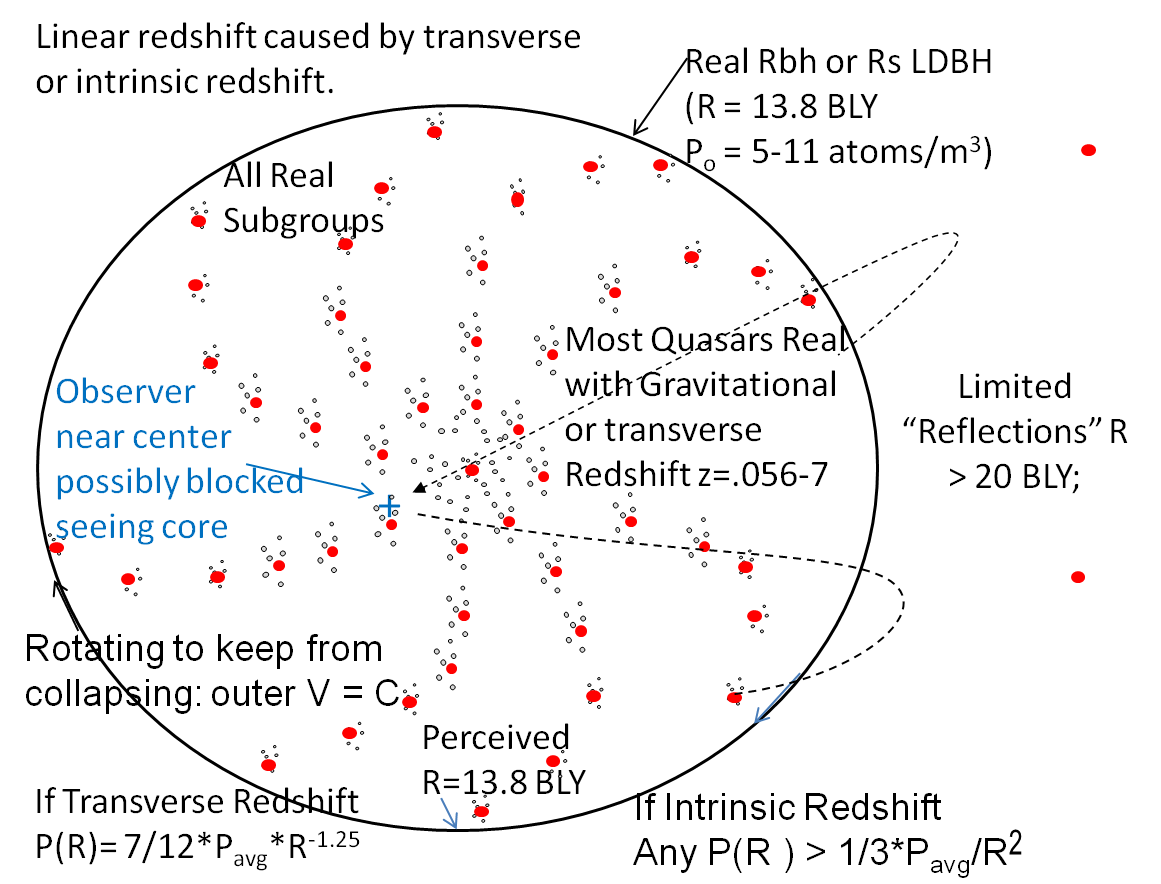}
	\caption{Most Quasars Real. }
        \end{subfigure}
	\qquad 
        \begin{subfigure}[b]{0.45\textwidth}
	\includegraphics[width=\textwidth]{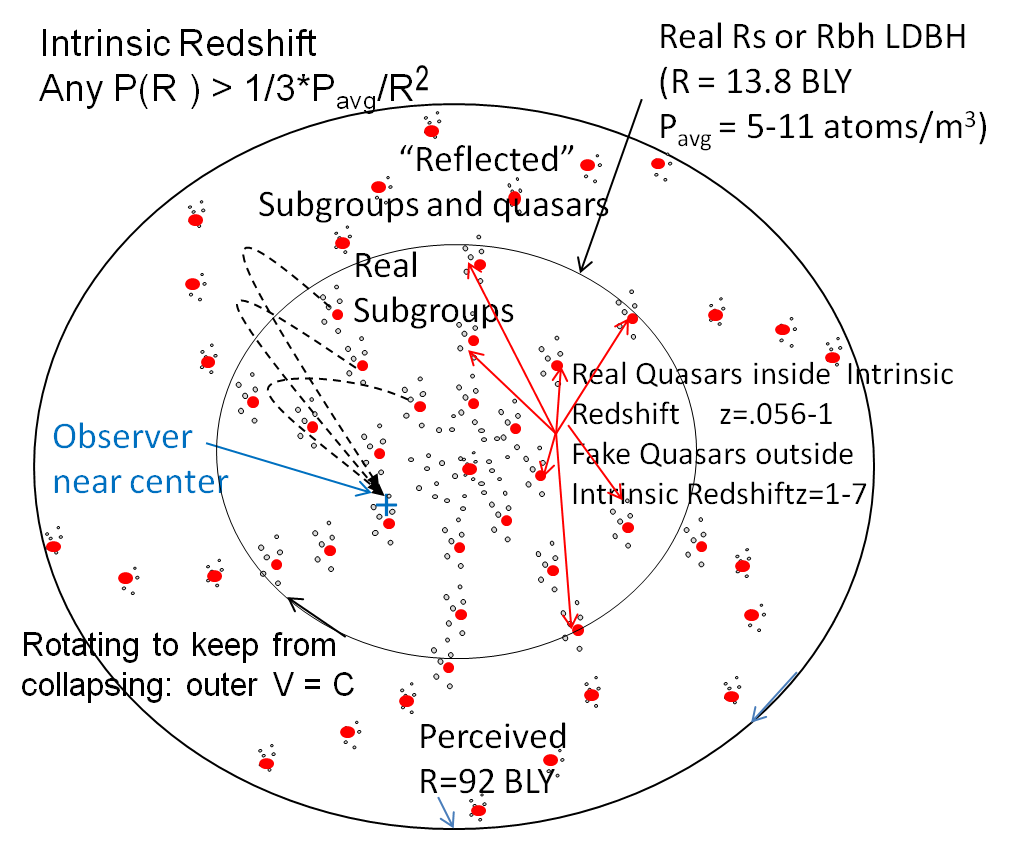}
	\caption{Most Quasars Reflecting.}
        \end{subfigure}
\captionsetup{width=0.9\textwidth}
\caption{Current Size Reflecting Universe Options. Our non-expanding universe could be about the perceived 13.8 BLY radius with limited reflections due to visibility or time since the beginning of the universe.  It would be rotating about its center to keep from collapsing at velocities up to C.  The linear redshift could be caused by either transverse redshift or intrinsic redshift.  Most of the quasars could be real with high redshifts caused by either transverse or gravitational redshift with z=.056-7 as shown on the left.  Alternately, if the quasar redshift is based on linear intrinsic redshift, most of the quasars would be reflections as shown on the right.}
\label{CurrentUniverse}
\end{figure}

Assume the universe is a Low Density Black Hole with $R_{s}\text{ or }R_{bh}=13.8$ BLY, with average density between 5.67 and 11 atoms/$m^3$.  The light from the real internal galaxies and quasars that travels outside the low density black hole boundary will be pulled backwards or be curved at the event horizon and create the illusion of a larger universe in Figure \ref{CurrentUniverse}.  

The “reflected” light would travel a longer path, get dimmer with distance, but be indistinguishable from real stars and galaxies.  
Intrinsic redshift over this longer path could accumulate over the actual distance traveled and appear indistinguishable from the Doppler redshift, at least out to about .5-1 BLY. 

Positioning the quasars with their larger redshifts that fit the observed data will be a challenge.  It will be hard to have these larger redshifts reconciled with the same redshift exhibited by the smaller redshifts in galaxies. The larger quasar redshift could be caused by gravitational or transverse redshift, but would not necessary continue to accumulate across reflections. Thus, if this is the case, there may be limited reflections in this universe perhaps due to limited visibility (hard to see things out past 20 BLYs) or due to limited time from the beginning of time.  This option is shown on the left with most of the quasars being real, and some possible quasar reflections outside as well.   Alternately, the large quasar redshift could be cause by accumulated intrinsic redshift over continuous reflection cycles.  If this is the case, most of the real quasars on the outside with $Z \textgreater 1$ would be reflections of the real quasars within the 13.8 BLY distance as shown on the right.

The primary location for the observer considered thus far, was having the observer near the middle.  However, being almost anywhere in the smaller mirror ball (e.g. in  Figure \ref{CurrentUniverse} on the left), would still be perceived as the middle of a potentially infinite looking universe.  Although this real universe would only go out to about 13.8 BLY, due to gravitational reflections it could be perceived to go outward to 13.8 BLY, or 92 BLY, or farther as our telescopes advance.

\subsection{Half Size Universe 6.9 BLY}
\begin{figure}[H]
\begin{center} 
\includegraphics[width=.53\textwidth]{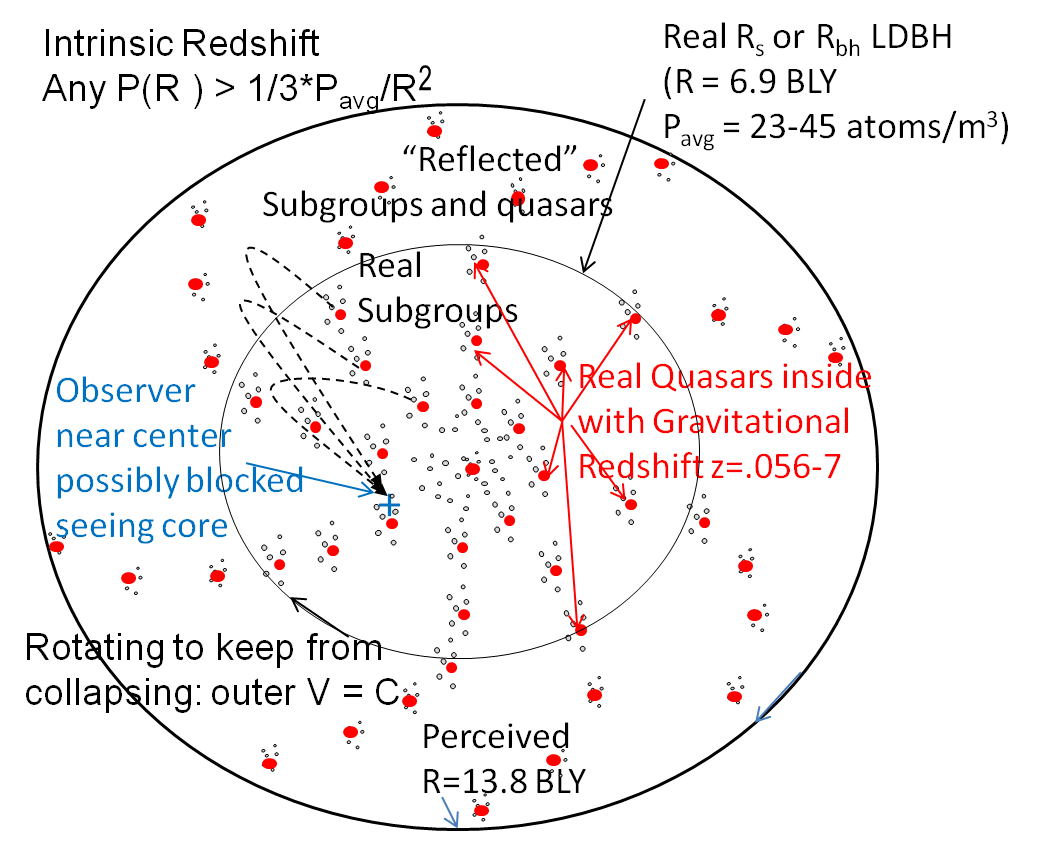}
\captionsetup{width=0.9\textwidth}

\caption{Half size Reflecting Universe. A half size universe would be possible if the average density is 45 atoms per cubic meter. The universe would only be 6.9 BLY, but due to gravitational reflections it would be perceived as 13.8 BLYs or larger.}
\label{ReflectingUniverse}
\end{center}
\end{figure}

Assuming a Low Density Black Hole with $R_{bh}=6.9$ BLY, the average density would be 45 atoms/$m^3$.  The light from the real internal galaxies and quasars that travels outside the low density black hole boundary will be pulled backwards and create the illusion of a larger universe in Figure \ref{ReflectingUniverse}.  The “reflected” light would travel a longer path, get dimmer with distance, but be indistinguishable from real stars and galaxies.  

Intrinsic redshift over this longer path could accumulate over the actual distance traveled and appear indistinguishable from the linear Doppler redshift.  It is not clear if transverse redshift would be accumulative across reflection cycles.  If one sees a reflected galaxy rotating about a perceived center, but at higher velocities due to its farther distance from the center, the perceived redshift may be appropriate for the perceived high speed as well.  In this sense, transverse redshift might be accumulative.  If transverse shifts are not accumulative, then accumulative intrinsic redshift mechanism would have to be used.  Thus, this paper will assume linear intrinsic redshift for the remaining reflecting universes, but accumulative transverse may also be an option.
Any redshift non-linearity, if any, that show up after the .5 BLY will be assumed to be associated with either additional gravitational redshift, bending of light, or a secondary effect due to a collapsing universe.  For example, the earlier collapsing universe would have more particles in its interstellar and intergalactic medium, and exhibit increased redshifts for the most distant galaxies.

Positioning the quasars with their larger redshifts that fit the observed data will be a challenge.  It will be hard to have these larger redshifts reconciled with the same redshift exhibited by the smaller redshifts in galaxies.  For this radius, this required using gravitational redshift for each quasar, and then adding the smaller linear redshift for everything.  The primary location for the observer considered thus far, was having the observer near the middle.  However, being almost anywhere in the smaller mirror ball, would still be perceived as the middle of an infinite looking universe.  Although this real universe would only go out to about 6.8 BLY, due to gravitational reflections it could be perceived to go outward to 13.8 BLY.  

\subsection{Quasar Reflecting Universe (750 MLY)}
\begin{figure}[H]
\begin{center}
\includegraphics[width=.5\textwidth]{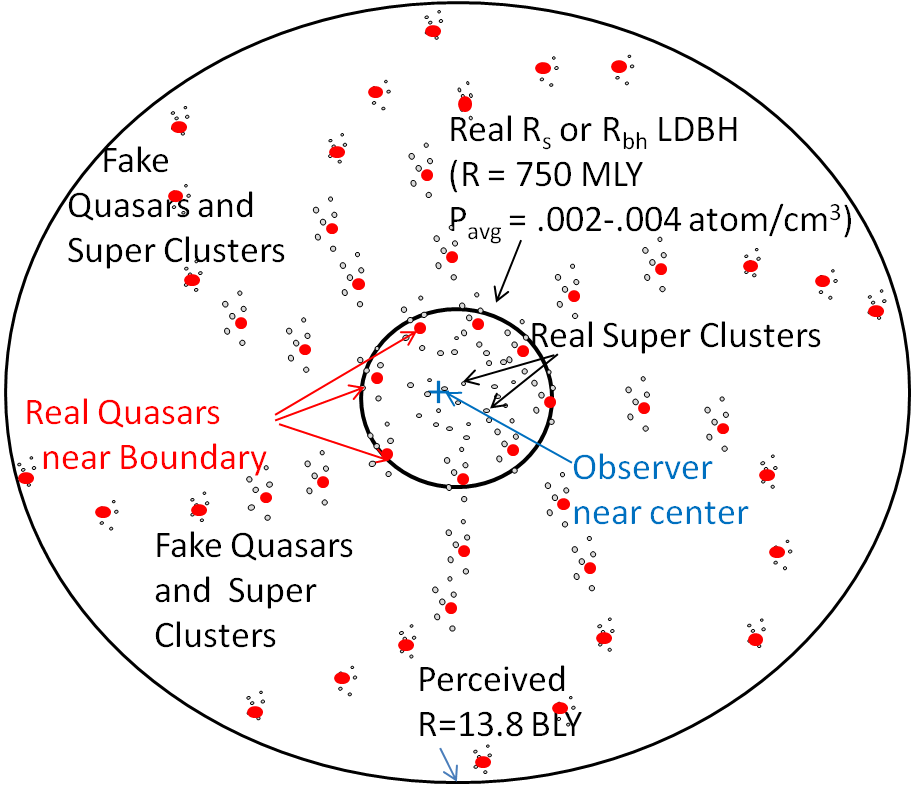}
\captionsetup{width=0.9\textwidth}
\caption{Quasar Reflecting Universe (real quasars at low density black hole boundary reflected). A quasar reflecting universe would be 750 MLY in radius, and have real quasars just inside its black hole boundary.
The quasar redshift would match the .062 redshift at that radius for the closest quasars due to just intrinsic redshift to the observer somewhere near the center.  The more distant quasars and galaxy clusters would be reflections, with transverse or intrinsic redshift continuing to accumulate linearly with increased distance traveled.  The average density of this universe would be equivalent to .0038 atoms per cubic centimeter.  The universe would only be 750 MLY, but due to gravitational reflections it would be perceived as 13.8 BLYs or larger.}
\label{QuasarReflectingUniverse}
\end{center}
\end{figure}

Recall that quasars have redshifts between 0.056 and 7.085 and are believed between 737 million and 12.5 billion light years away.  If the closest quasars are real, and the others are reflections, this could suggest a 750 MLY radius of the universe with a corresponding density of .0038 atoms/$cm^3$ as shown in Figure \ref{QuasarReflectingUniverse}. This model assumes that all redshifts, including that of the closest quasars and reflected quasars, come from an intrinsic redshift mechanism that would produce a Non-Doppler redshift equivalent to the Hubble constant redshift.  

The more distant quasars and galaxy super clusters would be “reflections” of the real, closer quasars and super clusters.  Like a disco mirror ball, the reflections would be perceived as traveling (linearly) farther due to their longer path. These simulated longer distances would look dimmer link more distant real galaxies and quasars, and have the associated higher redshifts.  Although this real universe would only go out to about 750 MLY, due to gravitational “reflections”  it would be perceived to go outward to 13.8 BLYs.

\subsection{Quasar Reflecting Low Density Black Hole Core (325 LY)}
\begin{figure}[H]
\begin{center}
\includegraphics[width=.5\textwidth]{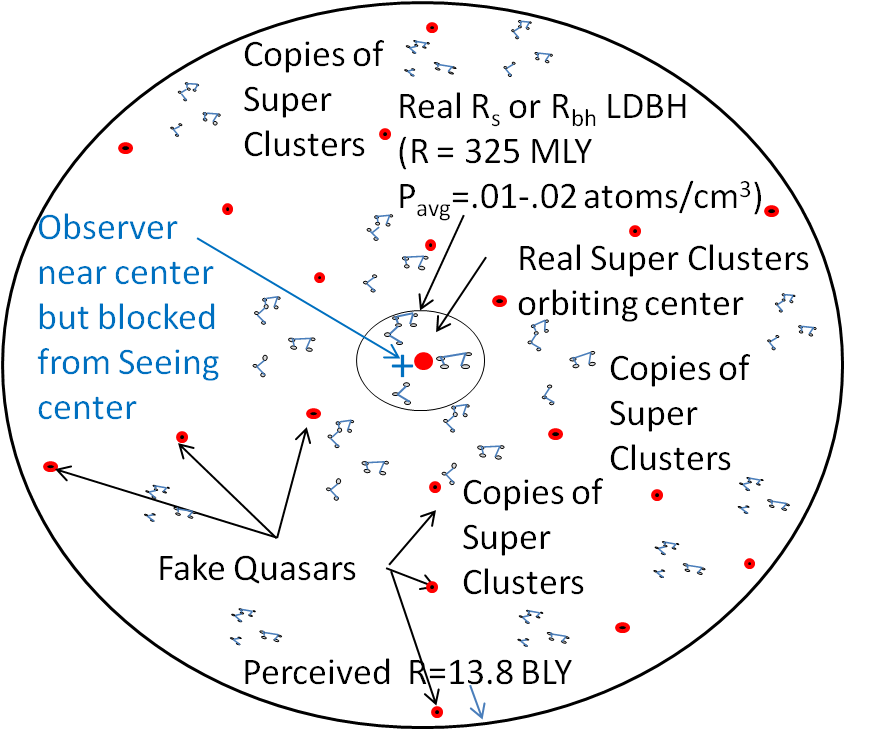}
\captionsetup{width=0.9\textwidth}
\caption{Fake Quasar Reflecting Universe (quasar as “reflections” of central core). A fake quasar reflecting universe would be 325 MLY in radius, and have real quasars at the galactic core, but direct viewing may be blocked from view by the galactic disk. All visible quasars would be reflections from the core.  The closest reflection would match the .062 redshift at that radius for the closest quasars due to just intrinsic redshift to the observer somewhere near the center.  The more distant quasars and galaxy clusters would be reflections, with intrinsic redshift continuing to accumulate linearly with the increased distance traveled.  The average density of this universe would be equivalent to .02 atoms per cubic centimeter.  The universe would only be 325 MLY, but due to gravitational reflections it would be perceived as 13.8 BLYs or larger.}
\label{FakeQuasarReflectingUniverse}
\end{center}
\end{figure}

If one assumes that the closest quasars are possible gravitational “reflections” themselves, from the central core of the universe low density black hole itself; then, the real radius of our reflecting low density black hole could be at about 325  MLYs with an average density about .02 atoms/ $cm^3$ (Figure \ref{FakeQuasarReflectingUniverse}).

The observer would be back near the center, but   direct viewing of the core would be blocked by the Milky Way’s accretion disk or the low density black hole accretion disk.  Light from the center core of the low density black hole is eventually “reflected” backwards and appears like a distant object 750 MLYs away, to get the needed redshift.   Subsequent cycles around or through the low density black hole, would travel farther and produce dimmer apparent magnitudes with more redshift.   

Assuming Bremsstrahlung intergalactic gas densities of 0.01 atoms/$cm^3$, half of the (0.02 atoms/$cm^3$) mass density would be in collapsed objects like suns and galaxies.  This would have an average density to .02 atoms/$cm^3$ on average, but still have $.01$ $atoms/cm^3$ intergalactic gas.  This 325 MLY universe would be perceived to go outward to 13.8 BLYs.
\subsection{Local Super Cluster Universe 66 MLY - 100 MLY}
\begin{figure}
        \centering
        \begin{subfigure}[b]{0.47\textwidth}
	\includegraphics[width=\textwidth]{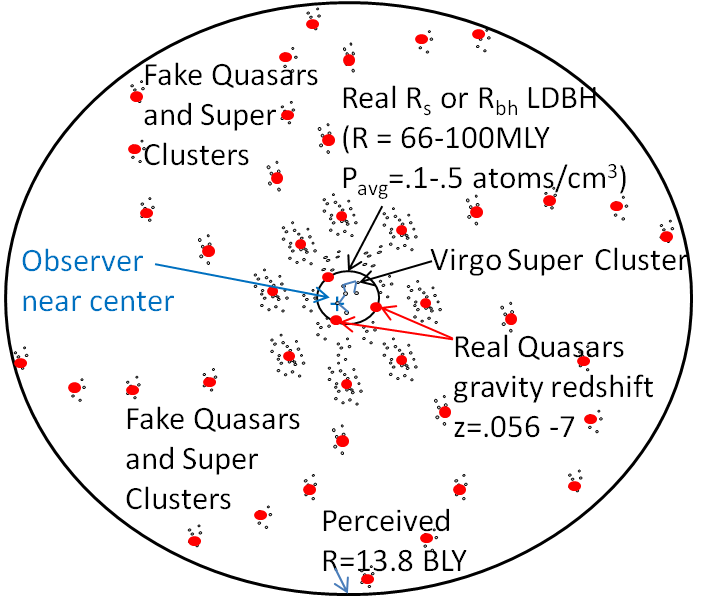}
	\caption{Real Quasars Near Low density black hole Boundary.}
        \end{subfigure}
	\qquad 
        \begin{subfigure}[b]{0.45\textwidth}
	\includegraphics[width=\textwidth]{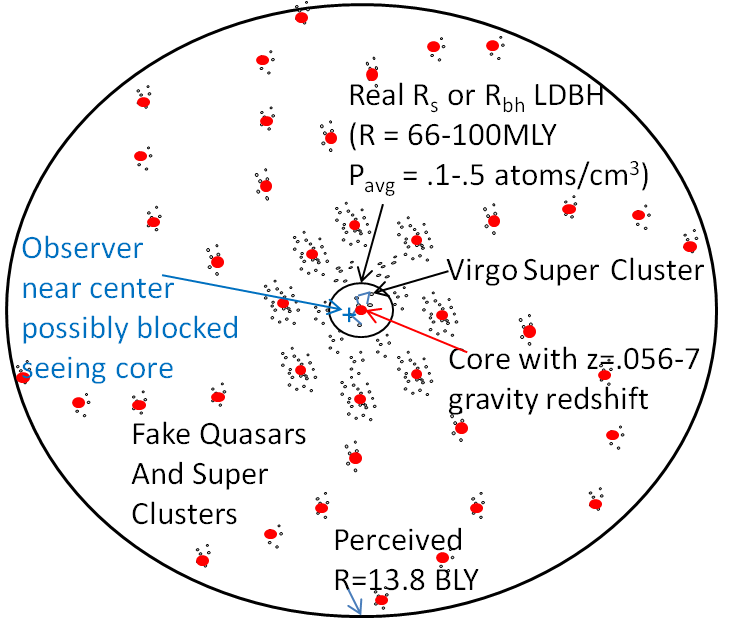}
	\caption{Real Quasar at low density black hole Core.}
        \end{subfigure}
\captionsetup{width=0.9\textwidth}
\caption{Virgo Super Cluster Universe. The local supercluster reflecting universe would be 66 to 100 MLY in radius. Real quasars could be either at (a) the black hole boundary or (b) at the galactic core, but direct viewing may be blocked from view by the galactic disk.  All more distant quasars and galactic clusters would be reflections from the real ones within the black hole. The quasar redshift would be caused primarily by gravitational redshift, with a smaller contribution coming from the linear redshift; or alternately by intrinsic redshift which accumulates over 92 BLY of travel.
The average density of this universe would be equivalent to .5 to .2 atoms per cubic centimeter.  The universe would only be 66 - 100 MLY, but due to gravitational reflections it would be perceived as 13.8 BLYs or larger.}
\label{VirgoSuperClusterUniverse}
\end{figure}

\begin{figure}[H] 
\begin{center} 
\includegraphics[width=.8\textwidth]{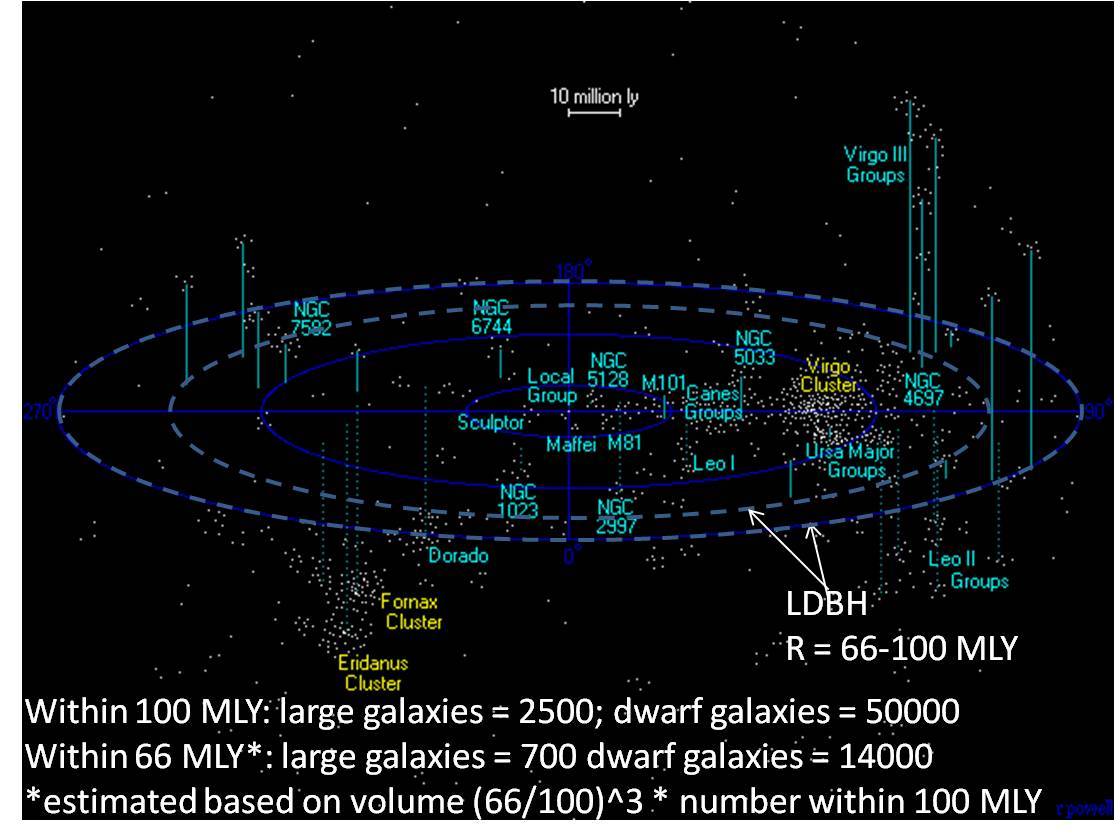}
\captionsetup{width=0.9\textwidth}
\caption{The only real galaxies and stars would be within the 66M to 100M LY radius as shown in the dashed lines. The number of real large galaxies available at 100 MLY is estimated at 2500 large galaxies, and 50000 dwarf galaxies.  This number is reduce at 66 MLY to about 700 large galaxies and about 14000 dwarf galaxies.  Note: overlays added to base 3D plots from \cite{atlas1}.}
\label{LocalSuperCluserLDBH}
\end{center}
\end{figure}

The next smaller logical reflecting universe would be around the Local Supercluster.  The average density would have to increase to .25 to .5 hydrogen atoms/$cm^3$.  The base redshift of $z=.000076$/MLY*66 MLY would be $z=0.005$.  The base redshift of $z=.000076$/MLY*100 MLY would be $z=0.0076$.  This base redshift would need to be multiplied by the number of reflection cycles because the optical illusion would appear to be traveling N times as far.  Other nearby superclusters would be reflected images of the Local Supercluster, but from a slightly different angle as shown in Figures \ref{VirgoSuperClusterUniverse} and \ref{LocalSuperCluserLDBH}.

The quasars could be real and located near the low density black hole boundary (as shown in Figure \ref{VirgoSuperClusterUniverse}a) or the quasars could be views of the center of the Local Supercluster black hole itself (as shown in Figure \ref{VirgoSuperClusterUniverse}b).  In both cases, these will need to be gravitationally redshifted to appear 750 MLY away. The quasar light would continue N time around the Local Supercluster, appearing deeper and deeper in space. All real galaxies and stars would be in the center of the low density black hole, see Figure \ref{LocalSuperCluserLDBH}. 
  
The incremental redshift could come from the following: Option (1), accumulative intrinsic redshift additions using the linear functions to reach $z$ values of about $z = 1.056$ at 13.8 BLY and  $z=7$ at 92 BLYs;  Option (2), the quasars near the low density black hole boundary just start out with gravitational redshifts between $z=.056$ to $z=6$; and then adding the linearly based intrinsic redshift up to 1 to reach $z=7$ at 13.8BLY; Option (3), the base quasar gravitational redshift (at the boundary or center) could vary over time and could have been higher in the past, perhaps due to more mass falling in at the quasar low density black hole in the past;  Option (4), the incremental intrinsic redshift could have been higher in the past due to more particles in the intergalactic medium in the past while the outer low density black hole was collapsing;  Option (5), the quasars could be external Local low density black hole outside the Local Cluster low density black hole, in a larger or infinite universe with $z=0.056$ to $z=7$.

Assuming Bremsstrahlung intergalactic gas densities of 0.01 atoms/$cm^3$, the majority of the mass (25-50 times) would be in collapsed objects like suns and galaxies.  This would raise the average density to 0.25-0.5 atoms/$cm^3$ on average, but still have .01 atoms/$cm^3$ intergalactic gas.  All other intrinsic redshift gravitational balls would be similar to the Bremsstrahlung shift gravitational ball above, but would not necessarily have to have the .01 atoms/$cm^3$ intergalactic gas density.  Although this real universe would only go out to 66 to 100 MLYs, due to gravitational reflections it could still be perceived to go outward to 13.8 BLY.  


\subsection{Local Cluster Universe 20-33 MLY}
\begin{figure}
        \centering
        \begin{subfigure}[b]{0.48\textwidth}
	\includegraphics[width=\textwidth]{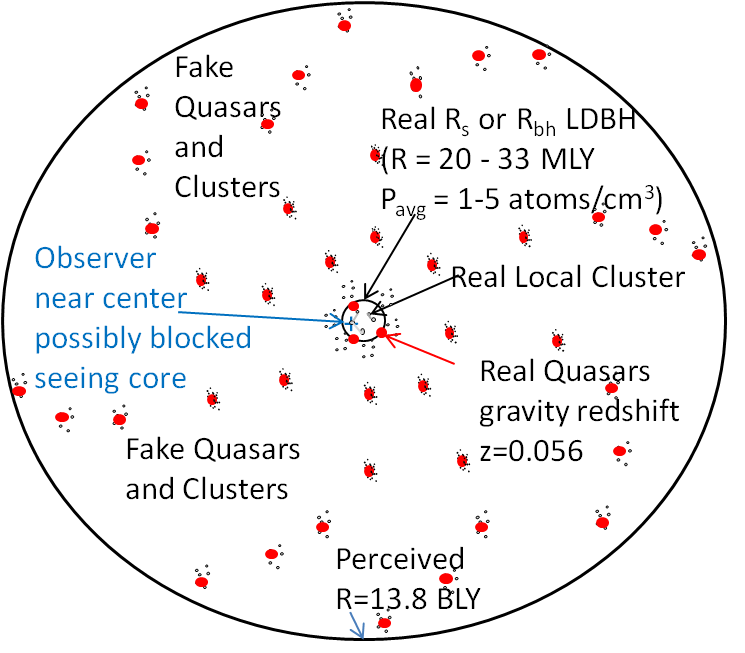}
	\caption{Real Quasars Near Low density black hole Boundary.}
        \end{subfigure}
	\qquad 
        \begin{subfigure}[b]{0.45\textwidth}
	\includegraphics[width=\textwidth]{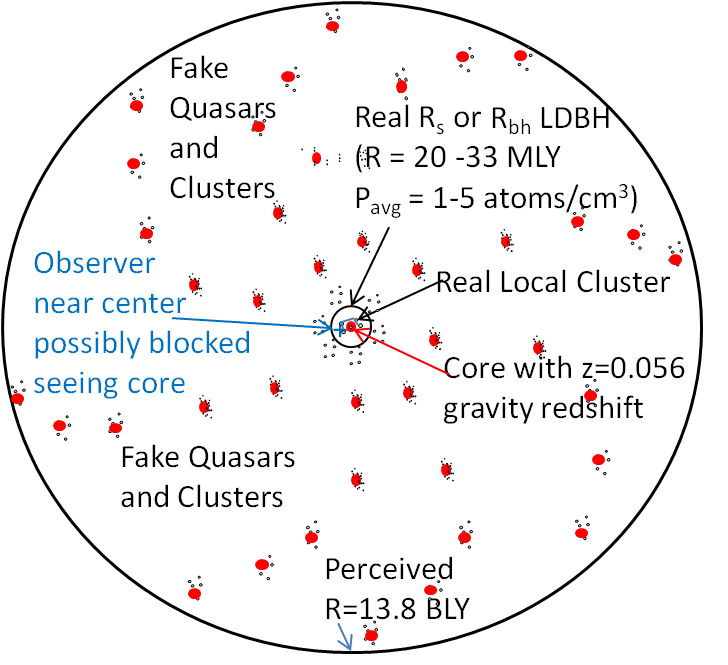}
	\caption{Real Quasar at low density black hole Core.}
        \end{subfigure}
\captionsetup{width=0.9\textwidth}
\caption{Local Cluster Universe. The local cluster reflecting universe would be 66 or 100 MLY in radius. Real quasars could be either at (a) the black hole boundary or (b) at the galactic core, but direct viewing may be blocked from view by the galactic disk.  All more distant quasars and galactic clusters would be reflections from the real ones within the black hole. The quasar redshift would be caused primarily by gravitational redshift, with a smaller contribution coming from the linear redshift; or alternatively by accumulated intrinsic redshift.
The average density of this universe would be equivalent to .5 to .2 atoms per cubic centimeter.  The universe would only be 66 - 100 MLY, but due to gravitational reflections it would be perceived as 13.8 BLYs or larger.}
\label{LocalClusterUniverse}
\end{figure}

The next smaller logical reflecting universe would be a low density black hole around the Local Cluster with a radius of 20-33 MLY.  The average density would have to increase to 1 to 5 hydrogen atoms/$cm^3$.  The base redshift of $z=.000076$/MLY* 20 MLY would be $z=0.00152$.  The base redshift of $z=.000076$/MLY* 33 MLY would be $z=0.0025$.  These base redshift would need to be multiplied by the number of reflection cycles because the optical illusion would appear to be traveling N times as far.  
Other nearby clusters would be reflected images of the Local Cluster, but from a slightly different angle as shown in Figure \ref{LocalClusterUniverse}.

\begin{figure}[H] 
\begin{center} 
\includegraphics[width=.9\textwidth]{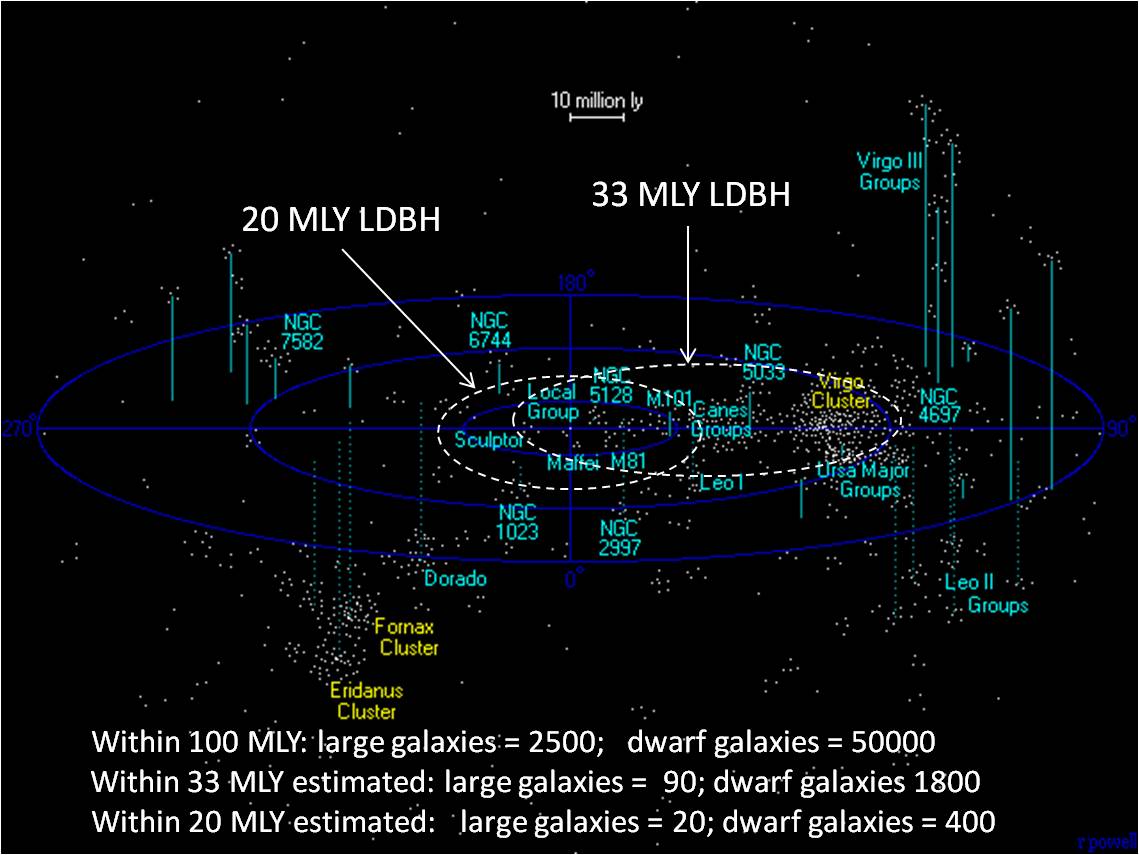}
\captionsetup{width=0.9\textwidth}
\caption{The only real galaxies and stars would be within the 20M LY or 33 M LY radii as shown in the dashed lines.  At 33 M LY radius, Fumax and Eridinus clusters would be illusions and the number of real large galaxies drop to 90 and real dwarf galaxies to about 1800.  At 20 M LY radius, the Virgo, Fumax, and Eridinus clusters would be illusions and the number of real large galaxies would drop to 20, and real dwarf galaxies to about 400.   Note: overlays added to base 3D plots from \cite{atlas1}.}
\label{20MLYLDBH}
\end{center}
\end{figure}

The Quasars could be real and located near the low density black hole boundary (as shown in Figure \ref{LocalClusterUniverse}a) or the quasars could be views of the center of the Virgo Super Cluster black hole itself (as shown in Figure \ref{LocalClusterUniverse}b).  In both cases, these will need to be gravitationally redshifted to appear 750 MLY away. The quasar light would continue N time around the Local Cluster, appearing deeper and deeper in space.  All real galaxies and stars would be in the center of the low density black hole, see Figure \ref{20MLYLDBH}. 
  
The incremental redshift could come from the following: Option (1), accumulative intrinsic redshift additions using the linear functions to reach $z$ values of about $z = 1.056$ at 13.8 BLY and  $z=7$ at 92 BLYs;  Option (2), the quasars near the low density black hole boundary just start out with gravitational redshifts between $z=.056$ to $z=6$; and then adding the linearly based intrinsic redshift up to 1 to reach $z=7$ at 13.8BLY; Option (3), the base quasar gravitational redshift (at the boundary or center) could vary over time and could have been higher in the past, perhaps due to more mass falling in at the quasar low density black hole in the past;  Option (4), the incremental intrinsic redshift could have been higher in the past due to more particles in the intergalactic medium in the past while the outer low density black hole was collapsing;  Option (5), the quasars could be external local low density black hole outside the Local Cluster low density black hole, in a larger or infinite universe with $z=0.056$ to $z=7$.

Assuming Bremsstrahlung intergalactic gas densities of 0.01 atoms/$cm^3$, the majority of the mass (100 to 500 times) would be in collapsed objects like suns and galaxies.  This would raise the average density to 1 - 5 atoms/$cm^3$ on average, but still have .01 atoms/$cm^3$ intergalactic gas.  

All other intrinsic redshift gravitational balls would be similar to the Bremsstrahlung shift gravitational ball above, but would not necessarily have to have the .01 atoms/$cm^3$ intergalactic gas density.

Although this real universe would go out to only 20 to 33 MLYs, due to gravitational reflections it could still be perceived to go outwards to 13.8 BLYs. 


\subsection{Local Group Reflecting Universe (5 MLY)}
The next smaller logical reflecting universe would be around the Local Group with radius of 5 MLYs.  The average density would have to increase to 50 hydrogen atoms/$cm^3$.  The base redshift of $z=.000076$/MLY* 5 MLY would be $z=0.0005$.  This base redshift would need to be multiplied by the number of reflection cycles because the optical illusion would appear to be traveling N times as far.  

\begin{figure}
        \centering
        \begin{subfigure}[b]{0.45\textwidth}
	\includegraphics[width=\textwidth]{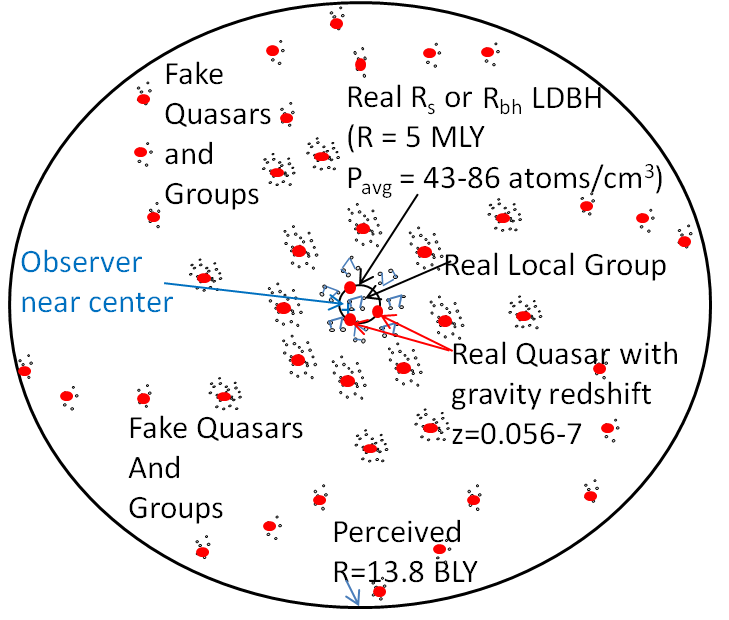}
	\caption{Real Quasars Near Low density black hole Boundary.}
        \end{subfigure}
	\qquad 
        \begin{subfigure}[b]{0.42\textwidth}
	\includegraphics[width=\textwidth]{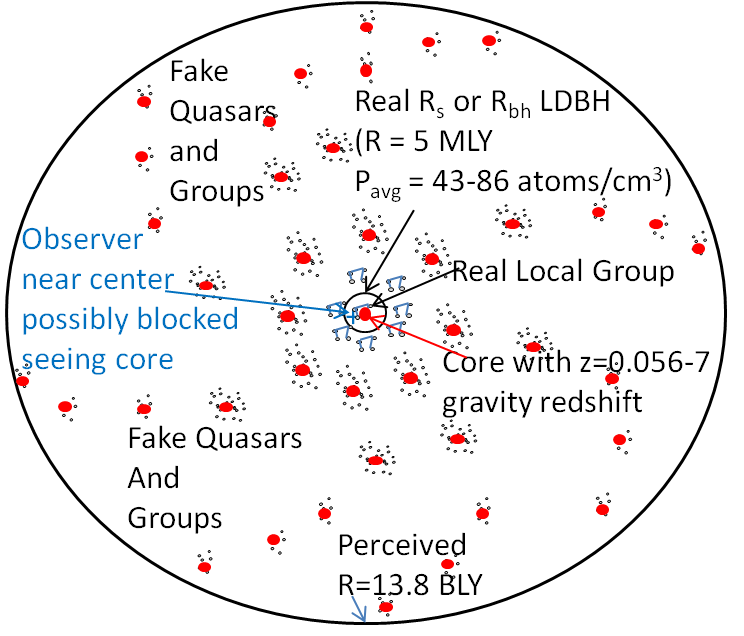}
	\caption{Real Quasar at low density black hole Core.}
        \end{subfigure}
\captionsetup{width=0.9\textwidth}
\caption{Local Group  Universe with 5 MLY Radius.  Real quasars could be at the black hole boundary (left) or at the galactic core (right).  Direct viewing may be blocked from view by the galactic disk.  All more distant quasars and galactic clusters would be reflections. The quasar redshift would be caused by gravitational redshift, with a smaller contribution coming from the linear redshift; or from accumulated linear redshift of quasars in the past. The average density would be $86$ $atoms/cm^3$.  Although only 5 MLY, due to gravitational reflections it would be perceived as 13.8 BLYs or larger.}
\label{LocalGroupClusterUniverse}
\end{figure}

Other nearby galaxy groups would be reflected images of the Local Group, but from a slightly different angle as shown in Figure \ref{LocalGroupClusterUniverse}.

\begin{figure}[H] 
\begin{center} 
\includegraphics[width=.85\textwidth]{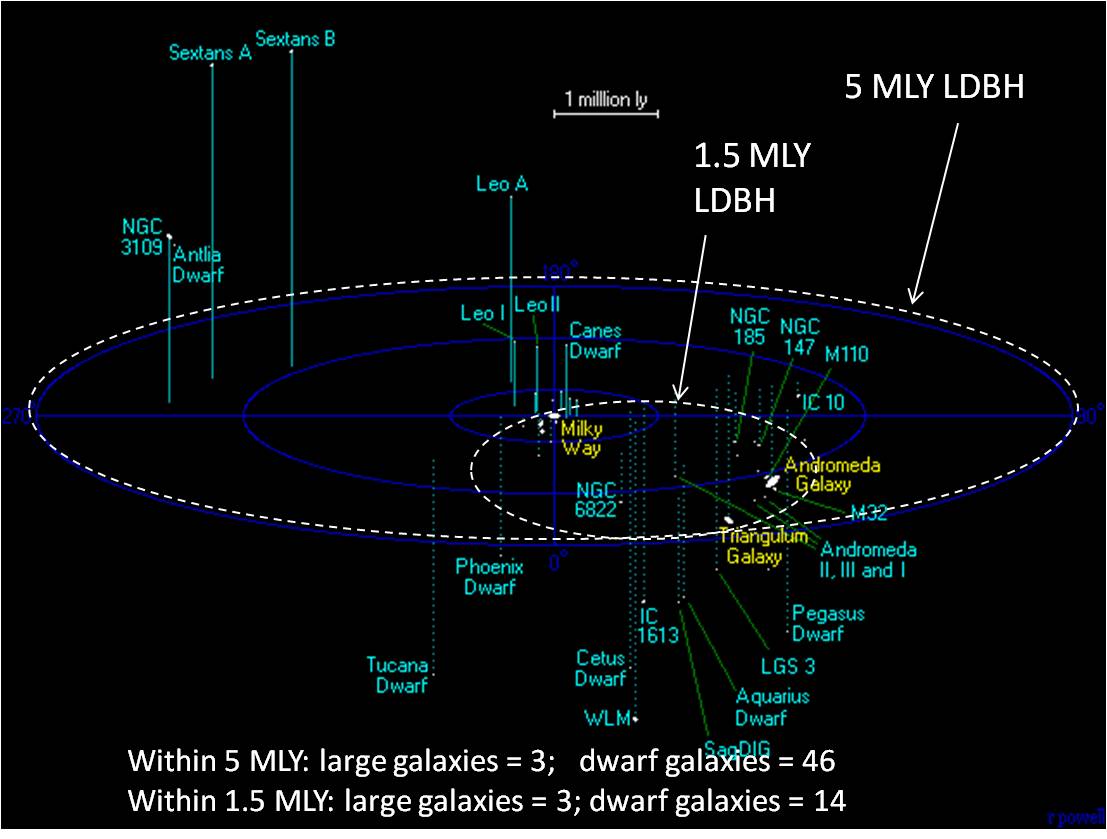}
\captionsetup{width=0.9\textwidth}
\caption{5 MLY Universe. The only real galaxies and stars that would be within the 1.5M or 5M LY radii are shown in the dashed lines.  The number of real large galaxies would be 3 (the Milky Way, Andromeda, and Triangulum galaxies).  The number of dwarf galaxies would be 14-46.  This small number of real galaxies does not seem consistent with the variety observed galaxies unless these galaxies look different from different angles, vary over time, or are distorted by a black hole effect.   Note: overlays added to base 3D plots from \cite{atlas2}.}
\label{5MLYLDBH}
\end{center}
\end{figure}

The quasars could be real and located near the low density black hole boundary (as shown in Figure \ref{LocalGroupClusterUniverse}a) or the quasars could be views of the center of the Local Group galaxy black hole itself (as shown in Figure \ref{LocalGroupClusterUniverse}b).  In both cases, these will need to be gravitationally redshifted to appear 750 MLY away. The quasar light would continue N time around the Virgo Super Cluster, appearing deeper and deeper in space. All real galaxies and stars would be in the center of the low density black hole, see the outer dashed ring in Figure \ref{5MLYLDBH}. 
  
The incremental redshift could come from the following: Option (1), accumulative intrinsic redshift using the linear functions to reach $z$ values of about $z = 1.056$ at 13.8 BLY and  $z=7$ at 92 BLYs;  Option (2), quasars near the low density black hole boundary just start out with gravitational redshifts between $z=.056$ to $z=6$; and then adding the linearly based intrinsic redshift up to 1 to reach $z=7$ at 13.8 BLY; Option (3), the base quasar gravitational redshift (at the boundary or center) could vary over time and could have been higher in the past, perhaps due to more mass falling in at the quasar low density black hole in the past;  Option (4), the incremental intrinsic redshift could have been higher in the past due to more particles in the intergalactic medium in the past while the outer low density black hole was collapsing;  Option (5), the quasars could be external Local low density black hole outside the Local Cluster low density black hole, in a larger or infinite universe with $z=0.056$ to $z=7$.

Assuming Bremsstrahlung interstellar gas densities of 0.01 atoms/$cm^3$, the majority of the mass (5000 times) would be in collapsed objects like suns and galaxies.  This would raise the average density 50 atoms/$cm^3$ on average, but still have .01 atoms/$cm^3$ intergalactic gas.

All other intrinsic redshift gravitational balls would be similar to the Bremsstrahlung shift gravitational ball above, but would not necessarily have to have the .01 atoms/$cm^3$ intergalactic gas density.

Although this real universe would go out to only 5 MLYs, due to gravitational reflections it could still be perceived to go outwards to 13.8 BLYs.

\subsection{Milky Way Subgroup Universe (1.5MLY)}
The next smaller logical reflecting universe would be around the Milky Way Subgroup with radius of 1.5 MLYs.  The average density would have to increase to 1 hydrogen atom/$mm^3$.  The base redshift of $z=.000076$/MLY* 1.5 MLY would be z=0.000114.  This base redshift would need to be multiplied by the number of reflection cycles because the optical illusion would appear to be traveling N times as far.  

\begin{figure}
        \centering
        \begin{subfigure}[b]{0.48\textwidth}
	\includegraphics[width=\textwidth]{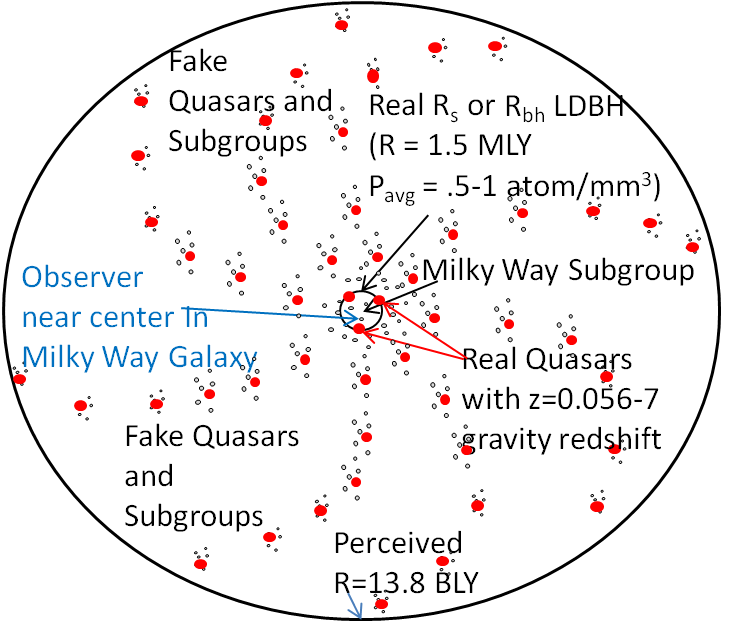}
	\caption{Real Quasars Near Low density black hole Boundary.}
        \end{subfigure}
	\qquad 
        \begin{subfigure}[b]{0.45\textwidth}
	\includegraphics[width=\textwidth]{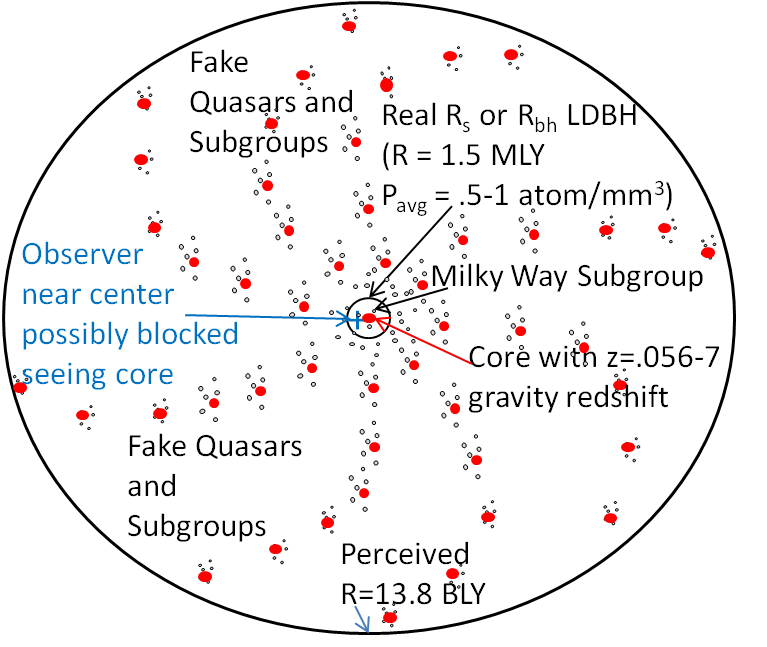}
	\caption{Real Quasar at low density black hole Core.}
        \end{subfigure}
\captionsetup{width=0.9\textwidth}
\caption{Milky Way Subgroup Reflecting Universe. The local group reflecting universe would be 1.5 MLY in radius. Real quasars could be either at (a) the black hole boundary or (b) at the galactic core, but direct viewing may be blocked from view by the galactic disk.  All more distant quasars and galactic clusters would be reflections from the real ones within the black hole. The quasar redshift would be caused primarily by gravitational redshift, with a smaller contribution coming from the linear redshift; or alternatively by accumulative intrinsic redshift. The average density of this universe would be equivalent to 1 atoms per cubic millimeter.  The universe would only be 1.5MLY, but due to gravitational reflections it would be perceived as 13.8 BLYs or larger.}
\label{MilkyWaySubgroupUniverse}
\end{figure}

Other nearby galaxy groups would be reflected images of the Milky Way Subgroup, but from a slightly different angle as shown in Figure \ref{MilkyWaySubgroupUniverse}.

The quasars could be real and located near the low density black hole boundary (as shown in Figure \ref{MilkyWaySubgroupUniverse}a) or the quasars could be views of the center of the Milky Way Subgroup black hole itself (as shown in Figure \ref{MilkyWaySubgroupUniverse}b).  In both cases, these will need to be gravitationally redshifted to appear 750 MLY away. The quasar light would continue N time around the Milky Way Subgroup, appearing deeper and deeper in space. All real galaxies and stars would be in the center of the low density black hole, see the inner dashed ring in Figure \ref{5MLYLDBH}. 
  
The incremental redshift could come from the following: Option (1), accumulative intrinsic redshift additions using the linear functions to reach $z$ values of about $z=1.056$ at 13.8 BLY and  $z=7$ at 92 BLYs;  Option (2), the quasars near the low density black hole boundary just start out with gravitational redshifts between $z=.056$ to $z=6$; and then adding the linearly based intrinsic redshift up to 1 to reach $z=7$ at 13.8BLY; Option (3), the base quasar gravitational redshift (at the boundary or center) could vary over time and could have been higher in the past, perhaps due to more mass falling in at the quasar low density black hole in the past;  Option (4), the incremental intrinsic redshift could have been higher in the past due to more particles in the intergalactic medium in the past while the outer low density black hole was collapsing;  Option (5), the quasars could be external Local low density black hole outside the Local Cluster low density black hole, in a larger or infinite universe with $z=0.056$ to $z=7$.

 Assuming Bremsstrahlung interstellar gas densities of 0.01 atoms/$cm^3$, the majority of the mass (10000 times) would be in collapsed objects like suns and galaxies.  This would raise the average density to 1 atom/$mm^3$ on average, but still have .01 atoms/$cm^3$ intergalactic gas.

All other intrinsic redshift gravitational balls would be similar to the Bremsstrahlung shift gravitational ball above, but would not necessarily have to have the .01 atoms/$cm^3$ intergalactic gas density.

Although this real universe would go out to only 1.5 MLYs, due to gravitational reflections it could still be perceived to go outwards to 13.8 BLYs.
 
\subsection{MilkyWay Galaxy Universe (200-50KLY)}
The next smaller logical reflecting universe would be around the Milky Way Galaxy with radius of 50-200 KLYs.  The average density would have to increase to 50-850 hydrogen atom/$mm^3$.  The base redshift of $z=.000076$/MLY* .05(to.85) MLY would be $z=0.0000646$ to $z=0.0000038$.  The base redshift would need to be multiplied by the number of reflection cycles because the optical illusion would appear to be traveling N times as far.  

\begin{figure}[H]
\begin{center} 
\includegraphics[width=.7\textwidth]{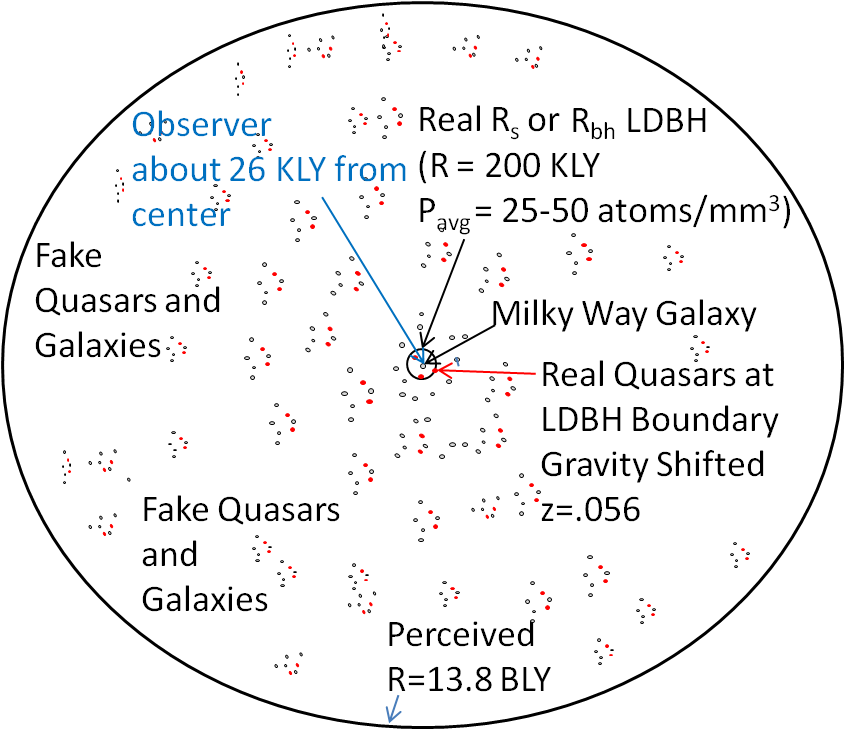}
\captionsetup{width=0.9\textwidth}
\caption{Milky Way Sphere Illusion. The Milky Way reflecting universe would be 200 KLY in radius.  Real quasars probably would be at the black hole boundary since we don’t see a quasar in the Milky Way core, but direct viewing may be blocked from view by the galactic disk.  All more distant quasars and galactic clusters would be reflections from the real ones within the black hole. The quasar redshift would be caused primarily by gravitational redshift, with a smaller contribution coming from the linear redshift; or alternatively via accumulative intrinsic redshift.  The average density of this universe would be equivalent to 50 atoms per cubic millimeter.  The universe would only be 200K LY, but due to gravitational reflections it would be perceived as 13.8 BLYs or larger.  This small number of real galaxies does not seem consistent with the variety of pictures that Hubble has shot unless these galaxies look different from different angles and varied in the past. The black hole gravitational field might also have to provide some type of radical distortion effect.}
\label{MilkyWaySphereIllusion}
\end{center}
\end{figure}

Other nearby galaxy groups would be reflected images of the Milky Way, but from a slightly different angle as shown in Figure \ref{MilkyWaySphereIllusion}.  Andromeda and Triangulum galaxies would be the first “reflections” of the Milky Way.  The Quasars could be real and located near the low density black hole boundary (as shown in Figure \ref{MilkyWaySphereIllusion}).  In this case, these will need to be gravitationally redshifted to appear 750 MLY away. The quasar light would continue N time around the Milky Way Galaxy, appearing deeper and deeper in space.

The incremental redshift could come from the following: Option (1), accumulative intrinsic redshift additions using the linear functions to reach $z$ values of about $z=1.056$ at 13.8 BLY and  $z=7$ at 92 BLYs;  Option (2), the quasars near the low density black hole boundary just start out with gravitational redshifts between $z=.056$ to $z=6$; and then adding the linearly based intrinsic redshift up to 1 to reach $z=7$ at 13.8BLY; Option (3), the base quasar gravitational redshift (at the boundary or center) could vary over time and could have been higher in the past, perhaps due to more mass falling in at the quasar low density black hole in the past;  Option (4), the incremental intrinsic redshift could have been higher in the past due to more particles in the intergalactic medium in the past while the outer low density black hole was collapsing;  Option (5), the quasars could be external Local low density black hole outside the Local Cluster low density black hole, in a larger or infinite universe with $z=0.056$ to $z=7$.

Assuming Bremsstrahlung interstellar gas densities of 0.01 atoms/$cm^3$, the majority of the mass (50000-850000 times) would be in collapsed mass like suns and galaxies.  This would raise the average density to 50-850 atoms/$mm^3$ on average, but still have .01 atoms/$cm^3$ intergalactic gas.

All other intrinsic redshift gravitational balls would be similar to the Bremsstrahlung shift gravitational ball above, but would not necessarily have to have the .01 atoms/$cm^3$ intergalactic gas density.

Although this real universe would go out to only 200-50 KLYs, due to gravitational reflections it could still be perceived to go outwards to 13.8 BLYs

\subsection{Sub Galaxy universe (e.g.$<50$ KLY)}
The average density would have to increase further to produce a low density black hole at smaller than the galaxy boundaries.  The rotating low density black hole would then have to also create the illusion of a full galaxy (perhaps by frame dragging space in a spiral pattern).  Then it would have to multiply this galaxy image via the mirrored ball or gravitational redshifting, and have them look different from different angles.  

\section{Quantized redshift}
These last two Milky Way Subgroup and Galaxy low density black holes would actually fit the quantized galactic distances and quantized redshift phenomenon researched in the late 90s \cite{hawkins, tifft1,tifft2,tifft3,arp1987}.  Other scientists (including Hawking) tried to find this periodicity with the latest data but found no correlation, or that could not be explained by grouping or selection effects \cite {hawkins}.  It also seemed to disagree with the Big Bang universe at the time.   Maybe with the Big Bang Theory off the table, quantized redshifts might warrant a second look.  

Note that the primary source of the redshift quantization presented here is from the reflected galaxies coming in reflective waves at periodic radii, and thus, there would be more redshift at these radii and less (or none) at others.  There may also be some minor secondary affects due to variations in the source of the redshift outside the black hole or at its boundary.  Also note, the reflecting universes presented earlier do not depend on this quantized redshift.  But if quantized redshift is present, a reflecting black hole could be its source.  Additionally, the low density black holes are not like stellar black holes containing only one light source, but are very large with many galaxies at different depths.  Thus, the light can start reflecting at lower levels and travel in arbitrary arcs.  Thus, crisp distinctive quantized levels may not be blatantly observable, but may only show up statistically.

\subsection{Milk Way Group Quantized Universe}
Tifft and colleagues found periodic quantized redshifts in nearby galaxies data in primary increments around $z=.00024$ and $z = .00012$ with minor sub-harmonics at $z=.00008$ (Doppler velocities of 72, 36, and 24 km/s) and that nearby galaxies were spaced in periodic intervals of about 3 MLY.   

\begin{figure}[H]
\begin{center} 
\includegraphics[width=.6\textwidth]{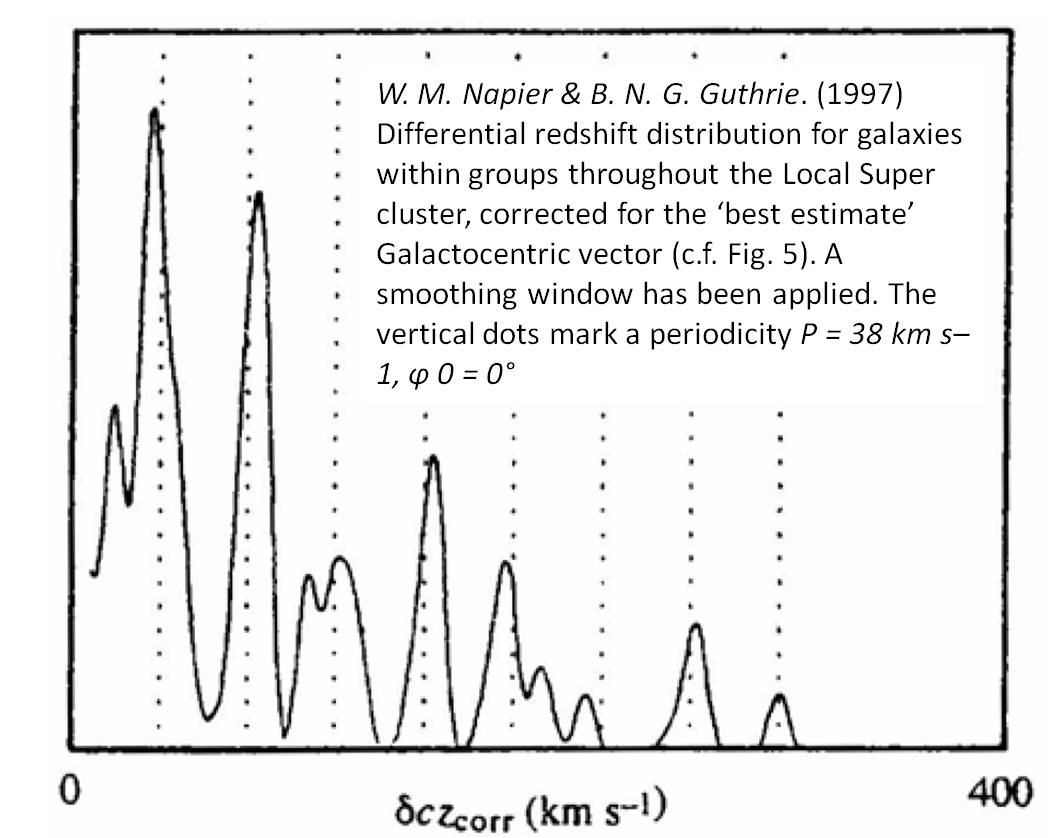}
\captionsetup{width=0.9\textwidth}
\caption{Quantized Redshift \cite{napier}. A periodic quantized redshift of 38 km/sec would correspond to a reflecting universe of 1.5 MLY radius.  Note: figure a composite from \cite{napier}.}
\label{QuantizedRedshift}
\end{center}
\end{figure}

In 1997, Napier and Guthrie performed an independent redshift analysis of nearby galaxies and found a periodicity at 38 km/sec  as shown in Figure \ref{QuantizedRedshift} from Napier et al. \cite{napier}.  Napier’s and Guthrie’s final statement: ``\textsl{Our conclusion is that extragalactic redshifts are quantized along the lines originally suggested by Tifft and coworkers, with galacto-centric periodicities of 37.5 km/sec in field galaxies and loose groupings, and 71.1 km/sec in the environment of dense clusters.}”

\begin{figure}
        \centering
        \begin{subfigure}[b]{0.45\textwidth}
	\includegraphics[width=\textwidth]{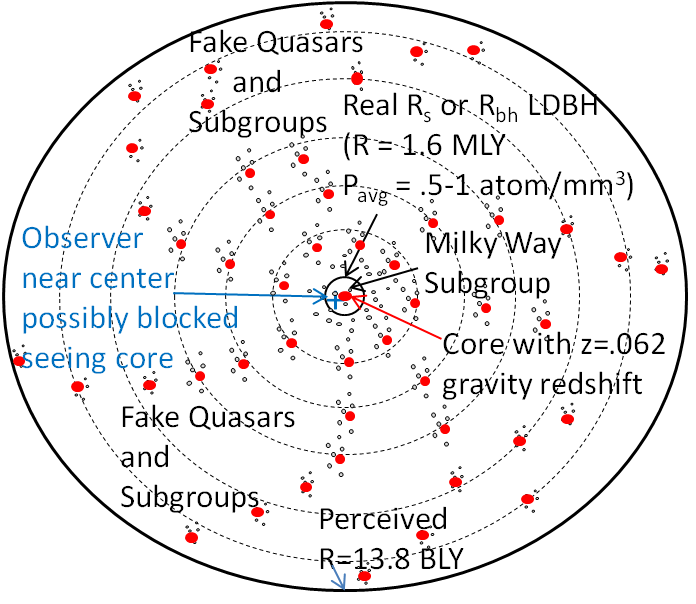}
	\caption{Quantized Redshift $z=.00024$, $d=3$MLY }
        \end{subfigure}
	\qquad 
        \begin{subfigure}[b]{0.48\textwidth}
	\includegraphics[width=\textwidth]{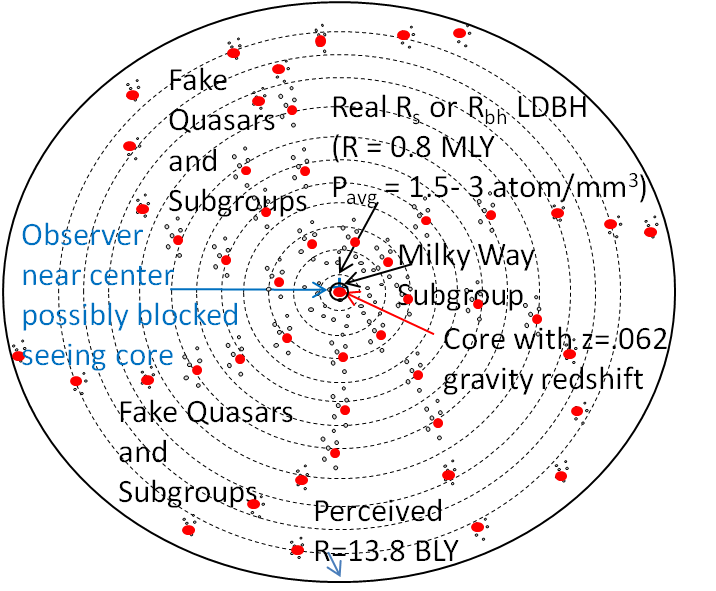}
	\caption{Quantizid Redshift $z=.00012$, $d=1.6$MLY }
        \end{subfigure}
\captionsetup{width=0.9\textwidth}
\caption{Milk Way Group Quantized Universe. The local group reflecting universe would be .8-1.6 MLY in radius. Real quasars could be either at (a) the black hole boundary or (b) at the galactic core, but direct viewing may be blocked from view by the galactic disk.  All more distant quasars and galactic clusters would be reflections from the real ones within the black hole. The quasar redshift would be caused primarily by gravitational redshift, with a smaller contribution coming from the linear redshift; or alternatively via accumulative intrinsic redshift.  The redshift would come in quantized increments due to the absence of mass between reflections.  The average density of this universe would be equivalent to 1-3 atoms per cubic millimeter.  The universe would only be .8-1.6MLY, but due to gravitational reflections it would be perceived as 13.8 BLYs or larger.}
\label{MilkyWayUniverse}
\end{figure}

These quantized redshifts and galaxy distances would be produce by a “reflecting” universe of radius at 1.6 MLY, at the edge of the Milky Way group as shown in Figure \ref{MilkyWayUniverse}a. Gravitational “reflections” would naturally line up in increments of 3 MLYs.  Figure \ref{MilkyWayUniverse}b repeats this for $z=.00012$, d=1.6 MLY distances with a R=0.8 MLY low density black hole radius.

In 1996, Tifft compensated for the 560 km/sec galactic rotation against the background radiation, then some less intense periodicities, at $z=6.1*10^{-5}$,  $3.1*10^{-5}$, and $8.7*10^{-6}$; and Doppler velocities of 18.3, 9.15, and 2.6 km/s \cite{tifft1996a, tifft1997b}.  These quantized redshifts are converted to redshifts in Table \ref{table6}, and then used to estimate the reflection Diameter and radii for the various shifts.  These additional smaller values could support a Milky Way Galaxy Size low density reflection radius or smaller. 

Note, that these quantized $z$ values could indicate a Milky Way Subgroup or Galaxy size universe; or just a Milky Way Subgroup or galaxy size localized low density black holes within a larger universe.
\begin{table}
\caption{Tifft Quantized redshift Radii}
\begin{center}
\begin{tabular}{| c | c | c | c | c | }
  \hline  
$\pbox{20cm}{Old Doppler\\V (km/s)}$& \pbox{20cm}{Quantize\\ $dz=V/C$}& \pbox{20cm}{Linear \\$d=z/.000076$\\(MLY)} & \pbox{20cm}{$R=d/2$ \\ (MLY)} &	\pbox{20cm}{$Cir=2R\pi$ \\$R=d/2\pi$ \\(MLY)} \\
\hline
72&	0.00024&	3.202&	1.579&	.5\\
36&	0.00012&	1.579&	0.789&	.25\\
24&	0.00008&	1.07&	0.53&	.17\\
18.3&	0.000061&	0.803&	0.401&	.128\\
9.15&	0.000031&	0.401&	0.201&	.064\\
2.6&	0.0000087&	0.114&	0.057&	.019\\
\hline
\end{tabular}
\end{center}
\label{table6}
\end{table}
\subsection{Quasar Quantization}
\begin{figure}[H]
\begin{center} 
\includegraphics[width=.6\textwidth]{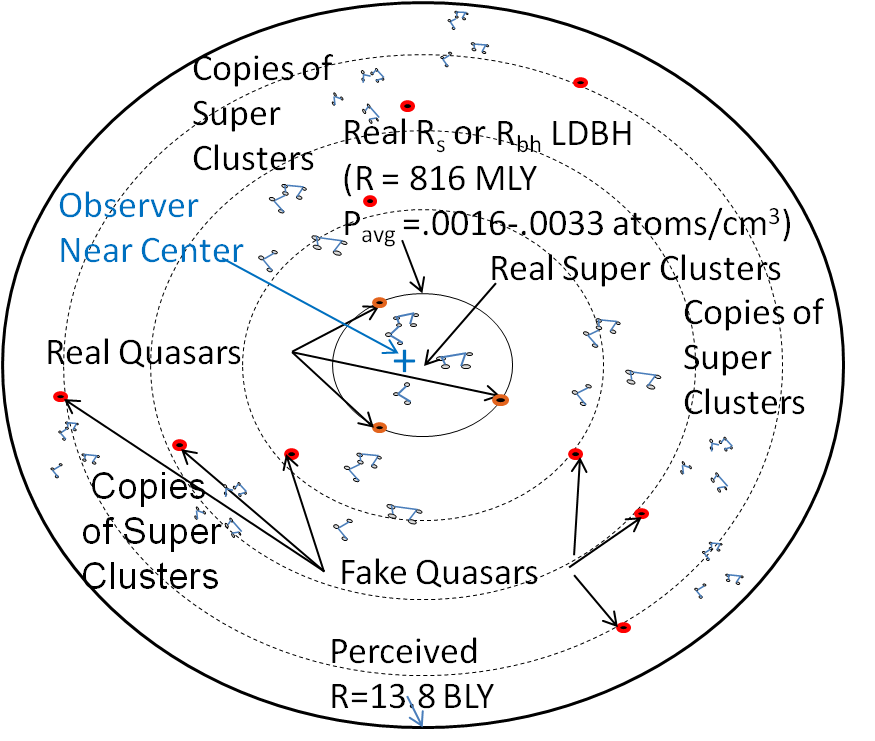}
\captionsetup{width=0.9\textwidth}
\caption{Quasar Quantized Universe. The quantized quasar reflecting universe would be 816 MLY in radius, and have real quasars just inside its black hole boundary.
The quasar redshift would match the .062 redshift at that radius for the closest quasars due to just intrinsic redshift to the observer somewhere near the center.  The more distant quasars and galaxy clusters would be reflections, with intrinsic redshift continuing to accumulate linearly with the increase distance traveled.  The average density of this universe would be equivalent to .0033 atoms per cubic centimeter.  The universe would only be 816 MLY, but due to gravitational reflections it would be perceived as 13.8 BLYs or larger.}
\label{QuasarQuantizedUniverse}
\end{center}
\end{figure}

In 2002, Bell claimed that all quasar redshifts may contain a discrete intrinsic component that is a harmonic of 0.062, and are tied directly to the discrete velocities found to be present in normal galaxies by Tiff (1996, 1997), and all are harmonically related to $z = 0.062 \pm 0.001$ \cite{bell}.  The .062 quantized redshifts, if real, could support Quasar reflection universes as discussed earlier in Section 9, but with a base shift of $z=0.062$ as shown in Figure \ref{QuasarQuantizedUniverse} ($d= z/.000076 = 0.063/.000076 = 816$ MLY).

\subsection{Nested Quantization}
\begin{figure}[H]
\begin{center} 
\includegraphics[width=.9\textwidth]{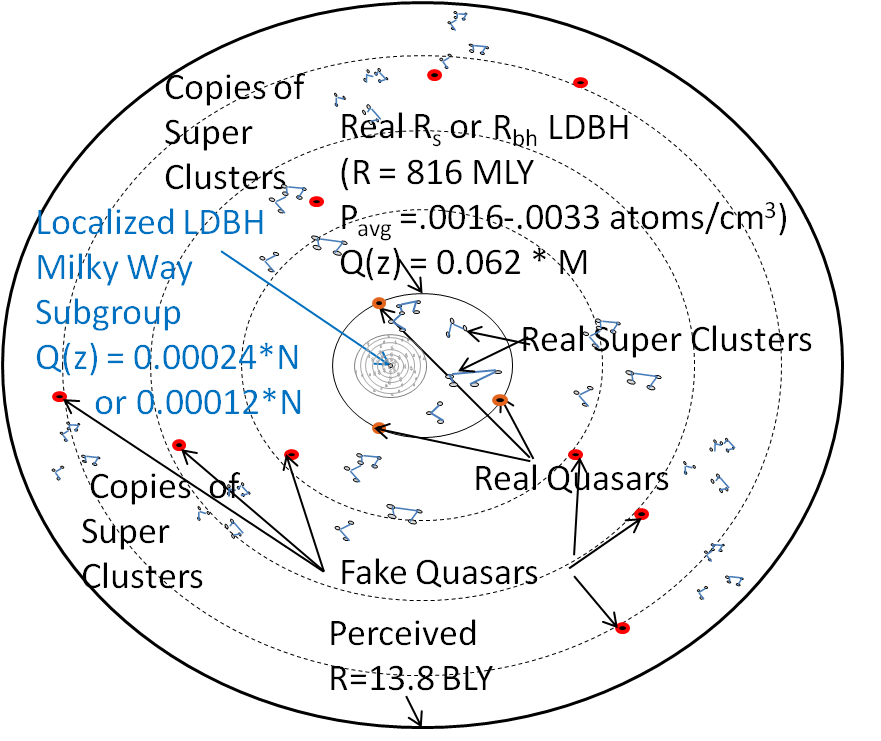}
\captionsetup{width=0.9\textwidth}
\caption{Nested Quantized Universe. The Nested Quantized Universe is a combination of the quantized quasar universe with a local group localized low density black hole causing smaller quantized reflections about the Milky Way Galaxy.}
\label{NestedQuantizedUniverse}
\end{center}
\end{figure}

The combination of both of the above quantized redshifts reflection could be possible at the same time with a quasar reflection universe and a Milky Way Subgroup or Milky Way Galaxy size localized reflecting low density black hole as shown in Figure \ref{NestedQuantizedUniverse}. In addition, there could be additional nested reflections spheres (not shown) at the super cluster layer as well.

\subsection{Mass Distribution Analysis} 

\begin{figure}[H]
\begin{center} 
\includegraphics[width=\textwidth]{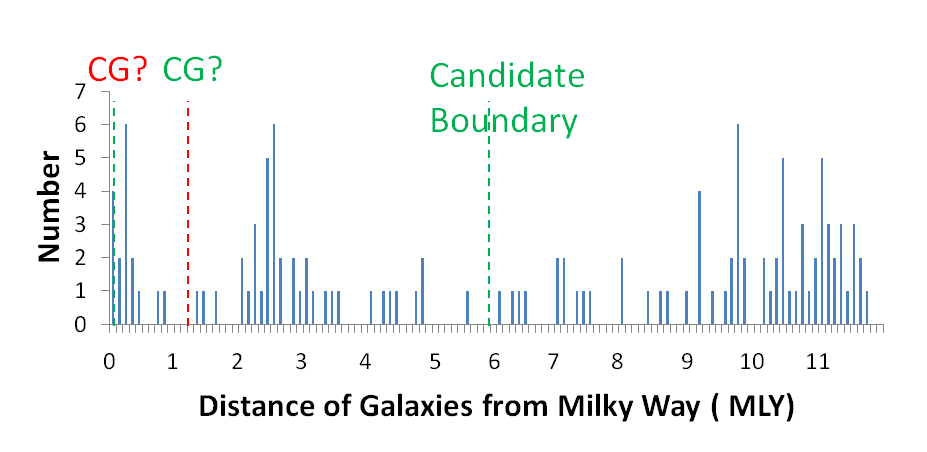}
\captionsetup{width=0.9\textwidth}
\caption{Plot of Number of Galaxies from Milky Way v.s. Distance.  If the Milky Way is the Center of Gravity (CG) of the black hole, the red dashed line cannot be the boundary of the black hole, since although it shows a primary reflection, the secondary reflections would be missing.  A candidate boundary would exist at about 5.8 MLY.  However, knowing that the galaxies at 2.5 MLY contain the Andromeda and Triangulum spiral galaxies have at least as much mass as the Milky Way Galaxy, the Milky Way galaxy is probably not the CG of the black hole, if the boundary is at 5.8 MLY.  Thus, a better CG would be between the Milky Way and Andromeda/Triangulum and the candidate boundary would be about 4.5 MLY.  Thus, this analysis pretty much rules out the Milky Way Black Hole universe, but still enable the Local Subgroup or Local Group black hole universe.  The universe or local black hole CG could be out even farther.  But, it looks like Andromeda and Triangulum galaxies are probably real.}
\label{MWmassdist}
\end{center}
\end{figure}

If the Milky Way Galaxy is the Center of Gravity (CG) of the universe or local low density black hole, somewhat periodic reflections should be visible in the plot of the number of galaxies near the Milky Way depicted in Figure \ref{MWmassdist}.  If the Milky Way is the Center of Gravity (CG) of the black hole,  the first plausible reflection line would be at 1.25 MLY.  If this was the case, the Andromeda and Triangulum galaxy would be the first reflections.  However, if this was the case, there would also be galaxy reflections at 5 MLY, but these are absent.  

A second plausible reflection boundary would be at 5.8 MLY.   But if this were the case, the Andromeda and Triangulum galaxies would be real, and since these are at least as massive as the Milky Way galaxy, the CG would probably not be the  center of the Milky Way.  Thus, a more likely scenario is that the CG would be between the Milky Way and Andromeda/Triangulum galaxies (between 1.25 and 1.75), and the first plausible reflection line becomes about 4.5 MLY from the CG.  Thus, this analysis pretty much rules out the Milky Way Black Hole universe, but still enables the Local Subgroup or Local Group black hole universe or localized black hole. However, the universe or local black hole CG could be out even farther, or not existent at all.  But, it looks like Andromeda and Triangulum galaxies are probably real, unless the missing stars are blocked from view by the Milky Way disk.    

This CG analysis could be extended farther, but will need to switch to 3D.   However, one would need good positional data for each galaxy in the local supercluster and need the mass of each galaxies to compute the CGs.   In addition, one may be able to use velocity data to improve that analysis.

\subsection{Giant Voids}
\begin{figure}[H]
\begin{center} 
\includegraphics[width=.76\textwidth]{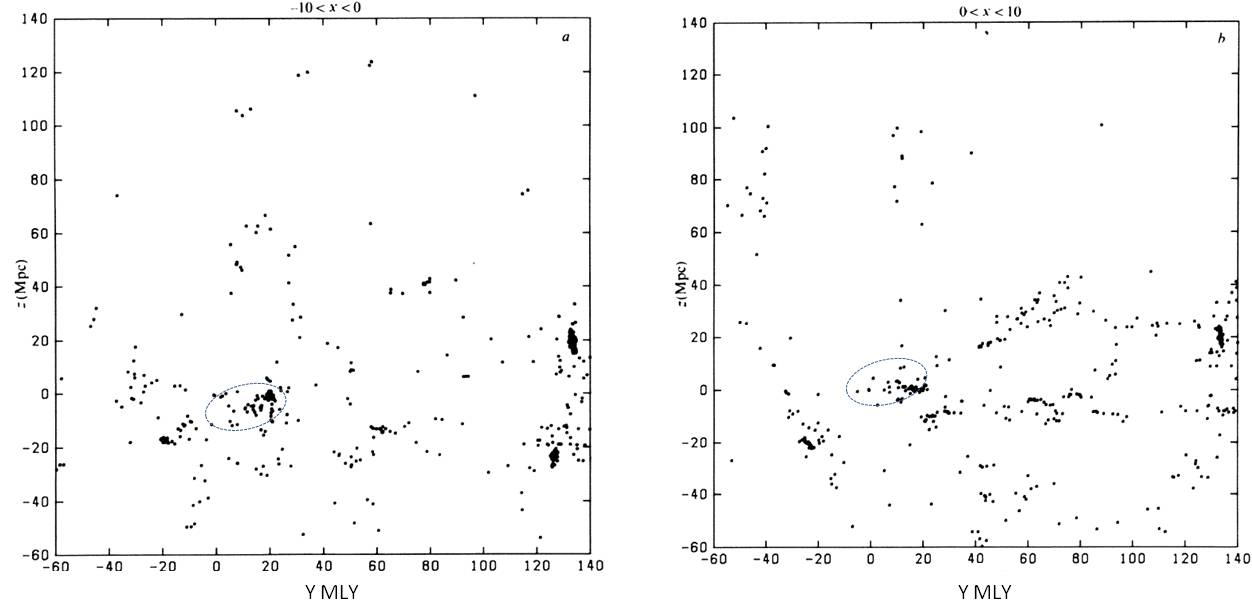}
\captionsetup{width=0.9\textwidth}
\caption{Giant Voids\cite{zeldovich}.   If the universe is a low density black hole, there should be large voids outside the boundary where no galaxies are present and the intra-galactic medium is of low density.   Plotting the voids around our Local Supercluster in the two horizontal slices provided above indicates that voids start showing up in about a 10 MPC radius and are up to 10 MPC in some directions but as little as 5 MPC.   This could argue for a elliptical shaped rotating black hole boundary with $R_{bh} = 5$ to $10 MPC (16-32.6LY)$.  These would be about the size of the partial local supercluster low density black holes.  Note: ellipse overlays added to figure void plots from  \cite{zeldovich}. }
\label{GiantVoid}
\end{center}
\end{figure}

Based on mass distribution analysis, the universe also includes giant voids in space between clusters, with up to 90\% of space being large empty voids.  The mass distribution for 200x200 MLY is shown in Figure \ref{GiantVoid} from \cite{zeldovich}.   The remaining 10\% contains mass in the form of clusters of connected galaxies in groups of about 5M PC radius.  These clusters  are possibly connected by strings of galaxies to other clusters, or separated by large voids in space.
 
\begin{figure}[H]
\begin{center} 
\includegraphics[width=.72\textwidth]{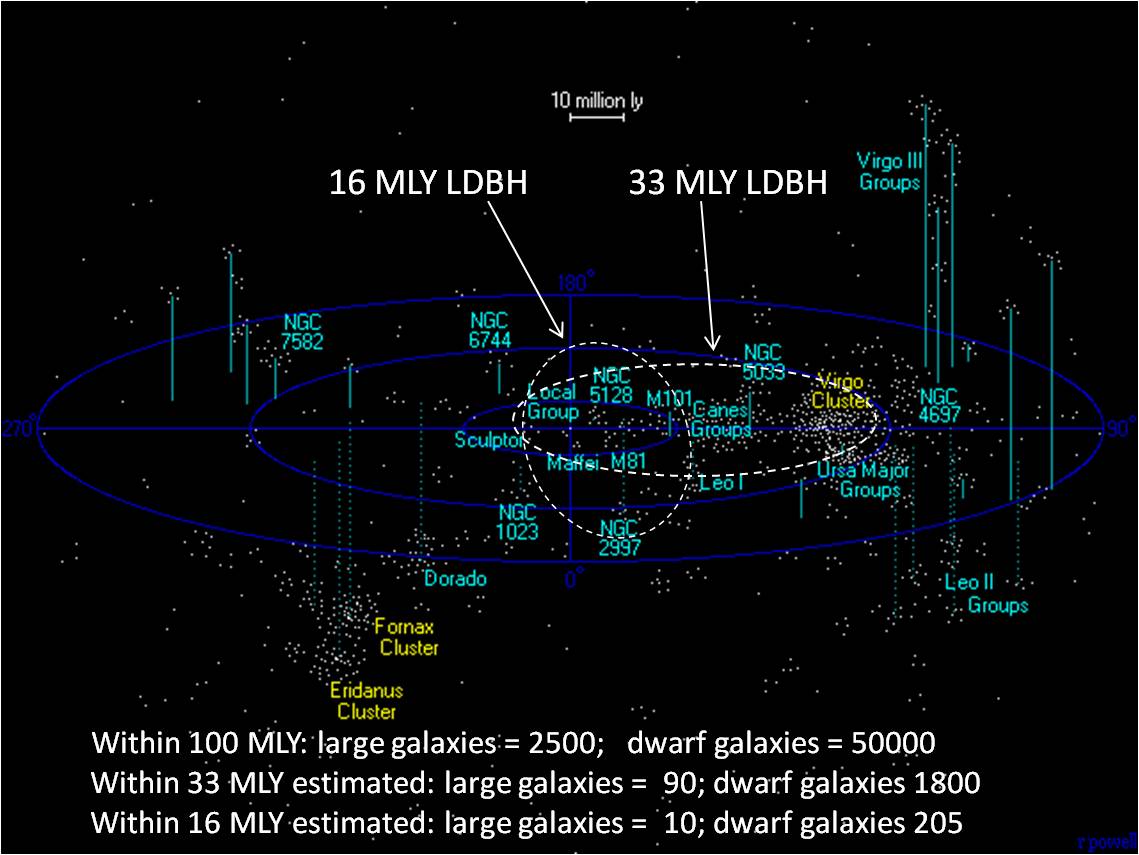}
\captionsetup{width=0.9\textwidth}
\caption{Local Supercluster Low Density Black Hole Consistent with Voids.  At that 33 M LY radius, the Fumax and Eridinus clusters would be illusions.  At the 16 M LY radius, the Virgo, Fumax, and Eridinus clusters would be illusions.  The number of real large galaxies in the 33 MLY sphere would drop to 90, and dwarf galaxies to 1800. The number of real large galaxies in the 16 MLY sphere would drop to 10, and dwarf galaxies to 205.  The inner local group could still be a localized low density black hole, and virgo could be another local density black holes; both within the larger quantized black hole.   Note: overlays added to base 3D plots from \cite{atlas1}.}
\label{GiantVoidU}
\end{center}
\end{figure}

\begin{figure}
        \centering
        \begin{subfigure}[b]{0.48\textwidth}
	\includegraphics[width=\textwidth]{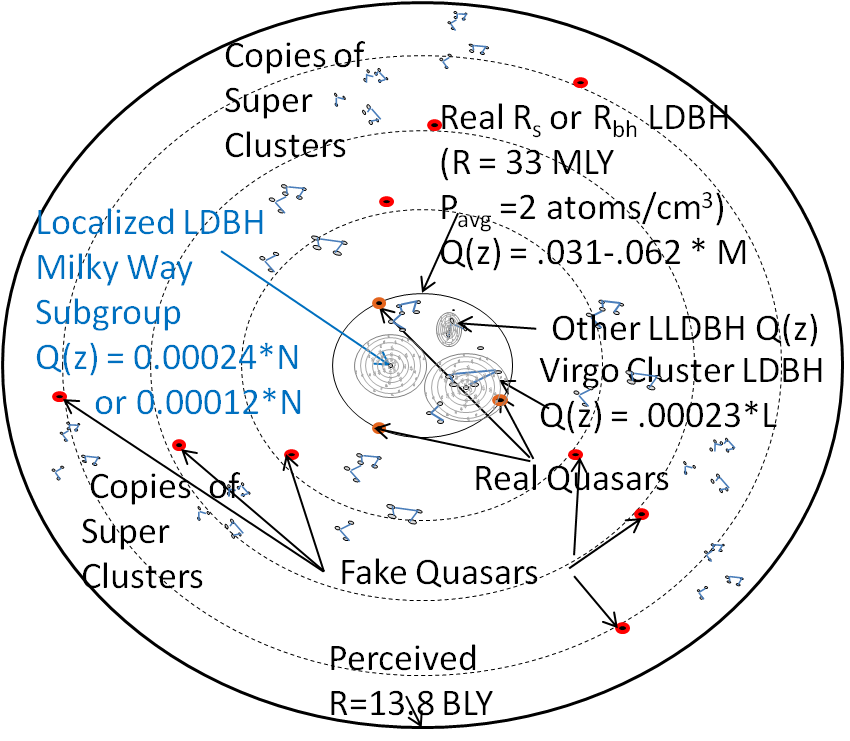}
	\caption{ 33 MLY universe}
        \end{subfigure}
	\qquad 
        \begin{subfigure}[b]{0.45\textwidth}
	\includegraphics[width=\textwidth]{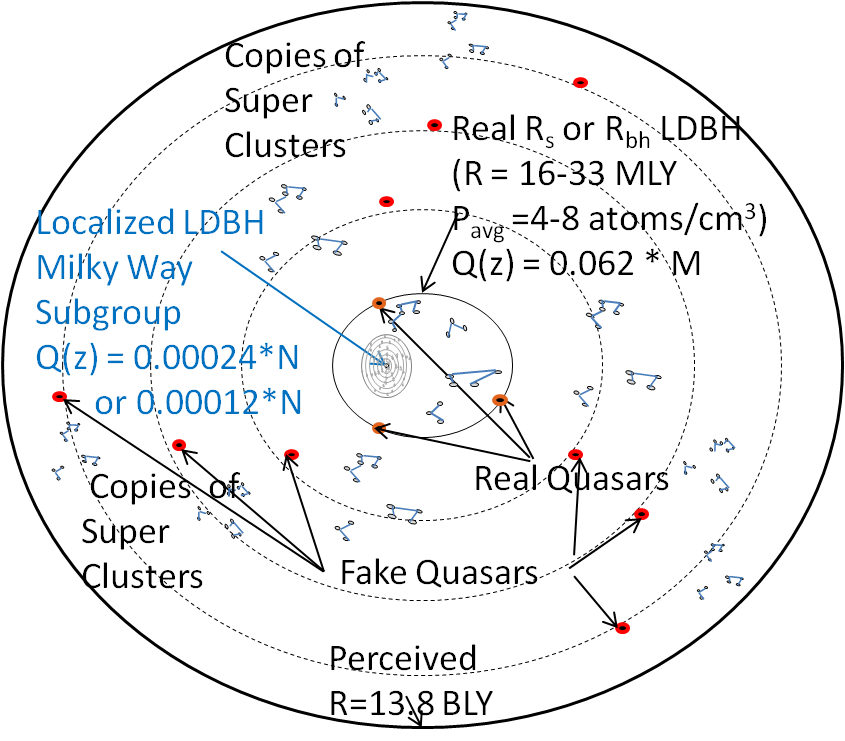}
	\caption{16 MLY Universe }
        \end{subfigure}
\captionsetup{width=0.9\textwidth}
\caption{Nested Quantized Universe with Supercluster Voids. This Nested Quantized Universe is sized to be consistent with the supercluster voids.  As presented here it is also a combination of the quantized quasar universe with a local group localized low density black hole causing smaller quantized reflections about the Milky Way Galaxy.  The outer radius would be spaced at 16-33 MLY to create quantized clusters and giant voids.  The quasar redshift could be accumulative intrinsic redshift or based on gravitational or transverse redshift that was larger in the past.  The higher gravitational redshift could be caused by more matter falling in to the black holes in the past closer to its gravity well, and be brighter as well.  The transverse redshift velocities could have been faster in the past and brighter because of they are running into more mass/sec, and are slowing due to the collapse.  Within the outer low density black hole, there could be one or more local low density black holes each with a different quantization value, based on their density and reflection radius.  In the 16 MLY universe, Virgo cluster would not be real and so the Milky Way quantized localized low density black hole would need to be slightly ellipsoid in shape to get the different quantization values in that direction. }
\label{Giant}
\end{figure}

If one were in a large low density black hole, although reflected galaxies and clusters look real, their light would essentially also produce a void as the light is being reflected through the vacuum of space outside the black hole boundary. Figure \ref{GiantVoid} has been overlaid with potential low density black hole boundaries around our galaxy.  The best match seems to be 5 M to 10 M PC (16 K to 33 M LY) radii.  These low density universes that would be produced are shown in Figures \ref{GiantVoidU} and \ref{Giant}.
 
\subsection{Quantization mechanisms}

The main reason for the quantization to occur is the gap or time that it takes the light to reflect through the void.   Most of the reflecting universes largely relied on intrinsic red shift.  The redshift function itself is not quantized, but the reflected mass just comes in waves at repeated interval radii.   
 
However, each redshift mechanism have additional secondary quantum mechanisms.
 \begin{itemize}
\item \ul{Quantum Gravity Redshift}:  This main quantization mechanism discussed above is different from quantum black holes, which may have finite legal values for mass and radii that perhaps correspond to harmonics of a gravity wave.   That said, quantum black holes could provide a secondary black hole quantization mechanism.
\item \ul{Transverse redshift}:  The quantized quasar universe did use transverse redshift to get linear redshift of distant galaxies out to 816 MLY.  At larger radii, the reflected light would appear to be moving even faster and thus may have incremental redshift.  In addition, the quasars could have been farther out and traveling faster in the past which would produce larger transverse redshifts before it collapsed to its current size.
\item \ul{Bremsstrahlung redshift}: Although Bremsstrahlung redshift is roughly linear with distance traveled due to scattering affects, it could be reduced somewhat as it passes through the void.  The void could still have gases due to solar winds at the end of the galaxy.  But this intra-galactic gas would have lower densities and may produce lower redshift in these regions.
\item \ul{Charge based redshift}:  Although the charge based redshift is roughly linear with distance traveled on average, there could be an absence  of charge in the void, or a large charge at the low density black hole boundary.  Thus, variations could occur.
\item \ul{Quasar redshift}:  Quasar redshift might have a second mechanism for producing its quantized redshift, such as its fundamental redshift changing over history due to the changes in collapsing material or changes in kinetic energy due to faster velocities in the past.  Additionally, since black holes are believed to be in the center of the quasars, black hole quantum effects may contribute to quasar quantization.
 \end{itemize}
Again, Hawking and other have tried to reproduce these quantized red shifts, but convinced themselves that they were not real or could be explained by other effects.  
\\[1ex]That said, as stated earlier, realize that the reflecting universes put forth in this paper do not require quantized redshift, but if quantized redshift is observed, the low density black hole reflection mechanism could be its source.  If quantized redshift are not observed, the more continuous reflections could be easily explained by the light being reflected more continuously.  That is, light does not only go straight out and in, but can easily travel in large elliptical orbits of arbitrary lengths.   The low density black holes are also not like stellar black holes with a single star in the center, but consist of many stars and galaxies at various depths.  The light from the lower depth would start to turn or be bent backwards starting at a much lower level.  Thus, crisp redshift quantization values may not be observed, but may only show up in statistical averages. Also, our measurement accuracies may not be good enough (yet) to tell.

\section{Conclusion}
This paper demonstrated the following related to low density black holes (LDBHs):
\begin{enumerate}
\item LDBHs could form at arbitrary average densities. ($R_s=\frac{.0013401}{\sqrt{P_{avg}}}$ LY, $R_{bh} = \frac{.0018952}{\sqrt{P_{avg}}}$ LY for $P_{avg}$ in $kg/m^3$). 
\item LDBHs could form with a 13.8 BLY radius if the universe is at its critical density of about 5 to 10 hydrogen atoms per cubic meter.
\item LDBHs could form with a 63MLY radius if the existing galactic densities are extended with a density equivalent to 1 hydrogen atom per 2 cubic cm.
\item LDBHs could form with 20, 6.3, or 2 MLY radius if dark matter is present or heavy dense inner or outer region averaging between 1 atom/$cm^3$ to 1 atom/$mm^3$.
\item One could be living in a low density black hole and not know it.
\begin{enumerate} \item G forces and tidal forces are imperceptible
				\item	Centripetal forces could keep the black hole from collapsing.
\end{enumerate}
\item We are not in an expanding universe: 1) if at critical density, would have encountered a finite black hole that would have stopped expansion when half the size, 2) if larger than the critical density, would have encountered a finite black hole (with additional mass) that would have stopped expansion when half the size,  3) if less than the critical density, would have encountered a black hole when the universe was smaller and denser.
\item The Big Bang probably did not happen.
\item The Hubble Doppler redshift interpretation is not valid and the 13.8 BLY creation date is no longer valid.
\item LDBHs can use spin's centripetal accelleration and/or charge to keep from collapsing.
\item LDBHs with density functions that fall off slower than $1/R^2$ can keep from collapsing with the appropriate speed curve.
\item We can be in a finite low density black hole that is rotating or collapsing. 
\item The Hubble redshift can no longer come from expansion, but now needs to come from transverse relativistic redshifts, gravitational redshifts, or one of the other space based intrinsic redshift mechanisms.
\item An infinite universe is possible, but will collapse into a disjoint low density black holes and escape the Olber’s paradox.
\item A non-reflecting universe could match the current expanding universes in size (e.g. 13.8 BLY) and mass, but formed inside a rotating Schwarzschild or finite low density black hole.
\item Reflecting universes could be much smaller and may be at 600, 66, 20, 6, or 1.5 M Light Years Radii.  (This would only require densities up to 1 hydrogen atom per cubic mm).  In these cases, the entire universe could only be as little as 1.5 M Light Years.
\item Andromeda and Triangulum galaxies are probably real and not just reflections of the Milky Way galaxy.
\item Quantized redshift may warrant a second look as LDBHs provide a new potential source via reflections.
\item If we are within a localized black hole, additional real external galaxies could exist and be seen from our black hole.
\item The only non-black hole universe would be a collapsing universe below critical mass.  But it could still contain localized black holes at the higher densities in the middle, would appear like a black hole, and will eventually collapse into a black hole.
\item Thus, we are probably in a black hole, or will be eventually.
\end{enumerate}
If the research from this paper does lead to a paradigm shift, this will be an exciting time to be a cosmologist, with everything a new.  However, most of what we know would not really change.  However, I think a little humility should be in order for a decade or two, especially if one can’t even trust what their eyes and telescopes are seeing.

\section{Future investigations}
Future directions are to:
\begin{itemize}
\item Vet the above theory with the scientific community and seek supporting or non-supporting evidence. 
\item Redo calculations using relativistic equations and KERR-NEWMANN rotating black hole equations \cite{kerr, kerrNewman}.
\item Re-analyze galaxy and quasar redshift data for quantized redshift 
\item Continue mass distribution analysis using 3D data for all galaxies in the local cluster.
\item Investigate reflecting potential within the Milky Way galaxy.
\end{itemize}

\section*{Acknowledgements}
Although I am an amateur cosmologist, I did take a few graduate classes as ASU to help me form a foundation in order to just begin to approach this problem.  I wanted to explore this idea on my own before getting the professionals involved.  I trust now they know the reason for my silly and sometimes unreasonable questions.     

Specifically, I would like to thank Astronomy Professor Aannestad who put up with me in several classes and Physics professor Nicole Herbots-Culbertson for teaching me relativity theory.  You gave me the tools to pursue this theory.   Without your knowledge, this would just be a nagging question on my “To Do” list that never got done.  And then, I would have learned nothing.  Thank you for all your help.

I would also like to recognize the Hubble Space Telescope, the Chandra Observatory, the Digital Sky Survey, NASA, Wikipedia, Google Search, Google Scholar, arXiv, and the other internet publishing services whose information was indispensible for this endeavor.  I would also be remiss if I did not mention Newton, Einstein, Hawking, Hubble, Schwarzschild, Reissner, Nordström, Kerr, and Newman.  

I would also like to thank my daughter Angie for helping me formalize my disproof of the Big Bang Theory (Theorem 1),  checking my derivations and converting my many equations into Latex,  and making my mathematical notations consistent.  She also formatted all my references, and managed my references and figures numbers throughout the document.  

I would also like to thank any reader that made it this far.  This was not an easy read, nor a simple subject.  The only thing worse than taking twenty five years to write a paper, is to do so and not have anyone willing to read it.  You just saved me from this most undesirable fate. I hope that you learned as much from this paper as I did writing it.

\newpage
\appendix
\section {Appendix Background on Black Holes}
\begin{figure}[H]
\begin{center}
\includegraphics[width=.9\textwidth]{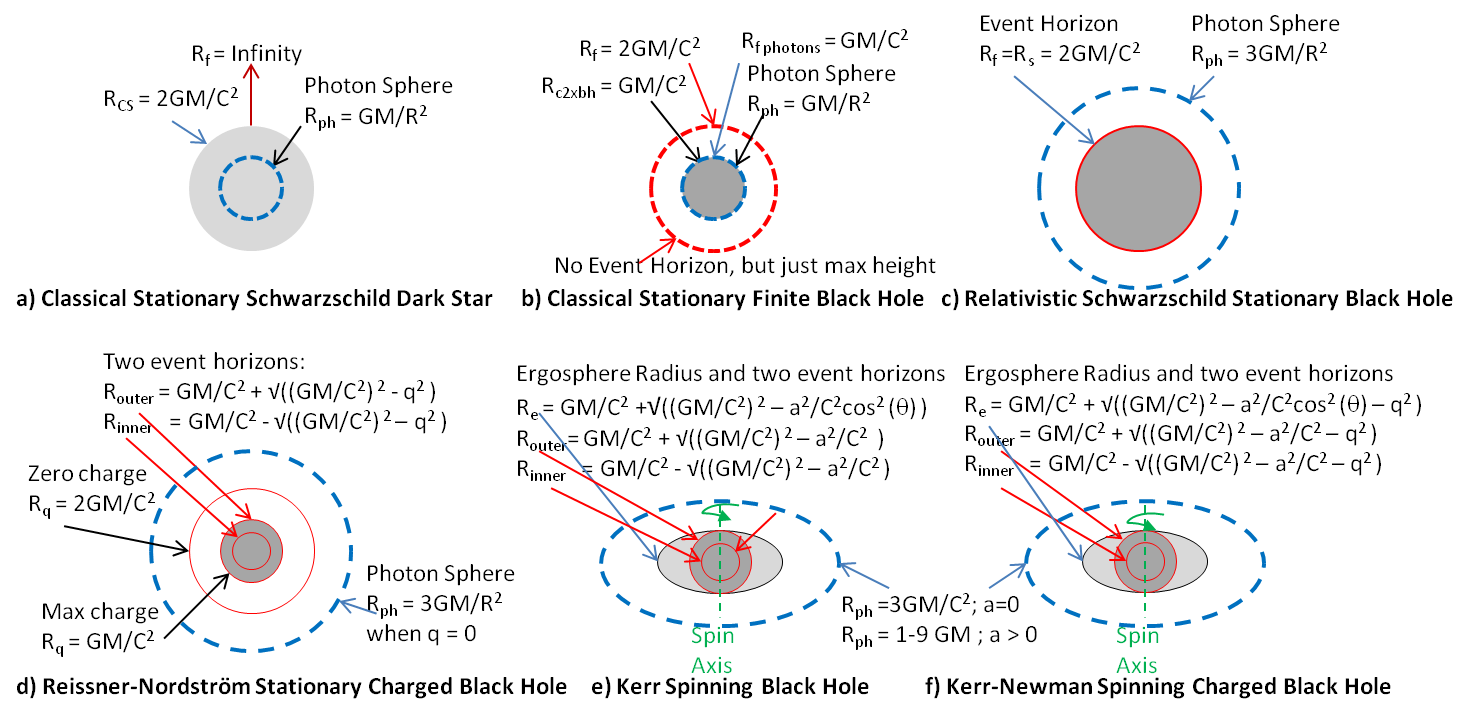}
\captionsetup{width=0.9\textwidth}
\caption{Black Hole Radii for Different Black Hole Types. All black holes have a radius between $GM/C^2$ and $2GM/C^2$\cite{kerr},\cite{KerrBH1},\cite{KerrBH2},\cite{kerrNewman},\cite{Zienikov},\cite{ChargedPH},\cite{Romero}, and \cite{QuantumBH}.  The classical Schwarzschild black hole most often derived in text books (was originally called a dark star by Michell/Laplace 1796) can keep light from escaping, but light can be seen at infinity ($R_f=\infty$).   A classical finite (twice) height black hole with final height $R_f=2GM/C^2$ has a radius $R_{c2xbh}=GM/C^2$.  All the remaining black holes would also not let light escape beyond a maximum height of $R_f = 2GM/C^2$.  All these black holes have photon sphere radii, $R_{ph}$ that bends light in circular orbits $R_{ph} = GM/C^2$ to $3GM/C^2$ (and larger) for spinning black holes.  All the stationary black holes (on the top) are expected to collapse into a singularity at the center. However, the spinning and charged black holes (on the bottom) can keep themselves from collapsing using spin or a rupulsive charge force.  A spinning black hole is larger at its equator (forming an ergosphere) due to centrifugal forces.  Spinning black holes also drags space around at its event horizon.   If one incorporates quantum mechanics into the black hole equations, one would get identical answers but some of the mass and radii values may need to come in discrete legal values, possibly at resonance frequencies with the gravitational waves.}
\label{BHRadii2}
\end{center}
\end{figure}

There are many, slightly different, definitions of black holes (Classical Schwarzschild, Classical finite height, relativistic Schwarzschild, Reissner-Nordström charged, Kerr Spinning (or rotating), and Kerr-Newman spinning (or rotating) charged, and quantum black holes).  Some of these definitions are radically different but use identical names and abbreviations.  Thus, it is a good idea to define terms to achieve a common language and understanding. 
A survey of black hole equations is shown in Figure \ref{BHRadii2} and Table \ref{TableBlackHoleRadiiEquations}, and are discussed in the subsections below.

\subsection{Classical Schwarzschild Black Hole Radius}
The "Classical Schwarzschild" black hole equation was originally derived from escape velocity, by John Michell in 1784 andPierre-Simon Laplace in 1796, and were originally referred to as Dark Stars \cite{Romero}. 
Escape velocity ($V_{esc}$) is given below:
\begin{equation}
V_{esc} = \left(\frac{2GM}{R} \right)^\frac{1}{2}
\end{equation}
where $G$ is the Gravitational constant and $M$ is the mass within radius $R$ \cite{zeilikP10}. Using \textsl{classical} physics, the classical Schwarzschild Black Hole radius ($R_{cs}$) is defined when $V_{esc}$ is set to the speed of light $C$ \cite{schwarzschild}, 
\begin{align}
C &= \left(\frac{2GM}{R_{cs}} \right)^\frac{1}{2} \\
R_{cs}    &=   \frac{2GM}{C^2}.                
\end{align}
Escape velocity is the velocity($V_i$) that a \textsl{ballistic} projectile needs at a given initial radius $R_i$ to just escape the gravitation force of the mass inside its radius.  When the escape velocity is the speed of light, C, light will just escape.  Increasing the mass slightly, (e.g. adding 1 gram) will prevent even light from escaping.

The escape velocity equation is derived from the initial and final kinetic energy ($KE_i, KE_f$) and the initial and final potential energy ($PE_i, PE_f$) of a projectile with mass $m$, 
\begin{align}
KE_f  + PE_f &=  KE_i + PE_i \label{energyEquation} \\
KE_f -KE_i &= PE_i - PE_f \\
\frac{1}{2}mV_f^2  -\frac{1}{2}mV_i^2 &= (-\frac{mGM}{R_i})-(-\frac{mGM}{R_f}) \\
\frac{1}{2}V_f^2  -\frac{1}{2}V_i^2 &= \frac{GM}{R_f} - \frac{GM}{R_i}  \label{finalEnergyEq}
\end{align}
Normally $R_f$ (the final height of the projectile) is chosen at an infinite height such that $PE_f = 0$ and $V_f$ (the final velocity) is chosen so infinitely small such that $KE_f=0$:
\begin{align}
0- \frac{1}{2}V_i^2 &= 0 - \frac{GM}{R_i} \\
\frac{1}{2}V_i^2 &= \frac{GM}{R_i} \\
V_i &= \left(\frac{2GM}{R_i}\right)^\frac{1}{2}
\end{align}
Here $V_i=V_{esc}$ since the projectile is traveling to an infinite height and is not returning. The derivation illustrates the assumption that $R_f=\infty$. Thus, this classical Schwarzschild definition of a black hole enables light or any mass to travel to infinity before it falls (or is pulled) back downward.  

Since these derived black hole equations were used to formally disprove the big bang theory, it is important to identify the simplest based axioms.  One can derive the Kinetic and Potential energy (Equation \ref{finalEnergyEq}) simply from Newton's laws.
\begin{align}
F&=ma \qquad \quad \text{(Newton's Law of Motion\cite{newtonM})}\\
F&=\frac{mGM}{r^2}  \qquad \text{(Newton's Law of Gravity \cite{newtonG})}\\
ma&=\frac{mGM}{r^2} \\
a&=\frac{GM}{r^2} \label{removemEQ}
\end{align}
Applying calculus yields the following.  
\begin{align}
a&=\frac{dV}{dt}= \frac{dV}{dr}\frac{dr}{dt} = \frac{dV}{dr}V = \frac{GM}{r^2}
\end{align}
Integration yields, 
\begin{align}
\int_{V_i}^{V_f} v  dv &=\int_{R_i}^{R_f} -\frac{GM}{r^2}dr \\
\frac{v^2}{2}  \bigg|_{V_i}^{V_f} &=\frac{GM}{r} \bigg|_{R_i}^{R_f} \\
\frac{V_f^2}{2}-\frac{V_i^2}{2} &=\frac{GM}{R_f}-\frac{GM}{R_i}
\end{align}

Black holes can also bend light up to 360 degrees and allow light to orbit at a radius $R_{ph}$ call the photon sphere.   Computing the photon sphere radius $R_{ph}$ where light would orbit this black hole can be obtained by setting the centripetal force $(F_c=\frac{mV^2}{R})$ equal to the gravitational force $(F_g = \frac{GmM}{R^2})$ and setting V = C:
\begin{align}
F_c &= F_g\\
\frac{mV^2}{R_{ph}} &= \frac{GmM}{R_{ph}^2}\\
R_{ph} &= \frac{GM}{C^2}
\end{align}
Since $R_{ph}$ is less than $R_{cs}$, this classical Schwarzschild black hole does not have a photon sphere, or more precisely, the mass would have to collapse below $R_{ph} = GM/R^2$ before it could use this photon sphere.  The classical Schwarzschild "Dark Star" characteristics are shown in Figure \ref{BHRadii2}a and Table \ref{TableBlackHoleRadiiEquations}.

\subsection{Classical Finite Height Black Hole Radii}
One can define a ``finite height" \textbf{Classical Black Hole}
by choosing $R_f$ to be some finite value (e.g. $R_f=2R_i$), and $V_f = 0$, and computing a new inner radius $R_i$.  
Plugging these values into Equation \ref{finalEnergyEq} yields the following. 
\begin{align}
0-\frac{1}{2}V_i^2 &= \frac{GM}{R_f}-\frac{GM}{R_i}  \\
V_i^2 &=\frac{2GM}{R_i}-\frac{2GM}{R_f}
\end{align}
Letting $V_i=C$ and $R_f=2R_i$ yields the following.
\begin{align}
C^2 &=\frac{2GM}{R_{i}}-\frac{2GM}{2R_i} \\
&=\frac{GM}{R_{i}} 
\end{align}
Hence 
\begin{align}
R_{i}&=\frac{GM}{C^2} 
\end{align}
$R_i$ is the radius of the classical ``finite height" black hole with $R_f=2R_i$. This $R_i$ is further noted as $R_{c2xbh}$. 
\begin{align}
R_{c2xbh}&=\frac{GM}{C^2}\\
R_{f2}&=2R_{c2xbh} 
\end{align}
These calculations could be redone using a finite height $R_f = n*Ri$ instead of $2*Ri$ to obtain an generic formula.
\begin{align}
R_{cnxbh} &=\frac{n-1}{n}*\frac{2GM}{C^2}\\
R_(fn) &=n*R_{cNxbh} 
\end{align}
A finite (10x) heigt black hole would have radius $R_{C10xbh}=9/10*2GM/C^2 = 1.8GM/C^2$ and $R_{f10} = 18GM/C^2$.

Computing where light would orbit these black holes can be obtained by setting the centripetal force $(F_c=\frac{mV^2}{R})$ to the gravitational force $(F_g = \frac{GmM}{R^2})$ and setting V to C:
\begin{align}
F_c &= F_g\\
\frac{V^2m}{R_{ph}} &= \frac{GMm}{R_{ph}^2}\\
R_{ph} &= \frac{GM}{C^2}
\end{align}
This classical finite black hole has a photon sphere at $R_{ph} = GM/C^2$ where light is bent 360 degrees.
The classical finite height black hole characteristics are shown in Figure \ref{BHRadii2}b and Table \ref{TableBlackHoleRadiiEquations}.

\vspace{3 mm}
A ballistic projectile \underline{mass} traveling upward from the surface of a finite black hole with at an initial velocity of $C$, will begin to slow down and stop (e.g. speed = 0) at $R_f$.  This mass will then begin to fall back in from $R_f$ and gain speed inward, increase in speed, and reach the black hole boundary traveling at the speed of $C$.  

Light traveling upwards doesn't slow down but decreases its frequency v, and lengthens its wavelength $\lambda$ to lose energy.  Some object to applying the mass based equations for light, since the rest mass of a photon is zero.  Specifically,  in deriving equations \ref{finalEnergyEq} and \ref{removemEQ}, dividing by mass (which may be zero for light) is frowned upon in some circles.   However, although the rest mass of a photon is zero, when it is moving at light speed, it is multiplied by $\gamma=1/\sqrt(1-V^2/C^2)$ \cite{Freedman} which at light speed would be infinity.  Infinity times zero is not necessarily zero.  In any event, the equations definitely are applicable to all non-light mass and the limit as m approaches zero is applicable for infinitely small masses, perhaps including light.

That said, some texts prefer using the  energy of a photon equation which is equal to Plank's constant times its frequency, v \cite{Beiser}:
\begin{align}
E &= h*v 
\end{align}
Although photons lack rest mass, they behave as though they have gratational mass with $m =\frac{p}{v} = \frac{hv}{C^2}$ \cite{Beiser}. Thus Gravitation potential energy,  
\begin{align}
PE_{photon} &=\frac{mGM}{R} \\
PE_{photon} &=\frac{hvGM}{RC^2} 
\end{align}
And the energy equation becomes:
\begin{align}
hv{_f} -\frac{hv_fGM}{R_fC^2} &= hv{_i} -\frac{hv_iGM}{R_iC^2}
\end{align}
A classical black hole that eliminates photons would be at an initial Radius such that the final frequency $v_f$ is zero:
\begin{align}
0 - 0 &= hv{_i} -\frac{hv_iGM}{R_iC^2} \\
0 &= 1 -\frac{GM}{R_iC^2} \\
R_i &= \frac{GM}{C^2} \\
R_{nophotons} &= \frac{GM}{C^2}
\end{align}
Thus, using the energy of a photon, the classical black hole radius is the same as the boundary and previously computed photon sphere.   Thus, this is a classical finite (twice height) black hole equation that appears to work for mass and light that would meet most definitions of a black hole.  However, there is no true event horizon, but just a finite maximum height, $R_f=2GM/C^2$.

\subsection{Relativistic Schwarzschild Black Hole Radius}
Relativistic physics theorize that gravity creates a black hole by curving space due to general relativistic effects of gravity.  Schwarzschild solved the general relativity equations for the stationary black hole and obtained a black hole radius identical to the classical Schwarzschild radius.
The Relativistic computations involve tensor calculus and are beyond the scope of this paper, but are available in the reference papers \cite{Rybczyk}.  
\begin{align}
R_{s} &= \frac{2GM}{C^2} \text{  and} \\
R_{f} &= \frac{2GM}{C^2} \text{  and} \\
R_{ph} &= \frac{3GM}{C^2} 
\end{align}
Note that although the radius is identical to the classical Schwarzschild radius, it was obtained using relativistic physics and the cause of the black hole is the curvature of space from the gravitational field.  Since space is curved completely at the black hole boundary, nothing can pass back outside the black hole event horizon boundary.  That is, ($R_f = R_s$).  The photon sphere of the relativistic black hole, where light will orbit outside the black hole, is one and a half times the radius ($R_{ph} =1.5R_{s}$).The relativistic Schwarzschild black hole characteristics are shown in Figures \ref{BHRadii2}c and Table \ref{TableBlackHoleRadiiEquations}.

\subsection{Relativistic Stationary Charged Black Hole Radius}
If one incorporates electric charge into the stationary relativity equations one would get slightly different equations known as the Reissner-Nordström stationary charged black hole equation \cite{Zienikov}\cite{Romero}.  This black hole has a repulsive charge, q, that helps it from collapsing. 
\begin{align}
R_{q} &= \frac{GM}{C^2} + \sqrt((\frac{GM}{C^2})^2 -q^2)
\end{align}
where the charge of the black hole is $q=\frac{GQ^2}{4*\pi*\epsilon_oC^4}$, with total charge Q, and $\epsilon_o = 8.854*10^{-12}$ farads/m \cite{Romero};

\vspace{3 mm}
There is an outer and inner event horizon:
\begin{align}
R_{outer} &= \frac{GM}{C^2} + \sqrt((\frac{GM}{C^2})^2 -q^2)  \\
R_{inner} &= \frac{GM}{C^2} - \sqrt((\frac{GM}{C^2})^2 -q^2)
\end{align}
and
\begin{align}
R_{f} &=  R_{q} \\
R_{ph} &= \frac{3GM}{C^2} \text{ when q=0} \cite{ChargedPH}
\end{align}

The stationary charged black hole characteristics are shown in Figure \ref{BHRadii2}d and Table \ref{TableBlackHoleRadiiEquations}.    

When  $q=0$,  $R_{q} = R_s = \frac{2GM}{C^2}$.

When $q$ \textgreater $0$, $R_q = \frac{GM}{C^2}$ to $\frac{2GM}{C^2}$

When q is at the maximum value, $R_{outer} = R_{inner} =  \frac{GM}{C^2}$.

\subsection{Relativistic Spinning (or Rotating) Kerr Black Hole Radius}
If one incorporates rotation with general relativity equations one would get slightly different equations known as the \textbf{Spinning (or Rotating) Kerr Black Hole}  \cite{kerr},\cite{KerrBH1}, \cite{KerrBH1}, \cite{Zienikov}, and \cite{Romero}. 
\begin{align}
R_{spin} &= \frac{GM}{C^2} + \sqrt((\frac{GM}{C^2})^2 - (a^2/C^2*cos(\theta)^2))   
\end{align}
where $a = \frac{JC}{GM}^2$  and J is the angular momentum \cite{Romero}

\vspace{3 mm}
There is an outer and inner event horizon:
\begin{align}
R_{out} &= \frac{GM}{C^2} + \sqrt((\frac{GM}{C^2})^2 - a^2/C^2) \\
R_{in} &= \frac{GM}{C^2} - \sqrt((\frac{GM}{C^2})^2 - a^2/C^2)
\end{align}
and
\begin{align}
R_{f} &=  R_{spin} []\\
R_{ph} &= \frac{3GM}{C^2} \text{ when a = 0} \cite{Zienikov}\cite{Romero}.
\end{align}
The spinning black hole characteristics are shown in Figure \ref{BHRadii2}e and Table \ref{TableBlackHoleRadiiEquations}.    

When a = 0, the black hole is not spinning and the radius becomes $R_{spin} = R_s = \frac{2GM}{C^2}$.

When $a$ \textgreater $0$, and $\theta = 0$ or $180$ degrees (at poles), the radius becomes  $R_{spin} = \frac{GM}{C^2}$.

When $a$ \textgreater $0$, where $\theta = 90$ or $270$ degrees (at equator), the radius becomes $R_{spin} = \frac{2GM}{C^2}$ due to $cos(\theta)=0$.  

When a is the maximum value, $R_e = R_{out} = R_{in} =  \frac{GM}{C^2}$.  The elongated radius, $R_e$ is called the ergosphere.

Light will orbit at a photon sphere $R_{ph}=\frac{GM}{C^2}$ in the same direction as the spin, and
 $R_{ph}=\frac{4GM}{C^2}$ to $\frac{9 GM}{C^2}$  in the opposite direction as the spin \cite{Zienikov}\cite{Romero}.

\vspace{3 mm}
Four additional distinguishing features of the spinning (or rotating) Kerr black hole are:

 1) The gravitational force is used as a centripetal force to can keep the black hole from collapsing,

 2) The centrifugal force can cause the black hole to elongate and become more like an elliptical sphere, and 

 3) There are two event horizons: outer and inner

 4) The black hole will drag space around as it rotates, which is referred to as frame dragging.

\subsection{Relativistic Spinning Charged Black Hole Radius}

If one incorporates electric charge into the spinning relativity equations one would get slightly different equations known as a Kerr-Newman black hole \cite{kerrNewman},\cite{Zienikov}, and \cite{Romero}.  This black hole is spinning and also has a repulsive charge, q, that helps it from collapsing. 
\begin{align}
R_{qspin} &= \frac{GM}{C^2} + \sqrt((\frac{GM}{C^2})^2 - (a^2/C^2*cos(\theta)^2) -q^2)  
\end{align}
where  $a = \frac{JC}{GM^2}$ with the angular momentum J

\vspace{3 mm}
There is an outer and inner event horizon:
\begin{align}
R_{out} &= \frac{GM}{C^2} + \sqrt((\frac{GM}{C^2})^2 - a^2/C^2 -q^2)  \\
R_{in} &= \frac{GM}{C^2} - \sqrt((\frac{GM}{C^2})^2 - a^2/C^2 -q^2)
\end{align}
and
\begin{align}
R_{f} &=  R_{qspin} \\
R_{ph} &= \frac{3GM}{C^2} \text{ when a=0 and q=0} 
\end{align}

Light will orbit at a photon sphere $R_{ph}=\frac{GM}{C^2}$ in the same direction as the spin, and
 $R_{ph}=\frac{4 GM}{C^2}$ to  $\frac{9 GM}{C^2}$ in the opposite direction as the spin \cite{Zienikov}\cite{Romero}.

The spinning charged black hole characteristics are shown in Figure \ref{BHRadii2}f and Table \ref{TableBlackHoleRadiiEquations}.    

When $a = 0$ and $q=0$,  $R_{qspin} = R_s = \frac{2GM}{C^2}$.

When $a = 0$ and$q$ \textgreater $0$, $R_{qspin} = R_q = \frac{GM}{C^2}$ to $\frac{2GM}{C^2}$

When $a$ \textgreater $0$ and $q$ \textgreater $0$, and $\theta = 0$ or $180$ degrees (at poles), the radius also becomes  $R_{qspin} = \frac{GM}{C^2}$.

When $a$ \textgreater $0$ and $q$ \textgreater $0$, where $\theta = 90$ or $270$ degrees (at equator), the radius becomes $R_{qspin} = \frac{2GM}{C^2}$.  

When a and q combination is at the maximum value, $R_e = R_{out} = R_{in} =  \frac{GM}{C^2}$.  

\subsection{Quantum Black Hole}
If one incorporates quantum mechanics into the black hole equations, one would get similar answers but some of the speeds and radii values may need to come in discrete legal values, possibly at resonance frequencies with the gravitational waves \cite{QuantumBH}. Thus, quantum black holes would have the same radii ranges as defined in Table \ref{TableBlackHoleRadiiEquations}, but just come at discrete legal values.  Further analysis of Quantum Black Holes is beyond the scope of this paper. 

\subsection{Black Hole Summary}
Although each black hole described above are computed differently under various conditions, their black hole radii are all between $GM/C^2$ and $2GM/C^2$; and their maximum final height Rf is less than or equal to $2GM/C^2$.  They each have a photon sphere between $GM/C^2$ and $3GM/C^2$ to $9GM/C^2$ where they can bend light up to 360 degrees.
Throughout the remainder of the text, Rs will refer to the Relativistic Schwarzschild radius and Rbh will used to represent all the black holes at $GM/C^2$ radius, but largely targeting the spinning or charged black holes.
\begin{table}
\centering
\caption{Black Hole Radii Equations}
\begin{center}
\bigskip
\noindent\makebox[\textwidth]{%
\begin{tabular}{| l | c | c | c | c | }
  \hline  
\pbox{10cm}{$Black$ Hole \\Description} & \pbox{10cm}{Event Horizon Radius \\(or Ergosphere Radius)} & \pbox{10cm}{Maximum Height \\$R_f$} & \pbox{10cm}{Photon Sphere\\$R_{ph}$} \\
\hline 
Classical Schwarzschild & $R_{cs} = 2GM/C^2$ &	$R_f=Infinity$ &	$R_{ph} = GM/C^2$\\ \hline 
Classical finite (twice) height & $R_{c2xbh} = GM/C^2$ & $R_{f} = 2GM/C^2$ & $R_{ph} = GM/C^2$\\ \hline 
Classical finite (n times) height & $R_{cnxbh} = (n-1)/n*2GM/C^2$ & $R_{f} = n*R_{cnxbh}$ & $R_{ph} = GM/C^2$\\ \hline 
Relativistic Schwarzschild &$R_{s} = 2GM/C^2$&$R_{f} = 2GM/C^2$&	$R_{ph} = 3GM/C^2$\\ \hline
Relativistic Charged &$R_{q} = GM/C^2 +/- \sqrt((GM/C^2)^2 -q^2)$&$R_{f} = 2GM/C^2$&	$R_{ph} = 3GM/C^2$\\ \hline
Relativistic Spinning Kerr &$R_{spin} = GM/C^2 + \sqrt((GM/C^2)^2 - a^2/C^2*cos^2(\theta))$&$R_{f} = 2GM/C^2$&	$R_{ph} = 1-9GM/C^2$\\ \hline
Relativistic Spinning Charged &$R_{spinq} = GM/C^2 + \sqrt((GM/C^2)^2 - a^2/C^2*cos^2(\theta) -q^2)$&$R_{f} = 2GM/C^2$&	$R_{ph} = 1-9GM/C^2$\\ \hline
\end{tabular}} 
\bigskip
\end{center}
\label{TableBlackHoleRadiiEquations}
\end{table}

\appendix
\section*{Appendix B}
\subsection*{Commonly Used Symbols }
Commonly used acronyms, symbols, and constants used throughout this paper are shown in Table \ref{tablesym}.
\begin{table}
\caption{Commonly Used Symbols }
\vspace{-.5cm}
\begin{center}
\begin{tabular}{| l | l |}
  \hline  
a&	acceleration in meters/sec$^2$\\
AU&	The average orbital radius of the earth around the sun = $1.496 *10^{11}$ meters\\
B SM&	Billion solar masses = 1,000,000,000 solar masses\\
BLY&	Billion Light Years = 1,000,000,000 Light Years\\
Bm&	Billion meters = 1,000,000,000 meters\\
C&	Speed of light (300,000,000 meters/sec)\\
Da&	Deflection angle, the angle that light will bend traveling past an object due to its gravity.\\
dGs&	Tidal force difference in Gs, $dGs = 2GMdr/R^3$ where dr is assume to be the $R_{earth}$\\
$F_{grav}$&	Gravitational Force = $GmM/R^2$ in Newtons (meters/sec$^2$* kg)\\
G&	Gravitational Constant $6.67384 *10^11N(m/kg)^2$\\
Gs&	Gravitation Acceleration, $Gs = GM/R^2/9.8$  in earth G units equal to 9.8 meters/sec$^2$\\
H&	Plank's constant $4.135 x 10^{-15}$ eV-sec = $6.625 x 10^{-27}$ erg-sec\\
$H_{0}$&	 Hubble constant (75 km/MPC = 23 km/MLY)\\
KE&	Kinetic Energy\\
kg&	Kilogram\\
KLY&	Thousand Light Years\\
km&	Kilometer = 1000 meters\\
$\lambda$&	 Wavelength\\
LDBH&	Low Density Black Hole\\
LY&	Light Year = 3e8*3600*24*365.256 meters\\
$M$&	Mass (kg)\\
$m$&	meter\\
M SM&	Million solar masses = 1,000,000 solar masses\\
$M(R )$&	Total Binding Mass with radius R\\
$m^3$&	meters cubed = volume 1 meter x 1 meter x 1 meter\\
$M_{critical}$&	Total mass of the universe that consists of the critical density.\\
MLY&	Million Light Years\\
$P$ &	Period usually in years, but sometimes seconds\\
$P(R)$&	Density at radius $R$ in $kg/m^3$\\
$P_{avg}$&	Average density ($kg/m^3$)\\
$P_{critical}$&	Critical Density of the Universe $9.47*10^{-27} kg/m^3$  (5.67 hydrogen atoms/$m^3$)\\
PE&	Potential Energy\\
$R_{bh}$&	Relativistic Finite height (1.5x) black hole radius, $R_{bh}=GM/C^2$\\
$R_{c2xbh}$&	Classical finite height (twice height) black hole radius, $R_{c2xbh} = GM/C^2$\\
$R_{current}$&	Current radius of the universe (currently estimated at 13.8 BLY) \\
$R_{earth}$&	Radius of the earth = 6353 kilometers\\
$R_f$&	Final Radius \\
$R_i$&	Initial Radius \\
$R_{min}$&	Radius of nearest approach of an elliptical orbit to its focus point.\\
$R_s$&	Schwarzschild Radius infinite height black hole, $R_s = 2GM/C^2$\\
$R_{source}$&	Distance between the emitting source and the center of the black hole.\\
SM&	 Solar Mass = $1.99*10^{30}$ kg\\
TLY&	Trillion Light Years = Million Million Light Years \\
$V$&	Velocity in meters/sec\\
$V(R)$&	Velocity at orbital radius $R$.\\
$V_{esc}$&	Escape velocity: $V_{esc} = (2GM/R)^{0.5}$ (needed to escape the gravitation force at radius $R$)\\
$Vol$&	Volume (usually of an assumed sphere = $\frac{4 \pi}{3}R^3$;  or disk = $\pi R^2$*height; or cube = $s^3$)\\
$V_s$&	Velocity at the surface\\
$z$&	Redshift ($z$) is the relative difference between observed wavelength and emitted wavelength\\
$z_d$&	Relativistic Doppler redshift due to Doppler shift of light moving to/from the observer.\\
$z_g$&	Relativistic gravitational shift due to time dilation in a gravity well.\\
$z_{Linear}$&	Linear redshift observable at non-relativistic speeds; $z_{Linear} = .000076 * d$ (with d in MLY)\\
$z_t$&	Relativistic Transverse redshift due to time dilation of light moving transversely to the observer\\
$z_{TL}$&	Linear transverse redshift approximate \\
$\epsilon$&	eccentricity of an ellipse\\
\hline
\end{tabular}
\end{center}
\label{tablesym}
\end{table}
\newpage
\section*{}
\bibliographystyle{ieeetr}
\bibliography{LowDensityBlackHoles}
\end{document}